%% file: paper.tex
\documentclass[aps,prd,twocolumn,superscriptaddress,groupedaddress,amsmath,amssymb,nofootinbib,preprintnumbers]{revtex4}

\usepackage{graphicx}  
\usepackage[mathlines]{lineno}
\usepackage[english]{babel} 
\usepackage{xcolor}
\usepackage{ulem}
\usepackage{dcolumn}   
\usepackage{bm}        
\usepackage{amssymb}   
\usepackage{amsbsy}   
\usepackage{comment}
\usepackage{multirow}
\usepackage{diagbox}
\usepackage[draft]{todonotes}
\usepackage{mathtools}

\usepackage{hyperref}
\hypersetup{
    colorlinks=true,
    linkcolor=cyan,
    filecolor=magenta,      
    urlcolor=cyan,
    citecolor=violet,
}

\def \sh {\mathrm{sh}}
\def \SD {\mathrm{SD}}
\def \FD {\mathrm{FD}}
\def \det {\mathrm{det}}
\def \EFD {$E_\text{FD}~$}
\def \ESD {$E_\text{SD}~$}

\makeatletter
\g@addto@macro\bfseries{\boldmath}
\makeatother

\newcommand{\dif}{\mathrm{d}}

\newcommand{\nn}{\mathbf{n}}

\newcommand{\thetamax}{\theta_\mathrm{max}}

\begin{document}


\title{Measurement of the cosmic-ray energy spectrum above 
$2.5{\times} 10^{18}$~eV  using the Pierre Auger Observatory}



\begin{abstract}
We report a measurement of the energy spectrum of cosmic rays for energies above $2.5 {\times} 10^{18}~$eV based on 215,030 events recorded with zenith angles below $60^\circ$.  A key feature of the work is that the estimates of the energies are independent of assumptions about the unknown hadronic physics or of the primary mass composition. The measurement is the most precise made hitherto with the accumulated exposure being so large that the measurements of the flux are dominated by systematic uncertainties except at energies above $5 {\times} 10^{19}~$eV. The principal conclusions are: \par \medskip

\addtolength{\leftskip}{1.5em} \setlength{\parindent}{-1.5em}

\makebox[1.5em][l]{1.}The flattening of the spectrum near $5 {\times} 10^{18}~$eV, the so-called ``ankle'', is confirmed. \par \medskip

\makebox[1.5em][l]{2.}The steepening of the spectrum at around $5 {\times} 10^{19}~$eV is confirmed.  \par \medskip 

\makebox[1.5em][l]{3.}A new feature has been identified in the spectrum: in the region above the ankle the spectral index $\gamma$ of the particle flux ($\propto E^{-\gamma}$) changes from $2.51 \pm 0.03~{\rm (stat.)} \pm 0.05~{\rm (sys.)}$ to $3.05 \pm 0.05~{\rm (stat.)} \pm 0.10~{\rm (sys.)}$ before changing sharply to $5.1 \pm 0.3~{\rm (stat.)} \pm 0.1~{\rm (sys.)}$ above $5 {\times} 10^{19}~$eV. \par \medskip

\makebox[1.5em][l]{4.}No evidence for any dependence of the spectrum
on declination has been found other than a mild excess from the
Southern Hemisphere that is consistent with the anisotropy observed above $8 {\times} 10^{18}~$eV.

\end{abstract}

\input{revtex_authorlist.tex}


\pacs{}
\maketitle

\section{\label{sec:into}Introduction}

Although the first cosmic rays having energies above $10^{19}~$eV were detected nearly 60 years ago~\cite{LSR1961,Linsley1963}, the question of their origin remains unanswered. In this paper we report a measurement of the energy spectrum of ultra-high energy cosmic rays (UHECRs) of unprecedented precision using data from the Pierre Auger Observatory. Accurate knowledge of the cosmic-ray flux as a function of energy is required to help discriminate between competing models of cosmic-ray origin. As a result of earlier work with the HiRes instrument~\cite{HiRes}, the Pierre Auger Observatory~\cite{Abraham:2008ru} and the Telescope Array~\cite{AbuZayyad:2012ru}, two spectral features 
were
identified beyond reasonable doubt (see, 
e.g.,~\cite{LSS2011,Watson_2014,KT2014,MV2015,DFS2017,ABDP2018} for recent reviews).  These are a hardening of the spectrum at about $5{\times} 10^{18}$~eV (\textit{the ankle}) and a strong suppression of the flux at an energy about a decade higher.  The results reported here are based on 215,030 events with energies above $2.5{\times}10^{18}$~eV.
The present measurement, together with recent observations of anisotropies in the arrival directions of cosmic rays on large angular scales above $8{\times} 10^{18}$~eV~\cite{Aab:2017tyv} and on intermediate angular scales above $3.9{\times} 10^{19}$~eV~\cite{Aab:2018chp}, and inferences on the mass composition~\cite{Aab:2014aea,Aab:2017cgk}, provide essential data against which to test phenomenological models of cosmic-ray origin.  As part of a broad study of directional anisotropies, the large number of events used in the present analysis allows examination of the energy spectrum as a function of declination as reported below.

The determination of the flux of cosmic rays is a non-trivial exercise at any energy.  It has long been recognised that $\simeq 70$ to 80\% of the energy carried by the primary particle is dissipated in the atmosphere through ionisation loss and thus, with ground detectors alone, one must resort to models of shower development to infer the primary energy.  This is difficult as a quantitative knowledge of hadronic processes in the cascade is required. While at about $10^{17}$~eV the centre-of-mass energies encountered in collisions of primary cosmic rays with air nuclei are comparable to those 
achieved at the Large Hadron Collider, details of the interactions of pions, which are key to the development of the cascade, are lacking, and the presence of unknown processes is also possible. Furthermore one has to make an assumption about the primary mass.  Both conjectures lead to systematic uncertainties that are difficult, if not impossible, to assess.  To counter these issues, methods using light produced by showers as they cross the atmosphere have been developed.  In principle, this allows a calorimetric estimate of the energy. Pioneering work in the USSR in the 1950s~\cite{Nesterova} led to the use of Cherenkov radiation for this purpose, and this approach has been successfully adopted at the Tunka~\cite{Tunka} and Yakutsk~\cite{Yakutsk} arrays. The detection of fluorescence radiation, first achieved in Japan~\cite{Hara:1969} and, slightly later, in the USA~\cite{Bergeson:1977nw}, has been exploited particularly effectively in the Fly's Eye and HiRes projects to achieve the same objective.  The Cherenkov method is less useful at the highest energies as the forward-beaming of the light necessitates the deployment of a large number of detectors while the isotropic emission of the fluorescence radiation enables showers to be observed at distances of $\simeq 30$~km from a single station.  For both methods, the on-time is limited to moonless nights, and an accurate understanding of the aerosol content of the atmosphere is needed.

The Pierre Auger Collaboration introduced the concept of a hybrid observatory in which the bulk of the events used for spectrum determination is obtained with an array of detectors deployed on the ground and the integral of the longitudinal profile, measured using a fluorescence detector, is used to calibrate a shower-size estimate made with the ground array.  This hybrid approach has led to a substantial improvement in the accuracy of reconstruction of fluorescence events and to a calorimetric estimate of the energy of the primary particles for events recorded during periods when the fluorescence detector cannot be operated.  The hybrid approach has also been adopted by the Telescope Array Collaboration~\cite{AbuZayyad:2012ru}. 

A consistent aim of the Auger Collaboration has been to make the derivation of the energy spectrum as free of assumptions about hadronic physics and the primary composition as possible.  The extent to which this has been achieved can be judged from the details set out below.  After a brief introduction in Sec.~\ref{sec:Auger} to relevant features of the Observatory and the data-set, the method of estimation of energy is discussed in Sec.~\ref{sec:rec}. In Sec.~\ref{sec:EnSp}, the approach to deriving the energy spectrum is described, including the procedure for evaluating the exposure and for unfolding the resolution effects, as well as a detailed discussion of the associated uncertainties and of the main spectral features. A search for any dependence of the energy spectrum on declination is discussed in Sec.~\ref{sec:declination}, while a comparison with previous works is given in Sec.~\ref{sec:ta}. The results from the measurement of the energy spectrum are summarized in the concluding Sec.~\ref{sec:discussion}.

\section{\label{sec:Auger} The Pierre Auger Observatory and the data sets}

\subsection{\label{sec:observatory} The Observatory}

The Pierre Auger Observatory is sited close to the city of Malarg\"ue, Argentina, at a latitude of 35.2$^{\circ}$ S with a mean atmospheric overburden of 875\,g/cm$^2$. A detailed description of the instrument has been published~\cite{ThePierreAuger:2015rma}, and only brief remarks concerning features relevant to the data discussed in this paper are given.

The surface detector (SD) array comprises about 1600 water-Cherenkov detectors laid out on a 1500~m triangular grid, covering an area of about 3000 km$^{2}$. Each SD has a surface area of 10 m$^{2}$ and a height of 1.2 m, holding 12 tonnes of ultra-pure water viewed by 3 ${\times} 9"$ photomultipliers (PMTs). The signals from the PMTs are digitised using 40 MHz 10-bit Flash Analog to Digital Converters (FADCs). Data collection is achieved in real time by searching on-line for temporal and spatial coincidences at a minimum of three locations. When this occurs, FADC data from the PMTs are acquired from which the pulse amplitude and time of detection of signals is obtained. The SD array is operated with a duty cycle close to $100\%$.  

The array is over-looked from four locations, each having six Schmidt telescopes designed to detect fluorescence light emitted from shower excitations of atmospheric nitrogen. In each telescope, a camera with 440 hexagonal PMTs is used to collect light from a 13~m$^{2}$ mirror. These instruments, which form the fluorescence detector (FD), are operated on clear 
nights with low background illumination
with an on-time of $\simeq 15\%$. 


Atmospheric conditions at the site of the Observatory must be known for the reconstruction of the showers. Accordingly, comprehensive monitoring of the atmosphere, particularly of the aerosol content and the cloud cover, is undertaken as described in~\cite{ThePierreAuger:2015rma}. Weather stations are located close to the sites of the fluorescence telescopes. 
Before the Global Data Assimilation system was adopted~\cite{Abreu:2012zg}, an extensive series of balloon flights was made to measure the humidity, temperature and pressure in the atmosphere as a function of altitude.

\subsection{\label{sec:data} The data sets}

The data set used for the measurement of the energy spectrum consists of extensive air showers (EAS) recorded by the SD array. EAS detected simultaneously by the SD and the FD play a key role in this work. Dubbed hybrid events, they are pivotal in the determination of the energy of the much more numerous SD events~\cite{Verzi2013}. We use here SD events with zenith angle $\theta<60^\circ$, as the reconstruction of showers at larger angles requires a different method due to an asymmetry induced in the  distribution of the shower particles by the geomagnetic field and geometrical effects (see \cite{AugerHASRec}). A brief description of the reconstruction of SD and hybrid events is given in~\cite{FDNIM2010}: a more detailed description is in ~\cite{AugerRecoPaper}. We outline here features relevant to the present analysis. 

The reconstruction of the SD events is used to determine the EAS geometry (impact point of the shower axis and arrival direction) as well as a shower-size estimator. To achieve this, the amplitude and the start-time of the signals, recorded at individual SD stations and quantified in terms of their response to a muon travelling vertically and centrally through it (a vertical equivalent muon or VEM), are used. The arrival direction is determined to about $1^\circ$ from the relative arrival times of these signals. The impact point and the shower-size estimator are in turn derived by fitting the signal amplitudes to a lateral distribution function (LDF) that decreases monotonically with distance from the shower axis. The shower-size estimator adopted is the signal at 1000~m from the axis, $S(1000)$. For the grid spacing of 1500~m, 1000~m is the optimal distance to minimize the uncertainties of the signal due to the imperfect knowledge of the functional form of the LDF in individual events~\cite{NKW2007}. The combined statistical and systematic uncertainty  decreases from 15\% at a shower size of 10~VEM to 5\% at the highest shower sizes. The uncertainty on the impact point is of order 50~m. $S(1000)$ is influenced by changes in atmospheric conditions that affect shower development~\cite{AugerJINST2017}, and by the geomagnetic field that impacts on the shower particle-density~\cite{AugerJCAP2011}. Therefore, before using the shower-size estimator in the calibration procedure (Sec.~\ref{sec:rec}), corrections of order 2\% and 1\% for the atmospheric and geomagnetic effects, respectively, are made.

For the analysis in this paper, the SD reconstruction is carried out only for events in which the detector with the highest signal is surrounded by a hexagon of six stations that are fully operational. This requirement not only ensures adequate sampling of the shower but also allows evaluation of the aperture of the SD in a purely geometrical manner in the regime where the array is fully efficient~\cite{Abraham:2010zz}. As shown in Sec.~\ref{sec:EnSp}, such a regime is attained for events with $\theta<60^\circ$ at an energy $2.5{\times} 10^{18}$~eV. With these selection criteria, the SD data set used below consists of 215,030 events recorded between 1 January 2004 and 31 August 2018.

For hybrid events the reconstruction procedure exploits the amplitude and timing of the signals detected by each PMT in each telescope as well as additional timing information from the SD station with the highest signal. Combining the timing information from FD and SD improves the directional precision to $\simeq 0.6^\circ$~\cite{ThePierreAuger:2015rma}. Hybrid reconstruction provides in addition the longitudinal profile from which the depth of the shower maximum ($X_{\rm max}$) and the primary energy are extracted. The light signals in the FD PMTs are converted to the energy deposited in sequential depths in the atmosphere, taking into account the fluorescence and Cherenkov light contributions~\cite{MUFDRec} and their attenuation due to scattering. The longitudinal profile of the energy deposit is reconstructed by means of a fit to a modified Gaisser-Hillas profile~\cite{AugerLR2019}. 

Integration of the longitudinal profile yields a calorimetric measure of the ionisation loss in the atmosphere which is supplemented by the addition of the undetected energy, or ``invisible energy'', carried into the ground by muons and neutrinos.  We denote the sum of these two contributions, our estimate of the energy carried by the incoming primary particle, as $E_{\mathrm{FD}}$. The invisible-energy correction is estimated with a data-driven analysis and is about 14\% at $2.5 {\times} 10^{18}~$eV falling to about 12\% at $10^{20}~$eV~\cite{InvisibleEnergy2019}. The resolution of \EFD is 7.4\% at $2.5 {\times} 10^{18}~$eV and worsens with energy to 8.6\% at $6 {\times} 10^{19}~$eV. It is obtained by taking into account all uncorrelated uncertainties between  different  showers. In addition to the statistical uncertainty arising from the fit to the longitudinal profile, this resolution includes uncertainties in the detector response, in the models of the state of the atmosphere, and in the expected fluctuations from the invisible energy which, parameterized as a function of the calorimetric energy, is assumed to be identical for any  primary of same energy. All the uncorrelated uncertainties are addressed in~\cite{DawsonICRC2019} with further details given in~\cite{MUFDRec}. 
We note that  at higher energies the showers are detected, on average, at larger distances from the FD telescopes because the detection and reconstruction efficiency at larger distances increases with energy. 
This causes a worsening of the energy resolution because of the interplay between the uncertainty from the aerosols increasing with energy and the uncertainty from photoelectrons decreasing with energy.

The hybrid trigger efficiency, i.e. the probability of detecting a fluorescence event in coincidence with at least one triggered SD station, is 100\% at energies greater than $10^{18}$~eV, independent of the mass of the nuclear primaries~\cite{ExpoHybrid2011}. The hybrid data set used for the calibration of the SD events comprises 3,338 events with $E>3{\times} 10^{18}$~eV collected between 1 January 2004 and 31 December 2017. Other criteria for event selection are detailed in Sec.~\ref{sec:rec}.

\section{\label{sec:rec} Energy Estimation From Events Recorded by the Surface Array}

The energy calibration of the SD shower-size estimator against the energy derived from measurements with the FD is a two-step process. For a cosmic ray of a given energy, the value of $S(1000)$ depends on zenith angle because of the different atmospheric depths crossed by the corresponding shower. As detailed in Sec.~\ref{sec:SDrec}, we first correct for such an attenuation effect by using the Constant Intensity Cut (CIC) method~\cite{Hersil1961}. The calibration is then made between the corrected shower-size estimator, denoted by $S_{38}$, and the energy measured by the FD in hybrid events, $E_{\mathrm{FD}}$: the procedure to obtain the SD energy, $E_{\mathrm{SD}}$, is explained in Sec.~\ref{sec:EnCalib}. The systematic uncertainties associated with the SD energy scale thus obtained are described in Sec.~\ref{sec:EnScale}. Finally, the estimation of $E_{\mathrm{SD}}$ from $E_{\mathrm{FD}}$ allows us to derive the resolution, $\sigma_\textrm{SD}(E)$, as well as the bias, $b_\textrm{SD}(E)$, down to energies below which the detector is not fully efficient. We explain in Sec.~\ref{sec:SDResponse} the method used to measure $b_\textrm{SD}(E)$ and $\sigma_\textrm{SD}(E)$, from which we build the resolution function for the SD to be used for the unfolding of the spectrum. 

\subsection{\label{sec:SDrec} From $S(1000)$ to $S_{38}$} 

\begin{figure}[h]
        \centering
        \includegraphics[width=0.5\textwidth]{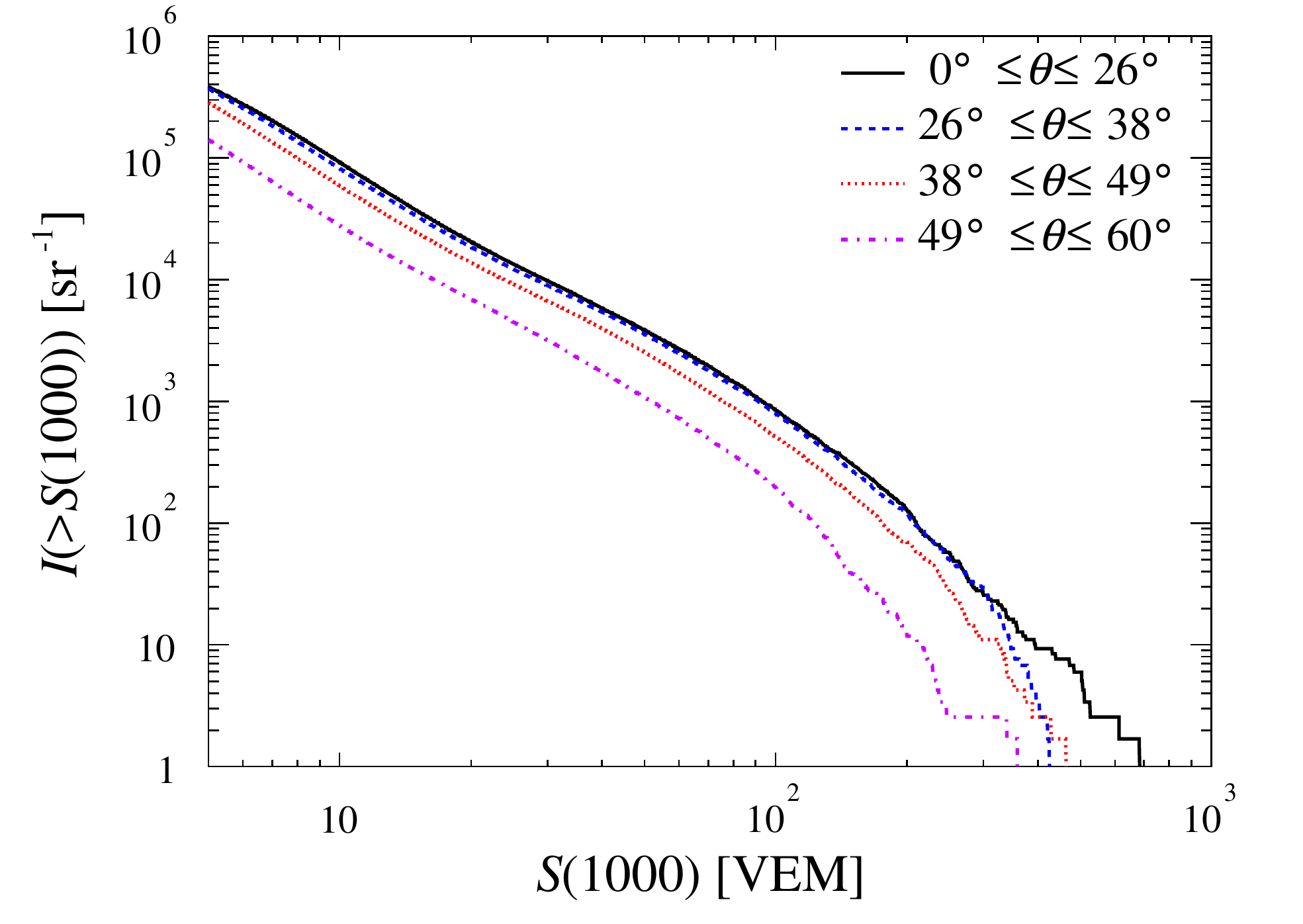}
        \caption{\small{Integral intensity above $S(1000)$ thresholds, for different zenithal ranges of equal exposure.}}
        \label{fig:IvsS}
\end{figure}

For a fixed energy, $S(1000)$ depends on the zenith angle $\theta$  because, once it has passed the depth of shower maximum, a shower is attenuated as it traverses the atmosphere.  The intensity of cosmic rays, defined here as the number of events per steradian above some $S(1000)$ threshold, is thus dependent on zenith angle as can be seen in Fig.~\ref{fig:IvsS}. 

Given the highly isotropic flux, the intensity is expected to be $\theta$-independent after correction for the attenuation. Deviations from a constant behavior can thus be interpreted as being due to attenuation alone. Based on this principle, an empirical procedure, the so-called CIC method, is used to determine the attenuation curve as function of $\theta$ and therefore a $\theta$-independent shower-size estimator ($S_{38}$). It can be thought of as being the $S(1000)$ that a shower would have produced had it arrived at $38^\circ$, the median angle from the zenith. The small anisotropies in the arrival directions and the zenithal dependence of the resolution on $S_{38}$ do not alter the validity of the CIC method in the energy range considered here, as shown in Appendix~\ref{app:sin2}. 

\begin{figure}[h]
        \centering
        \includegraphics[width=0.5\textwidth]{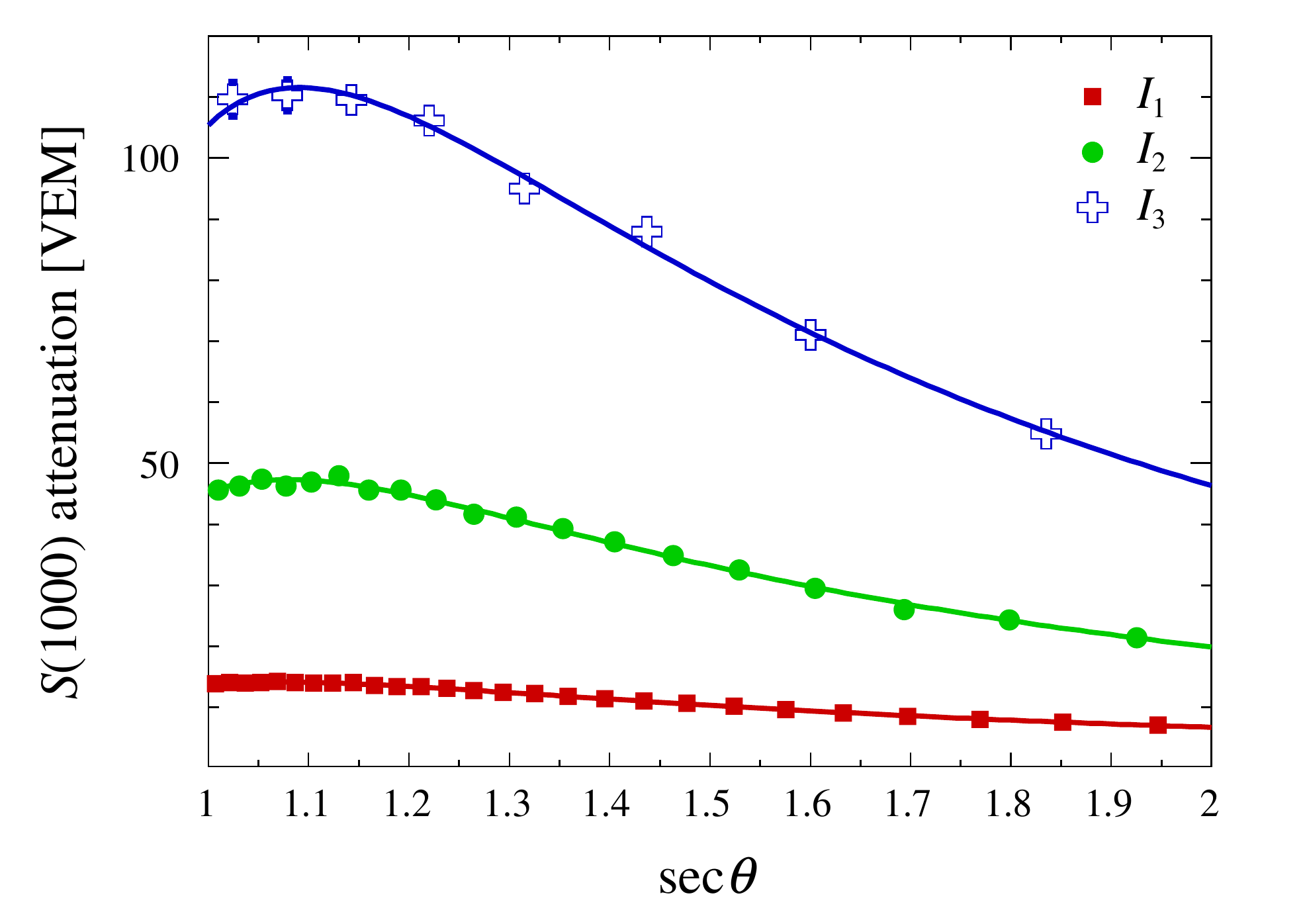}        \includegraphics[width=0.5\textwidth]{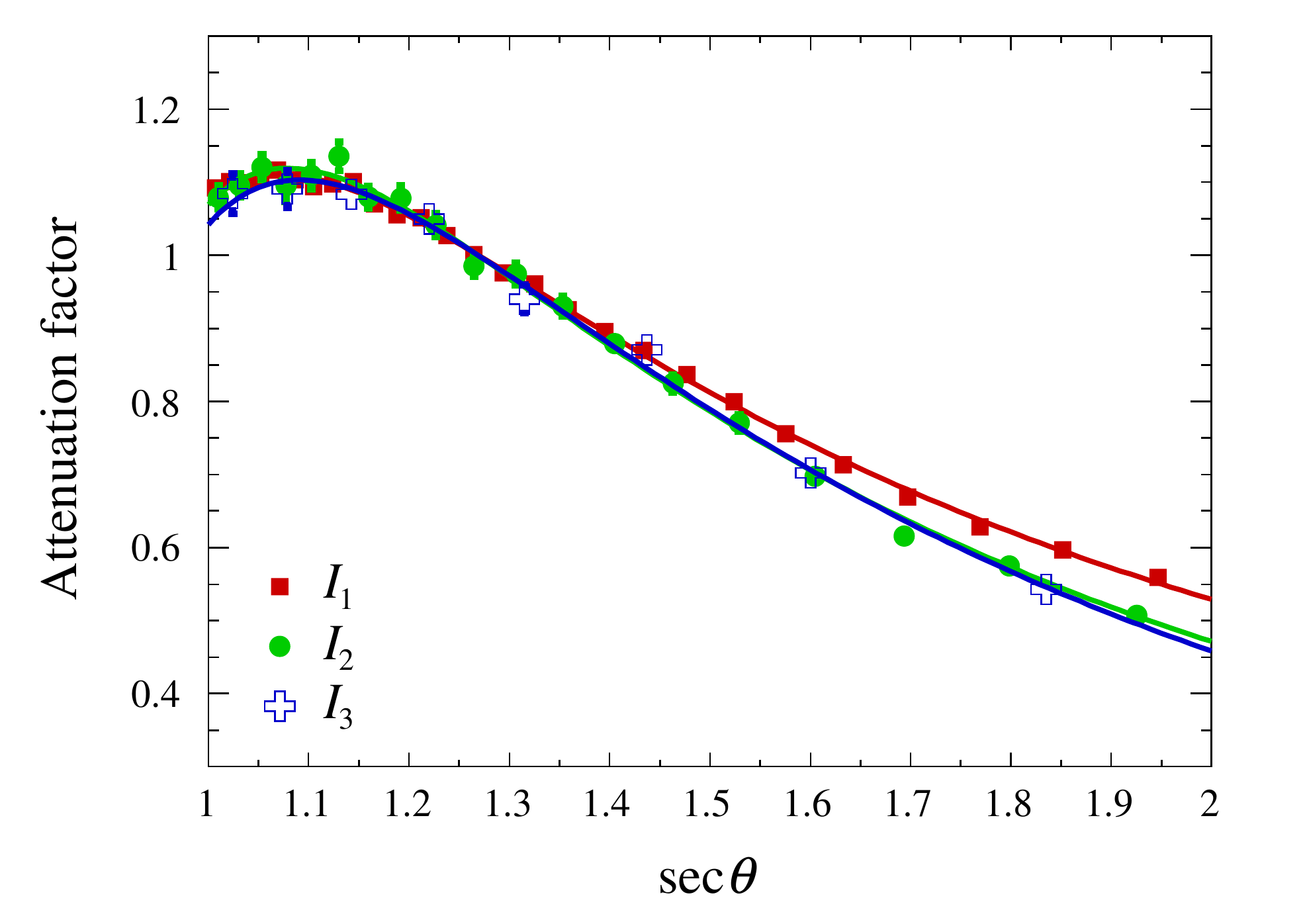}
        \caption{\small{Top: $S(1000)$ attenuation as a function of $\sec{\theta}$, as derived from the CIC method, for different intensity thresholds (see text). Bottom: Same attenuation curves, normalised to 1 at $\theta=38^\circ$ (note that $\sec{38^\circ}\approx 1.269$), to exhibit the differences for the three different intensity thresholds. The intensity thresholds are $I_1=2.91{\times}10^{4}~$sr$^{-1}$, $I_2=4.56{\times}10^{3}~$sr$^{-1}$ and $I_3=6.46{\times}10^{2}~$sr$^{-1}$ at 38$^\circ$. Anticipating the conversion from intensity to energy, these correspond roughly to $3{\times} 10^{18}$~eV, $8{\times} 10^{18}$~eV and $2{\times} 10^{19}$~eV, respectively.}}
        \label{fig:CIC}
\end{figure}

In practice, a histogram of the data is first built in $\cos^2{\theta}$ to ensure equal exposure; then the events are ordered by $S(1000)$ in each bin. For an intensity high enough to guarantee full efficiency, the set of $S(1000)$ values, each corresponding to the $N$th largest signal in the associated $\cos^2{\theta}$ bin, provides an empirical estimate of the attenuation curve. Because the mass of each cosmic-ray particle cannot be determined on an event-by-event basis, the attenuation curve inferred in this way is an effective one, given the different species that contribute at each intensity threshold. The resulting data points are fitted with a third-degree polynomial, $S(1000)=S_{38}(1+ax+bx^2+cx^3)$, where  $x=\cos^2{\theta}-\cos^2{38^\circ}$.  Fits are shown in the top panel of Fig.~\ref{fig:CIC} for three different intensity thresholds corresponding to $I_1=2.91{\times}10^{4}~$sr$^{-1}$, $I_2=4.56{\times}10^{3}~$sr$^{-1}$ and $I_3=6.46{\times}10^{2}~$sr$^{-1}$ at 38$^\circ$. 
The attenuation is plotted as a function of $\sec{\theta}$ to exhibit the dependence on the thickness of atmosphere traversed.  The uncertainties in each data point follow from the number of events above the selected $S(1000)$ values. The $N$th largest signal in each bin is a realization of a random variable distributed as an order-statistic variable where the total number of ordered events in the $\cos^2{\theta}$ bin is itself a Poisson random variable. Within a precision better than 1\%, the standard deviation of the random variable can be approximated through a straight-forward Poisson propagation of uncertainties, namely $\Delta S(N)\simeq (S(N+\sqrt{N})-S(N-\sqrt{N}))/2$. The number of bins is adapted to the available number of events for each intensity threshold, from 27 for $I_1$ so as to guarantee a resolution on the number of events of 1\% in each bin, to 8 for $I_3$ so as to guarantee a resolution of 4\%.

The curves shown in Fig.~\ref{fig:CIC} are largely shaped by the electromagnetic contribution to $S(1000)$ which, once the shower development has passed its maximum, decreases with the zenith angle because of attenuation in the increased thickness of atmosphere. The muonic component starts to dominate at large angles, which explains the flattening of the curves. In the bottom panel, the curves are normalized to 1 at $38^\circ$ to exhibit the differences for the selected intensity thresholds. Some dependence with the intensity thresholds, and thus with the energy thresholds, is observed at high zenith angles: high-energy showers appear more attenuated than low energy ones. This results from the interplay between the mass composition and the muonic-to-electromagnetic signal ratio at ground level. A comprehensive interpretation of these curves is however not addressed here. 

The  energy dependence in the CIC curves that is observed is accounted for by introducing an empirical dependence in terms of $y=\log_{10}(S_{38}/40~\textrm{VEM})$ in the coefficients $a$, $b$ and $c$ through a second-order polynomial in $y$. The polynomial coefficients derived are shown in Table~\ref{tab:cic_param}. They relate to $S_{38}$ values ranging from 15~VEM to 120~VEM. Outside these bounds, the coefficients are set to their values at 15 and 120~VEM. This is because below 15~VEM, the isotropy is not expected anymore due to the decreasing efficiency, while above 120~VEM, the number of events is low and there is the possibility of localized anisotropies. 
\begin{table}[h]
\caption{Coefficients of the second-order polynomial in terms of $y=\log_{10}(S_{38}/40~\textrm{VEM})$ for the CIC parameters $a$, $b$ and $c$.}
\label{tab:cic_param}
\begin{ruledtabular}
\begin{tabular}{l c c c}
 & $~y_0~$ & $~y_1~$ & $~y_2~$ \\
\colrule
$~a~$ & $0.952$ & $0.06$ & $-0.37$  \\
$~b~$ & $-1.64$ & $-0.42$ & $0.09$  \\
$~c~$ & $-0.9$ & $-0.04$ & $1.3$  
\end{tabular}
\end{ruledtabular}
\end{table}

\subsection{\label{sec:EnCalib} From $S_{38}$ to $E_{\mathrm{SD}}$}

The shower-size estimator, $S_{38}$, is converted into energy through a calibration with $E_{\mathrm{FD}}$ by making use of a subset of SD events, selected as described in Sec.~\ref{sec:Auger}, which have triggered the FD independently. For the analysis, we apply several selection criteria to guarantee a precise estimation of $E_{\mathrm{FD}}$ as well as fiducial cuts to minimise the biases in the mass distribution of the cosmic rays introduced by the field of view of the FD telescopes. 

The first set of cuts aims to select time periods during which data-taking and atmospheric conditions are suitable for collecting high-quality data~\cite{Aab:2014}. We require a high-quality calibration of the gains of the PMTs of the FD and that the vertical aerosol optical depth is measured within 1 hour of the time of the event, 
with its value integrated up to 3 km above the ground being less than 0.1.
Moreover, measurements from detectors installed at the Observatory to monitor atmospheric conditions~\cite{ThePierreAuger:2015rma} are used to select only those events detected by telescopes without clouds within their fields of view. Next, a set of quality cuts are applied to ensure a precise reconstruction of the energy deposit~\cite{Aab:2014}. 

We select events with a total track length of at least 200 ${\rm g/cm^2}$, requiring that any gap in the profile of the deposited energy be less than 20\% of the total track 
length and we reject events with an uncertainty in the reconstructed calorimetric energy larger than 20\%. We transform the $\chi^2$ into a 
variable with zero mean and unit variance, $z=\left(\chi^2-n_{\rm dof}\right)/\sqrt{2n_{\rm dof}}$ with $n_{\rm dof}$ the number of degrees of freedom, and require that the $z$ values be less than 3.
Finally, the fiducial cuts are defined by an appropriate selection of the lower and upper depth boundaries to enclose the bulk of the $X_{\rm max}$ distribution and by requiring that the maximum accepted uncertainty in $X_{\rm max}$ is $40~{\rm g/cm^2}$ and that the minimum viewing angle of light in the telescope is $20^\circ$~\cite{Aab:2014}. This limit is set to reduce contamination by Cherenkov radiation. A final cut is applied to $E_{\mathrm{FD}}$: it must be greater than $3{\times} 10^{18}$~eV to ensure that the SD is operating in the regime of full efficiency (see Sec.~\ref{sec:raw}).

\begin{figure}[h]
   	\centering
   	\includegraphics[width=0.5\textwidth]{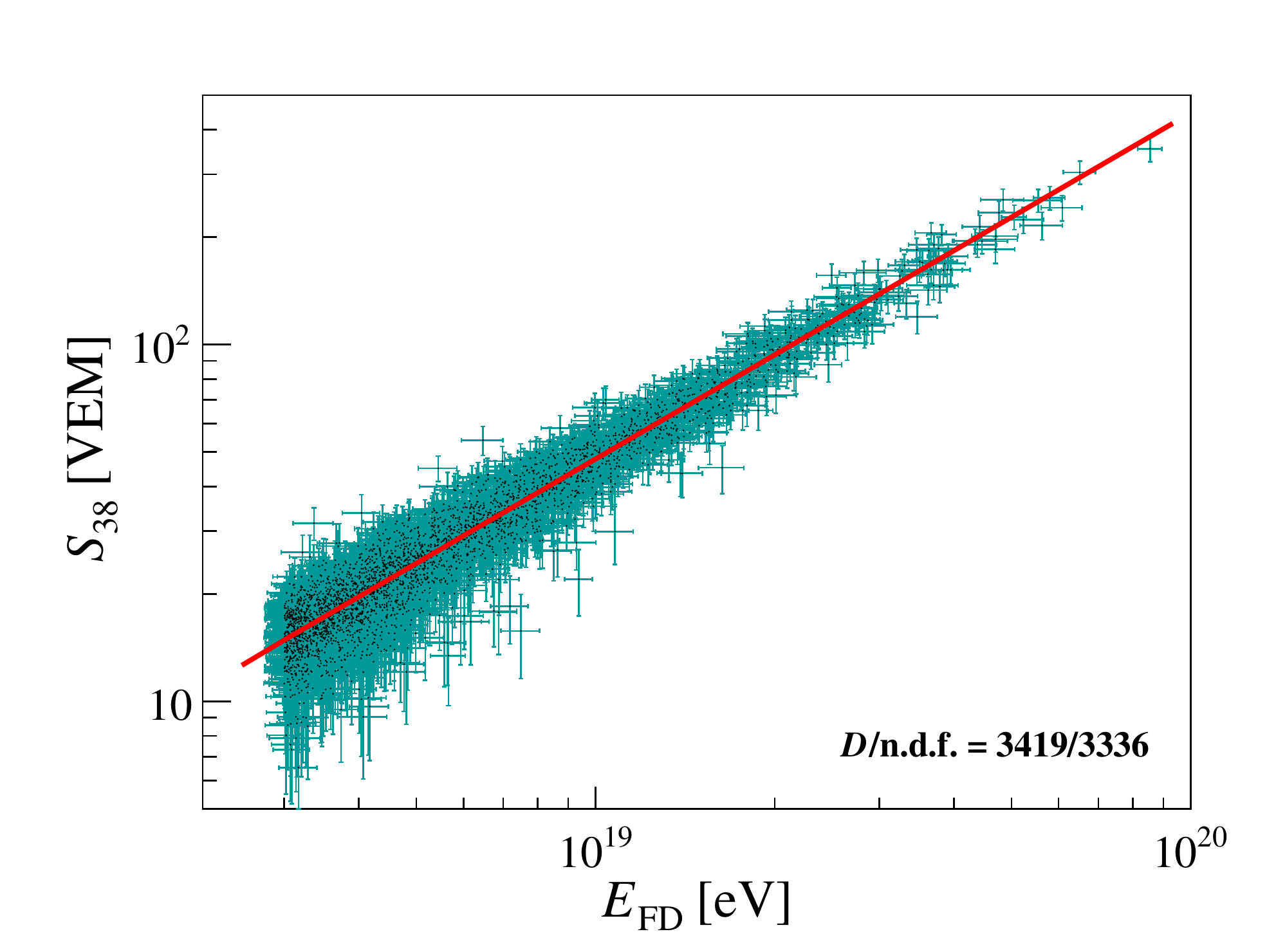}
   	\caption{\small{Correlation between the SD shower-size estimator, $S_{38}$, and the reconstructed FD energy, $E_{\mathrm{FD}}$, for the selected 3,338 hybrid events used in the fit. The uncertainties indicated by the error bars are described in the text. The solid line is the best fit of the power-law dependence \EFD$=A\,{S_{38}}^B$ to the data. 
The reduced deviance of the fit, whose calculation is detailed in Appendix~\ref{app:ES38fit}, is shown in the bottom-right corner.}}
   	\label{fig:EnCalib}
   \end{figure}

After applying these cuts, a data set of 3,338 hybrid events is available for the calibration process. With the current sensitivity of our $X_{\rm max}$ measurements in this energy range, a constant elongation rate (that is, a single logarithmic dependence of $X_{\rm max}$ with energy) is observed~\cite{Aab:2014}. In this case, a single power law dependence of $S_{38}$ with energy is expected from Monte-Carlo simulations. We thus describe the correlation between $S_{38}$ and $E_\textrm{FD}$, shown in Fig.~\ref{fig:EnCalib}, by a power law function,
\begin{equation}
E_\textrm{FD} = A~{S_{38}}^{B}, 
\label{eqn:ECalib}
\end{equation}
where $A$ and $B$ are fitted to data. In this manner the correlation captured through this power-law relationship is fairly averaged over the underlying mass distribution, and thus provides the calibration of the mass-dependent $S_{38}$ parameter in terms of energy in an unbiased way over the covered energy range. Due to the limited number of events in the FD data set at the highest energies, deviations from the inferred power law cannot be fully investigated currently. We note however that any indication for a strong change of elongation rate cannot be inferred at the highest energies from our SD-based indirect measurement reported in~\cite{Aab:2017cgk}.

The correlation fit is carried out using a tailored maximum-likelihood method allowing various effects of experimental origin to be taken into account~\cite{Dembinski:2015wqa}. The probability density function entering the likelihood procedure, detailed in Appendix~\ref{app:ES38fit}, is built by folding the cosmic-ray flux, observed with the effective aperture of the FD, with the resolution functions of the FD and of the SD. Note that to avoid the need to model accurately the cosmic-ray flux observed through the effective aperture of the telescopes (and thus to rely on mass assumptions), the observed distribution of events passing the cuts described above is used in this probability density function.  

The uncertainties in the FD energies are estimated, on an event-by-event basis, by adding in quadrature all uncertainties in the FD energy measurement which are uncorrelated shower-by-shower (see~\cite{DawsonICRC2019} for details). The uncertainties in $S_{38}$ are also estimated on an event-by-event basis considering the event-by-event contribution arising from the reconstruction accuracy of $S(1000)$. The error arising from the determination of the zenith angle is negligible. The contribution from shower-to-shower fluctuations to the uncertainty in $E_{\rm SD}$ is parameterized as a relative error in $S_{38}$ with $0.13 - 0.08x + 0.03x^2$ where $x = \log_{10}(E/\mathrm{eV}) - 18.5$. It is obtained by subtracting in quadrature the contribution of the uncertainty in $S_{38}$ from the SD energy resolution. The latter, as detailed in the following, is measured from data and the resulting shower-to-shower fluctuations are free from any reliance on mass assumption and model simulations. 

The best fit parameters are $A=(1.86 {\pm} 0.03){\times} 10^{17}$~eV and $B=1.031 {\pm} 0.004$ and the correlation coefficient between the parameters is $\rho = -0.98$.  The resulting calibration curve is shown as the red line in Fig.~\ref{fig:EnCalib}. The goodness of the fit is provided by the value of the reduced deviance, namely $D/n_\mathrm{dof}=3419/3336$. The statistical uncertainty on the SD energies obtained propagating the fit errors on $A$ and $B$ is 0.4 \% at $3 {\times} 10^{18}$ eV, increasing up to 1\% at the highest energies. The most energetic event used in the calibration is detected at all four fluorescence sites. Its energy is $(8.5 {\pm} 0.4){\times} 10^{19}$ eV, obtained from a weighted average of the four calorimetric energies and using the resulting energy to evaluate the invisible energy correction~\cite{InvisibleEnergy2019}. It has a depth of shower maximum of $(763 {\pm} 8)~{\rm g/cm^2}$, which is typical/close to the average for a shower of this energy~\cite{Aab:2014}. The energy estimated from $S_{38}=354~$VEM is $(7.9 {\pm} 0.6){\times} 10^{19}$~eV. 

\subsection{\label{sec:EnScale} $E_{\mathrm{SD}}$: systematic uncertainties}

The calibration constants $A$ and $B$ are used to estimate the energy for the bulk of SD events: $E_{\mathrm{SD}}\equiv A{S_{38}}^B$. They define the SD energy scale. The uncertainties in the FD energies are estimated, on an event-by-event basis, by adding in quadrature all uncertainties in the FD energy measurement which are correlated shower-by-shower~\cite{Verzi2013}. 

The contribution from the fluorescence yield is $3.6\%$ and is obtained by propagating the uncertainties in the high-precision measurement performed in the AIRFLY experiment of the absolute yield~\cite{FY-Airfly_AbsYield} and of the wavelength spectrum and quenching parameters~\cite{FY-Airfly_spectrum, FY-Airfly_T_h}. The uncertainty coming from the characterization of the atmosphere ranges from 3.4\% (low energies) to 6.2\% (high energies). It is dominated by the uncertainty associated with the aerosols in the atmosphere and includes a minor contribution related to the molecular properties of the atmosphere. The largest correlated uncertainty, associated with the calibration of the FD, amounts to 9.9\%. It includes a 9\% uncertainty in the absolute calibration of the telescopes and other minor contributions related to the relative response of the telescopes at different wavelengths and relative changes with time of the gain of the PMTs. The uncertainty in the reconstruction of the energy deposit ranges from 6.5\% to 5.6\% (decreasing with energy) and accounts for the uncertainty associated with the modelling of the light spread away from the image axis and with the extrapolation of the modified Gaisser-Hillas profile beyond the field of view of the telescopes. The uncertainty associated with the invisible energy is 1.5\%. The invisible energy is inferred from data through an analysis that exploits the sensitivity of the water-Cherenkov detectors to muons and minimizes the uncertainties related to the assumptions on hadronic interaction models and mass composition~\cite{InvisibleEnergy2019}.

\begin{table}[h]
\caption{Calibration parameters in three different zenithal ranges. $N$ is the number of events selected in each range.}
\label{tab:SDCalib_zenith}
\vspace{0.1cm}
\begin{ruledtabular}
\begin{tabular}{l c c c}
 & $0^\circ < \theta < 30^\circ$ 
 & $30^\circ < \theta < 45^\circ$ 
 & $45^\circ < \theta < 60^\circ$ \\
 \colrule
$N$ & 435 & 1641 & 1262 \\
$A/10^{17}\,\text{eV}$ & $1.89 {\pm} 0.08$  
                   & $1.86 {\pm} 0.04$   
                   & $1.83 {\pm} 0.04$  \\
$B$ & $1.029 {\pm} 0.012$  
    & $1.030 {\pm} 0.006$ 
    & $1.034 {\pm} 0.006$  
\end{tabular}
\end{ruledtabular}
\end{table}

We have performed several tests aimed at assessing the robustness of the analysis that returns the calibration coefficients $A$ and $B$. The correlation fit was repeated selecting events in three different zenithal ranges. The obtained calibration parameters are reported in Table~\ref{tab:SDCalib_zenith}. The calibration curves are within one standard deviation of the average one reported above, resulting in energies within 1\% of the average ones. Other tests performed using looser selection criteria for the FD events  give similar results. By contrast, determining the energy scale in different time periods leads to some deviation of the calibration curves with respect to the average one. Although such variations are partly accounted for in the FD calibration uncertainties, we conservatively propagate these uncertainties into a 5\% uncertainty on the SD energy scale. 

The total systematic uncertainty in the energy scale is obtained by adding in quadrature all of the uncertainties detailed above, together with the contribution arising from the statistical uncertainty in the calibration parameters. The total is about 14\% and it is almost energy independent as a consequence of the energy independence of the uncertainty in the FD calibration, which makes the dominant contribution.

 \subsection{\label{sec:SDResponse} {$E_{\mathrm{SD}}$: resolution and bias} }

Our final aim is to estimate the energy spectrum above $2.5{\times} 10^{18}~$eV. Still it is important to characterize the energies below this threshold because the finite resolution on the energies induces bin-to-bin migration effects that affect the spectrum. In this energy range, below full efficiency of the SD, systematic effects enter into play on the energy estimate. While the FD quality and fiducial cuts still guarantee the detection of showers without bias towards one particular mass in that energy range, this is no longer the case for the SD due to the higher efficiency of shower detection for heavier primary nuclei~\cite{Abraham:2010zz}. Hence the distribution of $S_{38}$ below $3{\times} 10^{18}~$eV may no longer be fairly averaged over the underlying mass distribution, and a bias on $E_{\mathrm{SD}}$ may result from the extrapolation of the calibration procedure, in addition to the trigger effects that favor positive fluctuations of $S_{38}$ at a fixed energy over negative ones. In this section, we determine these quantities, denoted as $\sigma_\textrm{SD}(E,\theta)/E$ for the resolution and as $b_\textrm{SD}(E,\theta)$ for the bias, in a data-driven way. These measurements allow us to characterize the SD resolution function that will be used in several steps of the analysis presented in the next sections. This, denoted as 
$\kappa(E_\textrm{SD}|E;\theta)$, 
is the conditional p.d.f. for the measured energy 
$E_\textrm{SD}$ 
given that the true value is $E$. It is normalized such that the event is observed at any reconstructed energy, that is, 
$\int\dif E_\textrm{SD}~\kappa(E_\textrm{SD}|E;\theta)=1$. 
In the energy range of interest, we adopt a Gaussian curve, namely:
\begin{widetext}
\begin{equation}
\label{eqn:kappa}
\kappa(E_\textrm{SD}|E;\theta) = \frac{1}{\sqrt{2\pi}\sigma_\textrm{SD}(E,\theta)} \exp{\left[-\frac{(E_\textrm{SD}-E(1+b_\textrm{SD}(E,\theta)))^2}{2\sigma^2_\textrm{SD}(E,\theta)}\right]}.
\end{equation}
\end{widetext}

\begin{figure}[h]
	\centering
	\includegraphics[width=0.5\textwidth]{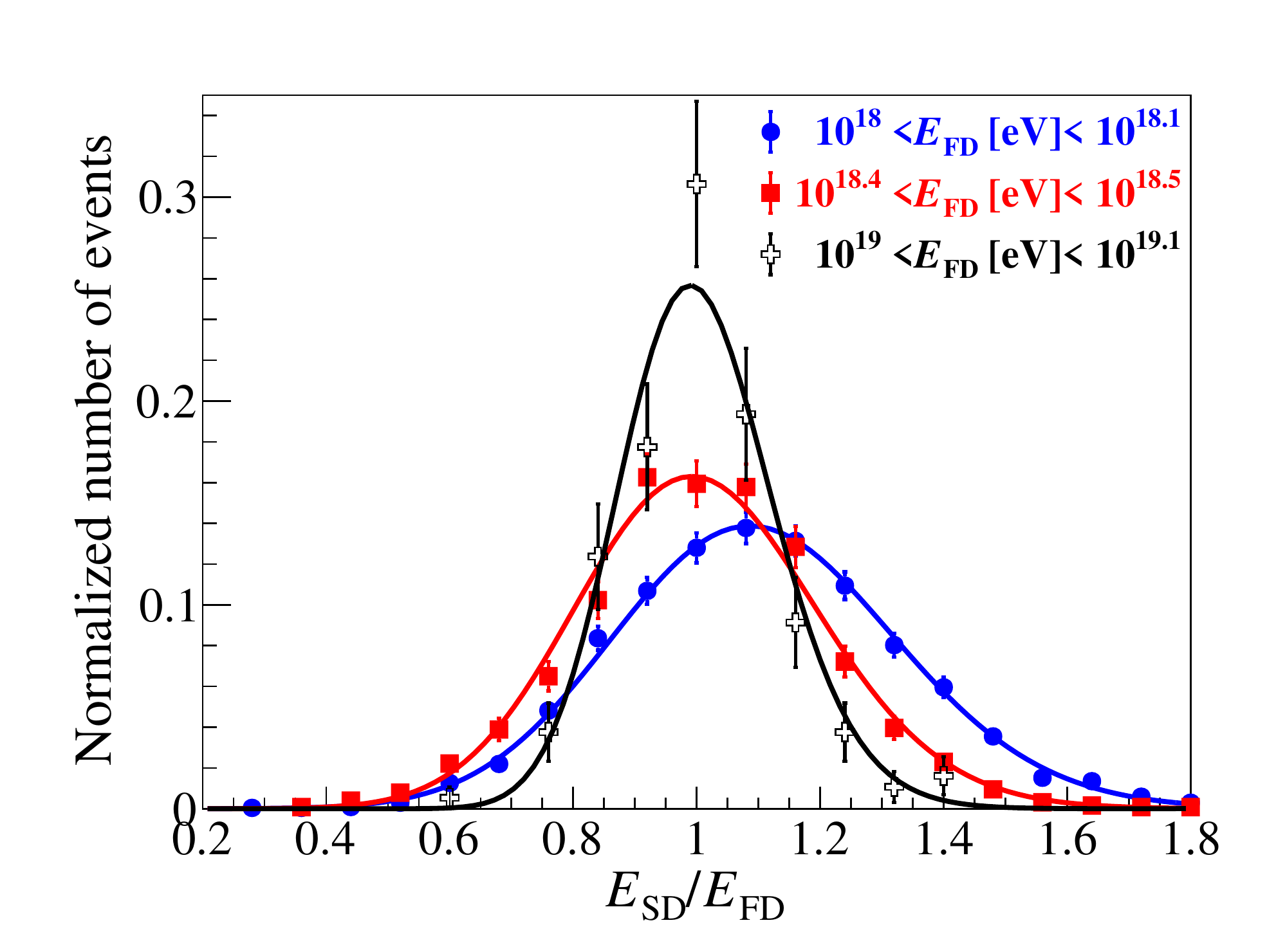}
	\caption{\small{Ratio distribution of the SD energy, $E_{\mathrm{SD}}$, to the FD energy, $E_{\mathrm{FD}}$, from the selected data sample, for three energy ranges. The distributions are all normalized to unity to better underline the difference in their shape. The total number of events for each distribution is 2367, 1261 and 186 from the lower to the higher energy bin, respectively.
	}}
	\label{fig:hist_resolution}
\end{figure}

\begin{figure}[h]
	\centering
	\includegraphics[width=0.5\textwidth]{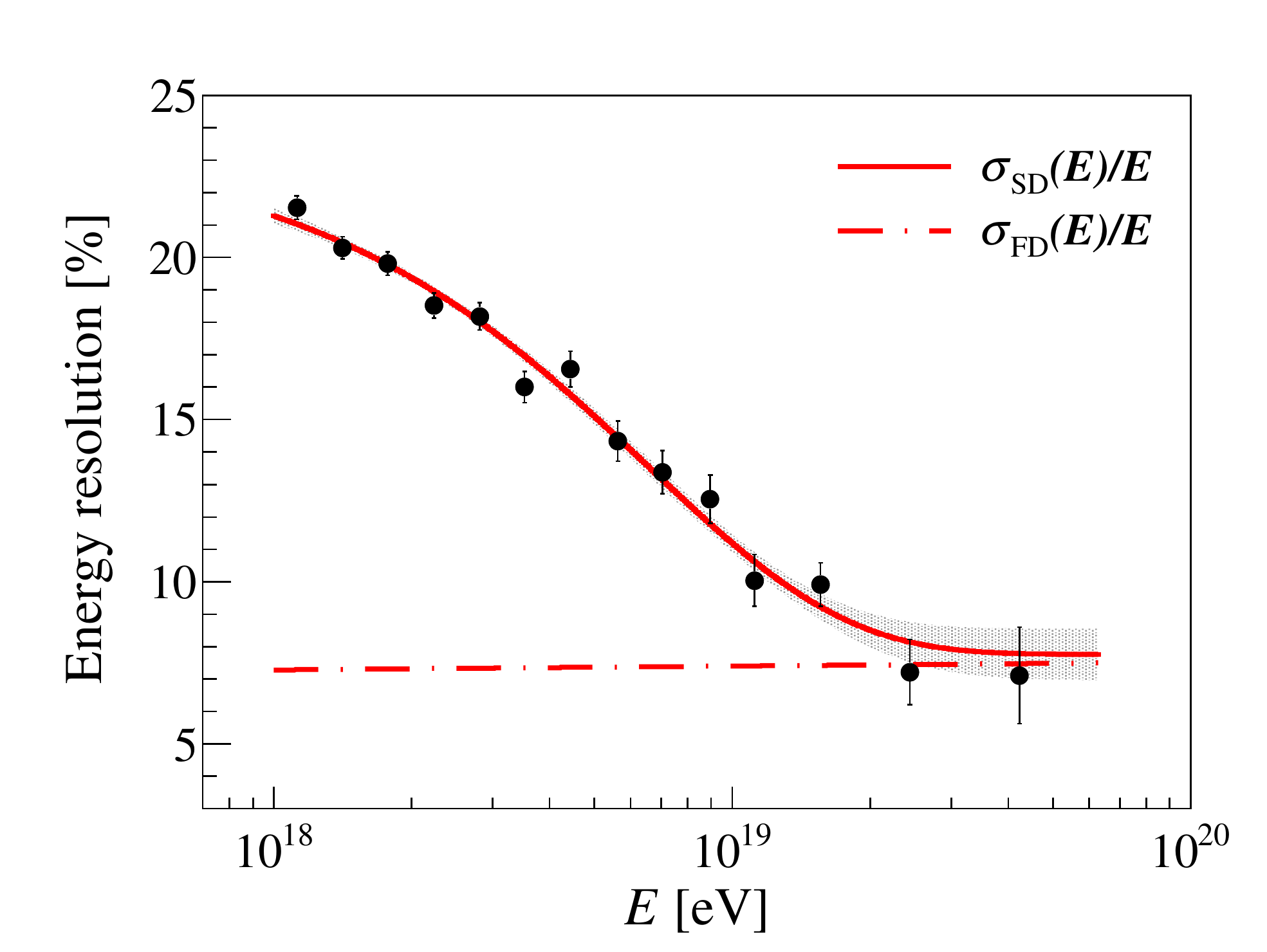}
	\caption{\small{Resolution of the SD as a function of energy. The measurements with their statistical uncertainties are shown with points and error bars. The fitted parameterization 
	is depicted with the continuous line and its 
	statistical uncertainty is shown as a shaded band. The FD resolution is also shown for reference (dotted-dashed line).}}
	\label{fig:sd_resolution}
\end{figure}

The estimation of $b_\textrm{SD}(E,\theta)$ and $\sigma_\textrm{SD}(E,\theta)$ is obtained by analyzing the $E_\textrm{SD}/E_\textrm{FD}$ histograms as a function of $E_\textrm{FD}$, extending here the $E_\textrm{FD}$ range down to $10^{18}~$eV. For Gaussian-distributed \EFD and \ESD variables, the $E_\textrm{SD}/E_\textrm{FD}$ variable follows a Gaussian ratio distribution. For a FD resolution function with no bias and a known resolution parameter, the searched  $b_{\textrm{SD}}(E,\theta)$ and $\sigma_{\textrm{SD}}(E,\theta)$ are then obtained from the data. The overall FD energy resolution is $\sigma_{\textrm{FD}}(E)/E\simeq 7.4\%$. 
In comparison to the number reported in Sec.~\ref{sec:data}, $\sigma_{\textrm{FD}}(E)/E$ is here almost constant over the whole energy range because it takes into account that, at the highest energies, the same shower is detected from different FD sites. In these cases, the energy used in analyses is the mean of the reconstructed energies (weighted by uncertainties) from the two (or more) measurements. This accounts for the improvement in the statistical error. 

\begin{figure}[h]
	\centering
	\includegraphics[width=0.5\textwidth]{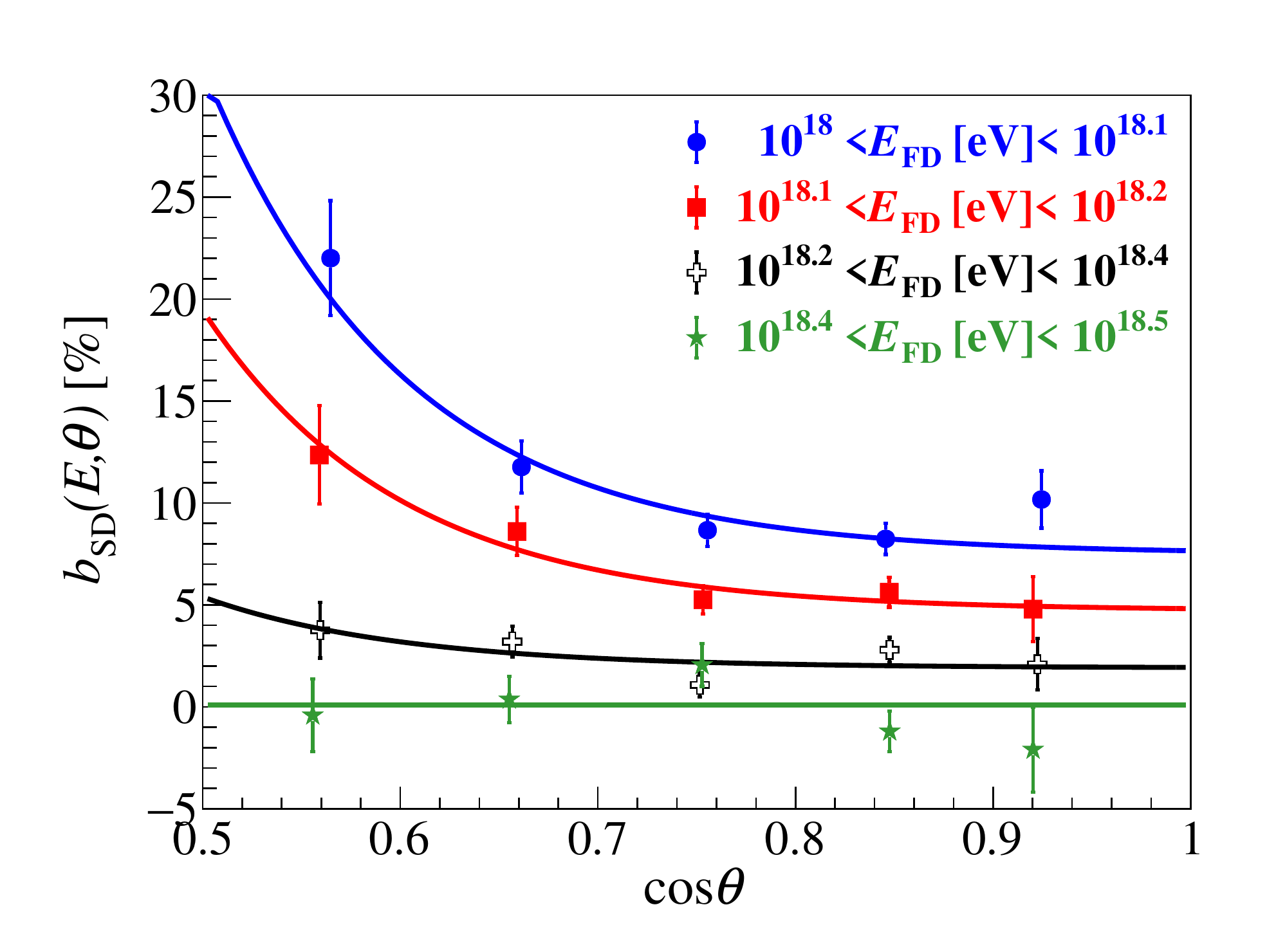}
	\caption{\small{Relative bias parameters of the SD as a function of the zenith angle, for four different energy ranges. The results of the fit of the $E_{\rm SD}/E_{\rm FD}$ distributions with the statistical uncertainties are shown with symbols and error bars, while the fitted parameterization is shown with lines.}}
	\label{fig:sd_bias}
\end{figure}

Examples of measured and fitted distributions of $E_\textrm{SD}/E_\textrm{FD}$ are shown in Fig.~\ref{fig:hist_resolution} for three energy ranges: the resulting SD energy resolution is $\sigma_{\textrm{SD}}(E)/E = (21.5 \pm 0.4)\%$, $(18.2 \pm 0.4)\%$ and $(10.0 \pm 0.8)\%$ between $10^{18}$ and $10^{18.1}$~eV, $10^{18.4}$ and $10^{18.5}$~eV, $10^{19}$ and $10^{19.1}$~eV, respectively. The parameter $\sigma_{\textrm{SD}}(E)/E$ is shown in Fig.~\ref{fig:sd_resolution} as a function of $E$: the resolution is $\simeq 20\%$ at $2{\times} 10^{18}~$eV and tends smoothly to $\simeq 7\%$ above $2{\times} 10^{19}~$eV. Note that no significant zenithal dependence has been observed. The bias parameter $b_{\textrm{SD}}(E,\theta)$ is illustrated in Fig.~\ref{fig:sd_bias} as a function of the zenith angle for four different energy ranges. The net result of the analysis is a bias larger than $10\%$ at $10^{18}~$eV, going smoothly to zero in the regime of full efficiency.

Note that the selection effects inherent in the FD field of view induce different samplings of hybrid and SD showers with respect to shower age at a fixed zenith angle and at a fixed energy. These selection cuts also impact the zenithal distribution of the showers.
Potentially, the hybrid sample may thus not be a fair sample of the bulk of SD events. This may lead to some misestimation of the SD resolution determined in the data-driven manner presented above. We have checked, using end-to-end Monte-Carlo simulations of the Observatory operating in the hybrid mode, that the particular quality and fiducial cuts used to select the hybrid sample do not introduce significant distortions to the measurements of $\sigma_{\textrm{SD}}(E)$ shown in Fig.~\ref{fig:sd_resolution}: the ratio between the hybrid and SD standard deviations of the reconstructed energy histograms remain within 10\% (low energies) and 5\% (high energies) whatever the assumption on the mass composition. There is thus a considerable benefit in relying on the hybrid measurements,to avoid any reliance on mass assumptions when determining the bias and resolution factors. 

From the measurements, a convenient parameterization of the resolution is
\begin{equation}
\label{eqn:resolution}
\frac{\sigma_{\mathrm{SD}}(E)}{E} = \sigma_{0}+\sigma_{1}\exp{(-\frac{E}{E_\sigma})},
\end{equation}
where the values of the parameters are obtained from a fit to the data: $\sigma_0=0.078$, $\sigma_1=0.16$, and $E_\sigma=6.6{\times} 10^{18}~$eV. The function and its statistical uncertainty from the fit are shown in Fig.~\ref{fig:sd_resolution}. It is worth noting that this parameterization accounts for both the detector resolution and the shower-to-shower fluctuations. Finally, a detailed study of the systematic uncertainties on this parameterization leads to an overall relative uncertainty of about 10\% at $10^{18}$~eV and increasing with energy to about 17\% at the highest energies. It accounts for the selection effects inherent to the FD field of view previously addressed, the uncertainty in the FD resolution and the statistical uncertainty in the fitted parameterization.

The bias, also parameterized as a function of the energy, includes an additional angular dependence:
\begin{equation}
\label{eqn:bias}
b_{\mathrm{SD}}(E,\theta)
= \left(b_0+b_1 \exp{(-\lambda_b(\cos{\theta}-0.5))}\right)\log_{10}{\left(\frac{E_*}{E}\right)},
\end{equation}
for $\log_{10}{(E/\text{eV})}\leq\log_{10}{(E_*/\text{eV})}=18.4$, and $b_{\mathrm{SD}}=0$ otherwise. Here, $b_0=0.20$, $b_1=0.59$ and $\lambda_b=10.0$. The parameters are obtained in a two steps process: we first perform a fit to extract the zenith-angle dependence in different energy intervals prior to determining the energy dependence of the parameters. Examples of the results of the fit to the data are shown in Fig.~\ref{fig:sd_bias}. The relative uncertainty in these parameters is  estimated to be within 15\%, considering the largest uncertainties of the data points displayed in the figure. This is a conservative estimate compared to that obtained from the fit, but this enables us to account for systematic changes that would have occurred had we chosen another functional shape for the parameterization.

The two parameterizations of equations~\eqref{eqn:resolution} and ~\eqref{eqn:bias} are sufficient to characterize the Gaussian resolution function of the SD in the energy range discussed here.

\section{\label{sec:EnSp} Determination of the energy spectrum}

In this section, we describe the measurement of the energy spectrum, $J(E)$. Over parts of the energy range, we will describe it using $J(E) \propto E^{-\gamma}$, where $\gamma$ is the spectral index. In Sec.~\ref{sec:raw}, we present the initial estimate of the energy spectrum, dubbed the ``raw spectrum'', after explaining how we determine the SD efficiency, the exposure and the energy threshold for the measurement. In Sec.~\ref{sec:unfolded}, we describe the procedure used to correct the raw spectrum for detector effects, which also allows us to infer the spectral characteristics. The study of potential systematic effects is summarised in Sec.~\ref{sec:syst}, prior to a discussion of the features of the spectrum in Sec.~\ref{sec:features}.
 
\begin{figure*}[t]
	\centering
	\includegraphics[width=0.49\textwidth]{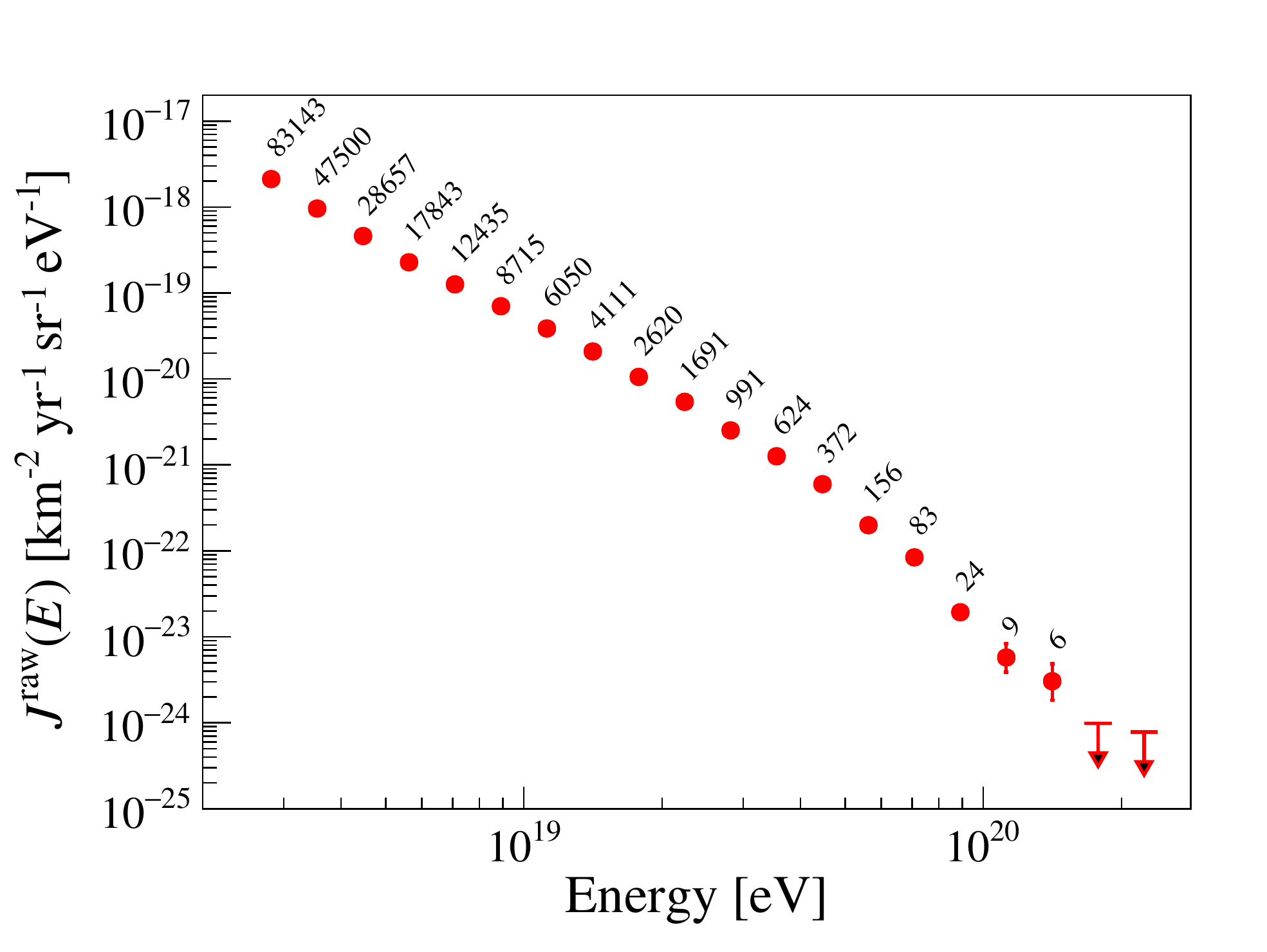}
	\includegraphics[width=0.49\textwidth]{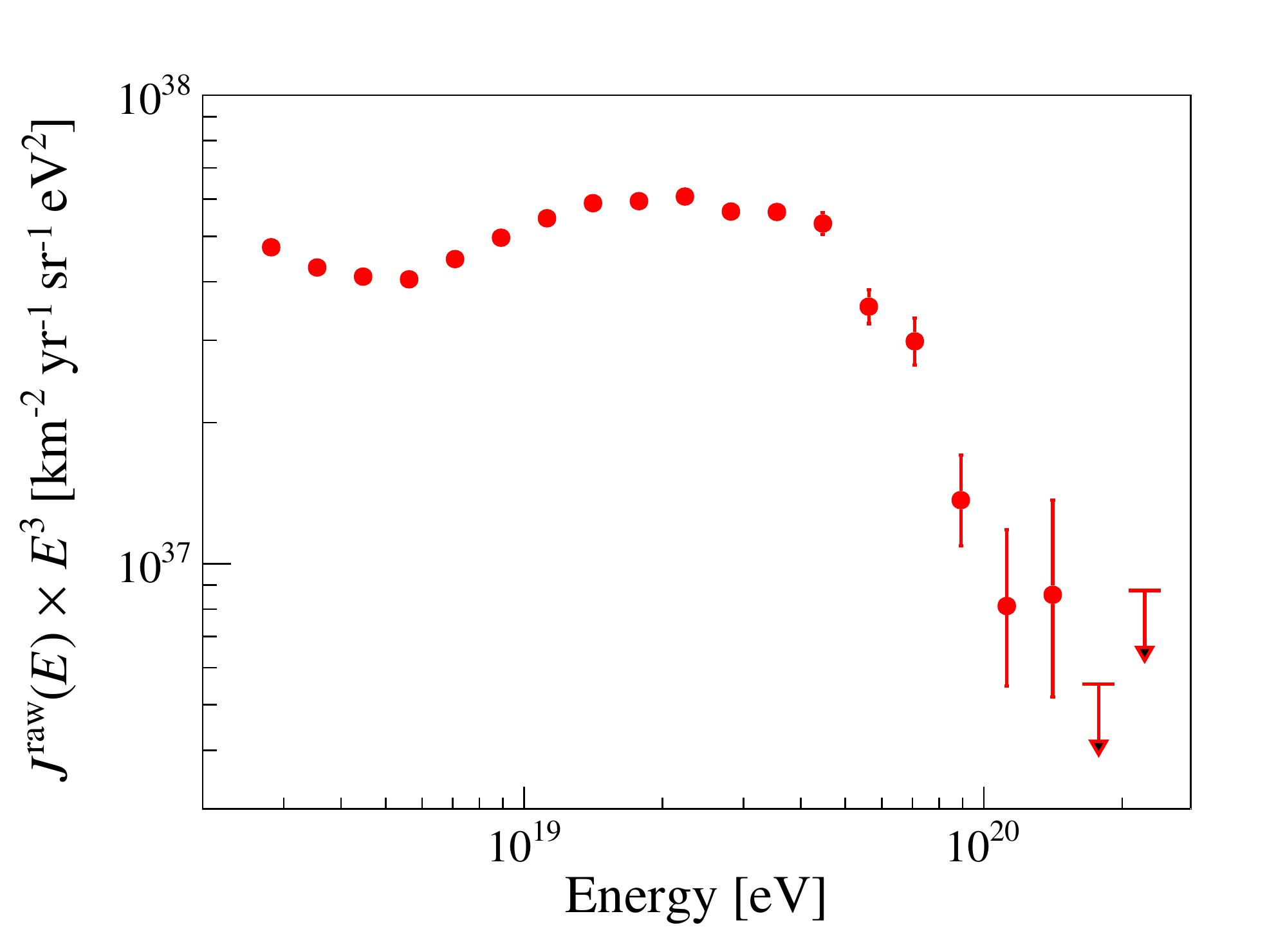}
	\caption{\small{Left: Raw energy spectrum $J_i^{\mathrm{raw}}$. The error bars represent statistical uncertainties. The number of events in each logarithmic bin of energy is shown above the points. Right: Raw energy spectrum scaled by the cube of the energy.}}
	\label{fig:RawSpectrum}
\end{figure*}

\subsection{\label{sec:raw} The raw spectrum}

An initial estimation of the differential energy spectrum is made by counting the number of observed events, $N_i$, in differential bins (centered at energy $E_i$, with width $\Delta E_i$) and dividing by the exposure of the array, $\mathcal{E}$,
\begin{equation}
J^{\mathrm{raw}}_i=\frac{N_i}{\mathcal{E}~\Delta E_i}.
\label{eqn:Jraw}
\end{equation}
The bin sizes, $\Delta E_i$, are selected to be of equal size in the logarithm of the energy, such that the width, $\Delta \log_{10}E_i=0.1$, corresponds approximately to the energy resolution in the lowest energy bin. The latter is chosen to start at $10^{18.4}$~eV, as this is the energy above which the acceptance of the SD array becomes  purely geometric and thus independent of the mass and energy of the primary particle. Consequently, in this regime of full efficiency, the calculation of $\mathcal{E}$ reduces to a geometrical problem dependent only on the acceptance angle, surface area and live-time of the array. 

The studies to determine the energy above which the acceptance saturates are described in detail in~\cite{Abraham:2010zz}. Most notably, we have exploited the events detected in hybrid mode as this has a lower threshold than the SD. Assuming that the detection probabilities of the SD and FD detectors are independent, the SD efficiency as a function of energy and zenith angle, $\epsilon(E,\theta)$, has been estimated from the fraction of hybrid events that also satisfy the SD trigger conditions. Above $10^{18}$~eV, the form of the detection efficiency (which will be used in the unfolding procedure described in Sec. \ref{sec:unfolded}) can be represented by an error function:
\begin{equation}
\label{eqn:trigeff}
\epsilon(E,\theta)=\frac{1}{2}\left[1+\mathrm{erf}\left(\frac{\log_{10}\left(E/\text{eV}\right)-p_0\left(\theta\right)}{p_1}\right)\right].
\end{equation}
where $p_1=0.373$ and $p_0(\theta) = 18.63 - 3.18\cos^2\theta + 4.38\cos^4\theta - 1.87\cos^6\theta$.

For energies above $E_\textrm{sat}=2.5{\times} 10^{18}~$eV, the detection efficiency becomes larger than 97\% and the exposure, $\mathcal{E}$, is then obtained from the integration of the aperture of the array over the observation time~\cite{Abraham:2010zz}. The aperture, $\mathcal{A}$, is in turn obtained as the effective area under zenith angle $\theta$, $A_0\cos\theta$, integrated over the solid angle
$\Omega$ within which the showers are observed. $A_0$ is well-defined as a consequence of the hexagonal structure of the layout of the array combined with the confinement criterion described in Sec.~\ref{sec:data}. Each station that has six adjacent, data-taking neighbors, contributes a cell of area $A_\text{cell} = 1.95~$km$^2$; 
the corresponding aperture for showers with $\theta\leq 60^\circ$ is $\mathcal{A_\text{cell}} = 4.59~$km$^2$~sr. The number of active cells, $n_\mathrm{cell}(t)$, is monitored second-by-second. The array aperture is then given, second-by-second, by the product of $\mathcal{A_\text{cell}}$ by $n_\mathrm{cell}(t)$. Finally, the exposure is calculated as the product of the array aperture by the number of live seconds in the period under study, excluding the time intervals during which the operation of the array is not sufficiently stable~\cite{Abraham:2010zz}. This results in a duty cycle larger than 95\%. 

Between 1 January 2004 and 31 August 2018 an exposure $(60{,}400 \pm 1{,}810)$ \,km$^2$\,sr\,yr was achieved. The resulting raw spectrum, $J_i^{\mathrm{raw}}$, is shown in Fig.~\ref{fig:RawSpectrum}, left panel. 
The energies in the x-axis correspond to the ones defined by the center of the logarithmic bins ($10^{18.45},  10^{18.55}, \cdots$~eV). 
The number of events $N_i$ used to derive the flux for each energy bin is also indicated. Upper limits are at the $90\%$ confidence level. 

The spectrum looks like a rapidly falling power law in energy with an overall spectral index of about $3$. To better display deviations from this function we also show, in the right panel, the same spectrum with the intensity scaled by the cube of the energy: the well-known ankle and suppression features are clearly visible at $\approx 5{\times}10^{18}$~eV and $\approx 5{\times}10^{19}$~eV, respectively.

\subsection{\label{sec:unfolded} The unfolded spectrum}

The raw spectrum is only an approximate measurement of the energy spectrum, $J(E)$, because of the distortions induced on its shape by the finite energy resolution. This causes events to migrate between energy bins: as the observed spectrum is steep, the migration happens especially from lower to higher energy bins, in a way that depends on the resolution function (see Sec.~\ref{sec:SDResponse}, Eq.~\eqref{eqn:kappa}). The shape at the lowest energies is in addition affected by the form of the detection efficiency (see Sec.~\ref{sec:raw}, Eq.~\eqref{eqn:trigeff}) in the range where the array is not fully efficient as events whose true energy is below $E_\textrm{sat}$ might be reconstructed with an energy above that. 

To derive $J(E)$ we use a bin-by-bin correction approach~\cite{Cowan}, where we first fold the detector effects into a model of the energy spectrum and then compare the expected spectrum thus obtained with that observed so as to get the unfolding corrections. The detector effects are taken into account through the following relationship, 
\begin{equation}
\label{eqn:Jfolded}
J^{\mathrm{raw}}(E_\textrm{SD};\mathbf{s})=\frac{\int\dif\Omega \cos{\theta}\int\dif  E \epsilon(E,\theta)J(E;\mathbf{s})\kappa(E_\textrm{SD}|E;\theta)}{\int\dif\Omega \cos{\theta}}
\end{equation}
where $\mathbf{s}$ is the set of parameters that characterizes the model. The model is used to calculate the number of events in each energy bin, $\mu_i(\mathbf{s})=\mathcal{E}\int_{\Delta E_i} \mathrm{d}E~J(E;\mathbf{s})$. 
The bin-to-bin migrations of events, induced by the finite resolution through Eq.~\eqref{eqn:Jfolded}, is accounted for by calculating the number of events expected between $E_i$ and $E_i+\Delta E_i$, $\nu_i(\mathbf{s})$, through the introduction of a matrix that depends only on the SD response function obtained from the knowledge of the $\kappa(E_\textrm{SD}|E)$ and $\epsilon(E)$ functions. 
To estimate $\mu_i$ and $\nu_i$, we use a likelihood procedure, aimed at deriving the set of parameters $\mathbf{s}_0$ allowing the best match between the observed number of events, $N_i$, and the expected one, $\nu_i$. Once the best-fit parameters are derived, the correction factors to be applied to the observed spectrum, $c_i$, are obtained from the estimates of $\mu_i$ and $\nu_i$ as $c_i=\mu_i/\nu_i$. More details about the likelihood procedure, the elements used to build the  matrix and the calculation of the $c_i$ coefficients are provided in Appendix~\ref{app:unfold}.

Guided by the raw spectrum, we infer the possible functional form for $J(E;\mathbf{s})$ by choosing parametric shapes naturally reproducing the main characteristics visible in Fig.~\ref{fig:RawSpectrum}. As a first step, we set out to reproduce a rapid change in slope (the ankle) followed by a slow suppression of the intensity at high energies. To do so, we use the 6-parameter function:
\begin{eqnarray}
J(E;\mathbf{s})&=&J_0\left(\frac{E}{E_{0}}\right)^{-\gamma_1}\left[1+\left(\frac{E}{E_{12}}\right)^{\frac{1}{\omega_{12}}}\right]^{(\gamma_1-\gamma_2)\omega_{12}} \nonumber \\ 
&{\times}&\frac{1}{1+(E/E_{\mathrm{s}})^{\Delta\gamma}}.
\label{eqn:J1}
\end{eqnarray}
In addition to the normalization, $J_0$, and to the arbitrary reference energy $E_0$ fixed to $10^{18.5}$~eV, the two parameters $\gamma_1$ and $\gamma_2$ approach the spectral indices around the energy $E_{12}$, identified with the energy of the ankle. The parameter $E_{\mathrm{s}}$ marks the suppression energy around which the spectral index slowly evolves from $\gamma_2$ to $\gamma_2+\Delta\gamma$. 
More precisely, it is the energy at which the flux is one half of the value obtained extrapolating the power law after the ankle. 
It is worth noting that the rate of change of the spectral index around the ankle is here determined by the parameter $\omega_{12}$ fixed at 0.05, which is the minimal value adopted to describe the transition given the size of the energy intervals.\footnote{With $\omega$=0.05, the transition between the two spectral indexes is roughly completed in $\Delta \log_{10}E=0.1$.}
Unlike a model forcing the change in spectral index to be infinitely sharp, such a choice of transition also makes it possible, subsequently, to test the speed of transition by leaving the parameters free.

We have used this function (Eq.~\eqref{eqn:J1}) to describe our data for over a decade. However, we find that with the exposure now accumulated, it no longer provides a satisfactory fit, with a deviance $D/n_\mathrm{dof}=35.6/14$. A more careful inspection of Fig.~\ref{fig:RawSpectrum} suggests a more complex structure in the region of suppression, with a series of power laws rather than a slow suppression. Consequently, we adopt as a second step a functional form describing a succession of power laws with smooth breaks:
\begin{eqnarray}
J(E;\mathbf{s})&=&J_0\left(\frac{E}{E_{0}}\right)^{-\gamma_1}\prod_{i=1}^3\left[1+\left(\frac{E}{E_{ij}}\right)^{\frac{1}{\omega_{ij}}}\right]^{(\gamma_i-\gamma_j)\omega_{ij}} 
\label{eqn:J2}
\end{eqnarray}
with $j=i+1$. This functional shape is routinely used to characterize the cosmic-ray spectrum at lower energies (see~\cite{Lipari2018} and references therein).  
\begin{figure}[b]
        \centering
        \includegraphics[width=0.5\textwidth]{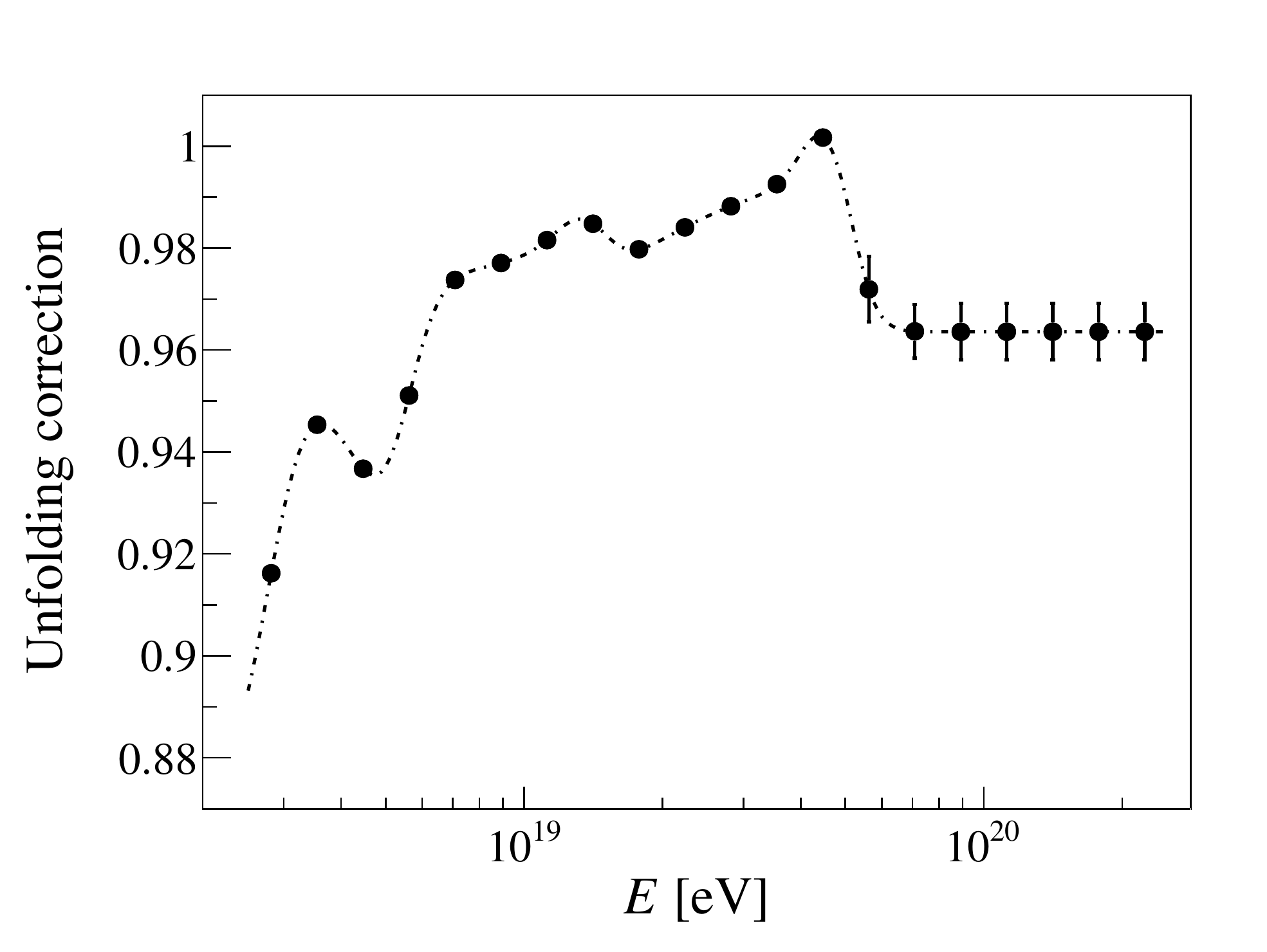}
        \caption{\small{ {Unfolding correction factor applied to the measured spectrum to account for the detector effects as a function of the cosmic-ray energy. 
        }}}
        \label{fig:UnfoldingFactors}
\end{figure}
\begin{figure*}[t]
        \centering
        \includegraphics[width=0.49\textwidth]{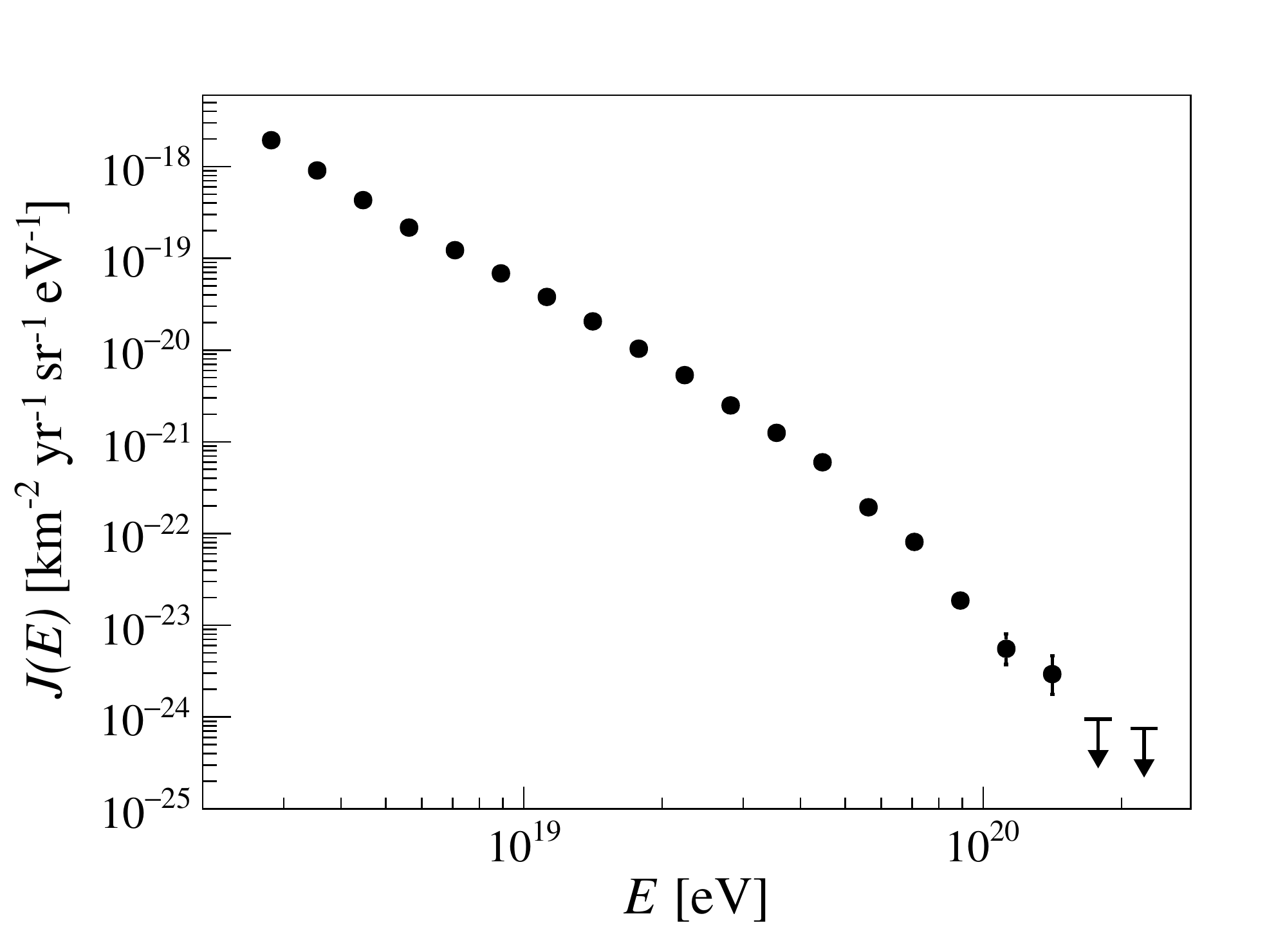}
        \includegraphics[width=0.49\textwidth]{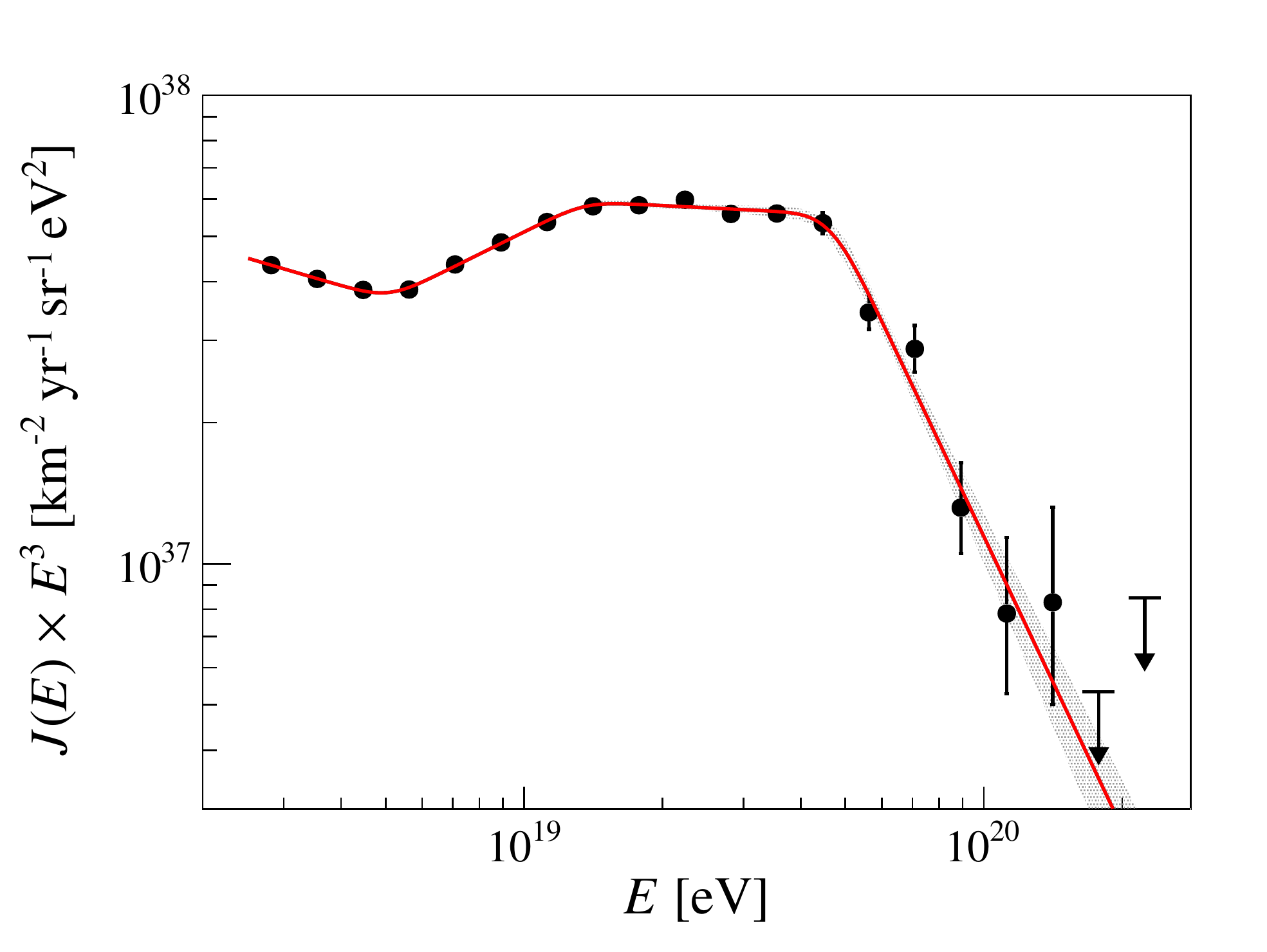}
        \caption{\small{Left: Energy spectrum. The error bars represent statistical uncertainties. Right: Energy spectrum scaled by $E^{3}$ and fitted with the function given by Eq.~\eqref{eqn:J2} with $\omega_{ij}=0.05$ (solid line). The shaded band indicates the statistical uncertainty of the fit.}}
        \label{fig:Spectrum}
\end{figure*}
The parameters $E_{23}$ and $E_{34}$ mark the transition energies between $\gamma_2$ and $\gamma_3$, and $\gamma_3$ and $\gamma_4$ respectively. The values of the $\omega_{ij}$ parameters are fixed, as previously, at the minimal value of 0.05. In total, this model has 8 free parameters and leads to a deviance of $D/n_\mathrm{dof}=17.0/12$. That this model better matches the data than the previous one is further evidenced by the likelihood ratio between these models which allows a rejection of Eq.~\eqref{eqn:J1} with $3.9\sigma$ confidence whose calculation is detailed in Appendix~\ref{app:unfold}.

As a third step, we release the parameters $\omega_{ij}$ one by one, two by two and all three of them so as to test our sensitivity to the speed of the transitions. 
Free parameters are only adopted as additions if the improvement to the fit is better than $2\sigma$. Such a procedure is expected to result in a uniform distribution of $\chi^2$ probability for the best-fit models, as exemplified in~\cite{Biasuzzi2019}.  
For every tested model, the increase in test statistics is insufficient to pass the $2\sigma$ threshold.

The adoption of Eq.~\eqref{eqn:J2} yields the coefficients $c_i$ shown as the black points in Fig.~\ref{fig:UnfoldingFactors} together with their statistical uncertainty. 
To be complete, we also show with a curve the coefficients calculated in sliding energy windows, to explain the behavior of the $c_i$ points.
This curve is determined on the one hand by the succession of power laws modeled by $J(E,\mathbf{s}_0)$, and on the other hand by the response function. The observed changes in curvature result from the interplay between the changes in spectral indices occurring in fairly narrow energy windows (fixed by the parameters $\omega_{ij}=0.05$) and the variations in the response function. At high energy, 
the coefficients tend towards a constant as a consequence of the approximately constancy of the resolution, because in such a regime, the distortions induced by the effects of finite resolution result in a simple multiplicative factor for a spectrum in power law. Overall, the correction factors are observed to be close to 1 over the whole energy range with a mild energy dependence. This is a consequence of the quality of the resolution achieved. 

We use the coefficients to correct the observed number of events to obtain the differential intensities as $J_i=c_i J_i^{\mathrm{raw}}$. This is shown in the left panel of  Fig.~\ref{fig:Spectrum}. The values of the differential intensities, together the detected and corrected number of events in each energy bin are given in Appendix~\ref{app:spectrumdata}. The magnitude of the effect of the forward-folding procedure can be appreciated from the following summary: above $2.5 {\times} 10^{18}~$eV, where there are 215,030 events in the raw spectrum, there are 201,976 in the unfolded spectrum; the corresponding numbers above $5 {\times} 10^{19}~$eV and $10^{20}~$eV are 278 and 269, and 15 and 14, respectively. Above $5{\times}10^{19}$~eV ($10^{20}$~eV), the integrated intensity of cosmic rays is $\left( 4.5 \pm 0.3 \right) {\times} 10^{-3}$~km$^{-2}$~yr$^{-1}$~sr$^{-1}$ 
($\left( 2.4 ^{+0.9}_{-0.6} \right) {\times} 10^{-4}$~km$^{-2}$~yr$^{-1}$~sr$^{-1}$).

\begin{table}[h]
\caption{Best-fit parameters, with statistical and systematic uncertainties, for the energy spectrum measured at the Pierre Auger Observatory.}
\label{tab:pars}
\begin{ruledtabular}
\begin{tabular}{l c}
 parameter & value $\pm \sigma_{\mathrm{stat.}} \pm \sigma_{\mathrm{sys.}}$ \\ 
\colrule
&\\[-1.0em]
 $J_{0}$ [km$^{-2}$sr$^{-1}$yr$^{-1}$eV$^{-1}$] & $ (1.315 \pm 0.004 \pm 0.400) {\times} 10^{-18}$ \\
 $\gamma_1$ & $3.29 \pm 0.02 \pm 0.10$\\
 $\gamma_2$ & $2.51 \pm 0.03 \pm 0.05$ \\
 $\gamma_3$ & $3.05 \pm 0.05 \pm 0.10$ \\
 $\gamma_4$ & $5.1  \pm 0.3  \pm 0.1$ \\
 $E_{12}$ [eV] (ankle) & $\left (5.0 \pm 0.1 \pm 0.8 \right) {\times} 10^{18}$ \\
 $E_{23}$ [eV] & $\left (13 \pm 1 \pm 2 \right) {\times} 10^{18}$ \\
 $E_{34}$ [eV] (suppression) & $\left (46 \pm 3 \pm 6 \right) {\times} 10^{18}$ \\
\hline
 $D/n_\mathrm{dof}$ & $17.0 / 12$ 
\end{tabular}
\end{ruledtabular}
\end{table}
In the right panel of Fig.~\ref{fig:Spectrum}, the fitted function $J(E,\mathbf{s}_0)$, scaled by $E^3$ to better appreciate the fine structures, is shown as the solid line overlaid on the data points of the final estimate of the spectrum. The characteristics of the spectrum are given in Table~\ref{tab:pars}, with both statistical and systematic uncertainties (for which a comprehensive discussion is given in the next section). These characteristics are further discussed in Sec.~\ref{sec:features}.

\begin{figure}[h]
        \centering
        \includegraphics[width=0.5\textwidth]{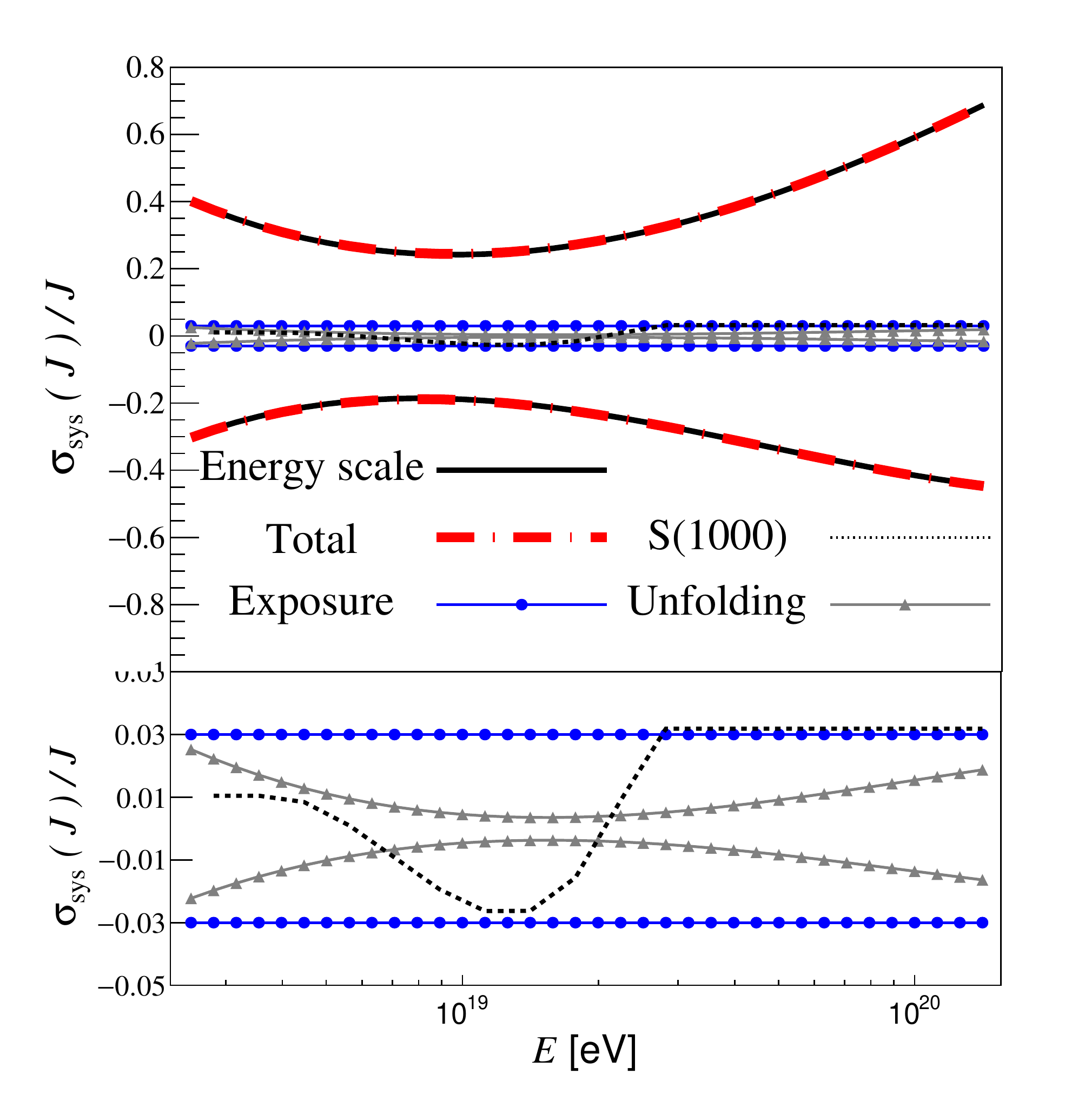}
        \caption{\small{Top panel: Systematic uncertainty in the energy spectrum as a function of the cosmic-ray energy (dash-dotted red line). The other lines represent the contributions of the different sources as detailed in the text: energy scale (continuous black), exposure (blue), $S(1000)$ (dotted black), unfolding procedure (gray). The  contributions  of  the  latter  three are zoomed in the bottom panel.   
        }}
        \label{fig:Syst}
\end{figure}

\subsection{\label{sec:syst} Systematic uncertainties}

There are several sources of systematic uncertainties which affect the measurement of the energy spectrum, as illustrated in Fig.~\ref{fig:Syst}. 

The  systematic uncertainty in the energy scale gives the largest contribution to the overall uncertainty.
As described in Sec.~\ref{sec:EnScale}, it amounts to about $14\%$ and is obtained by adding in quadrature 
all the systematic uncertainties in the FD energy estimation and the contribution arising from the statistical uncertainty in the calibration parameters. 
As the effect is dominated by the uncertainty in the calibration of the FD telescopes, the $14\%$ is almost energy independent. Therefore 
it has been propagated into the energy spectrum by changing the energy of all events by $\pm 14\%$ and then calculating 
a new estimation of the raw energy spectrum through Eq.~\eqref{eqn:Jraw}  and repeating the forward-folding procedure. 
When considering the resolution, the bias and the detection efficiency in the parameterization of the response function, the energy scale is shifted by $\pm 14\%$. The uncertainty in the energy scale 
translates into an energy-dependent uncertainty in the flux shown by a continuous black line in Fig.~\ref{fig:Syst}, top panel.
It amounts to $\simeq 30$ to $40\%$ around $2.5{\times}10^{18}$~eV, decreasing to $25\%$ around $10^{19}$~eV, and increasing again to $60\%$ at the highest energies.



A small contribution comes from the unfolding procedure. It stems from different sub-components: $(i)$ the functional form of the energy spectrum assumed, $(ii)$ the uncertainty in the bias and resolution parameterization determined in Sec.~\ref{sec:SDResponse} and $(iii)$ the uncertainty in the detection efficiency determined in Sec.~\ref{sec:raw}. The impact of contribution $(i)$ has been conservatively evaluated by comparing the output of the unfolding assuming Eq.~\eqref{eqn:J1} and Eq.~\eqref{eqn:J2}
and it is less than $1\%$ at all energies.
That of contribution $(ii)$ remains within $2\%$ and is maximal at the highest and lowest energies, while the one of contribution $(iii)$ is 
estimated propagating the statistical uncertainty in the 
fit function that parametrizes the detection efficiency (Eq.~\eqref{eqn:trigeff}) and it is 
within $\simeq$ $1\%$ below $4{\times}10^{18}$~eV and negligible above. The statistical uncertainties in the unfolding correction factors also contribute to the total systematic uncertainties in the flux and are taken into account. The overall systematic uncertainties due to unfolding are shown as a gray line in both panels of Fig.~\ref{fig:Syst} and are at maximum of $2\%$ at the lowest energies.

A third source is related to the global uncertainty of $3\%$ in the estimation of the integrated SD exposure~\cite{Abraham:2010zz}. This uncertainty, constant with energy, is shown as the blue line in both panels of Fig.~\ref{fig:Syst}.

A further component is related to the use of an average functional form for the LDF. The departure of this parameterized LDF from the actual one is source of a systematic uncertainty in $S(1000)$. This can be estimated using a subset of high quality events  for which the slope of the LDF~\cite{AugerRecoPaper} can be measured on an event by--event basis. The impact of this systematic uncertainty on the spectrum (shown as a black dotted line in Fig.~\ref{fig:Syst}) is around $2\%$ at $2.5 {\times} 10^{18}$~eV, decreasing to $-3\%$ at  $10^{19}$~eV, before rising again to $3\%$ above $\simeq$ $3{\times}10^{19}$~eV.
Other sources of systematic uncertainty have been  investigated and are negligible. 

\begin{figure}[h]
        \centering
        \includegraphics[width=0.5\textwidth]{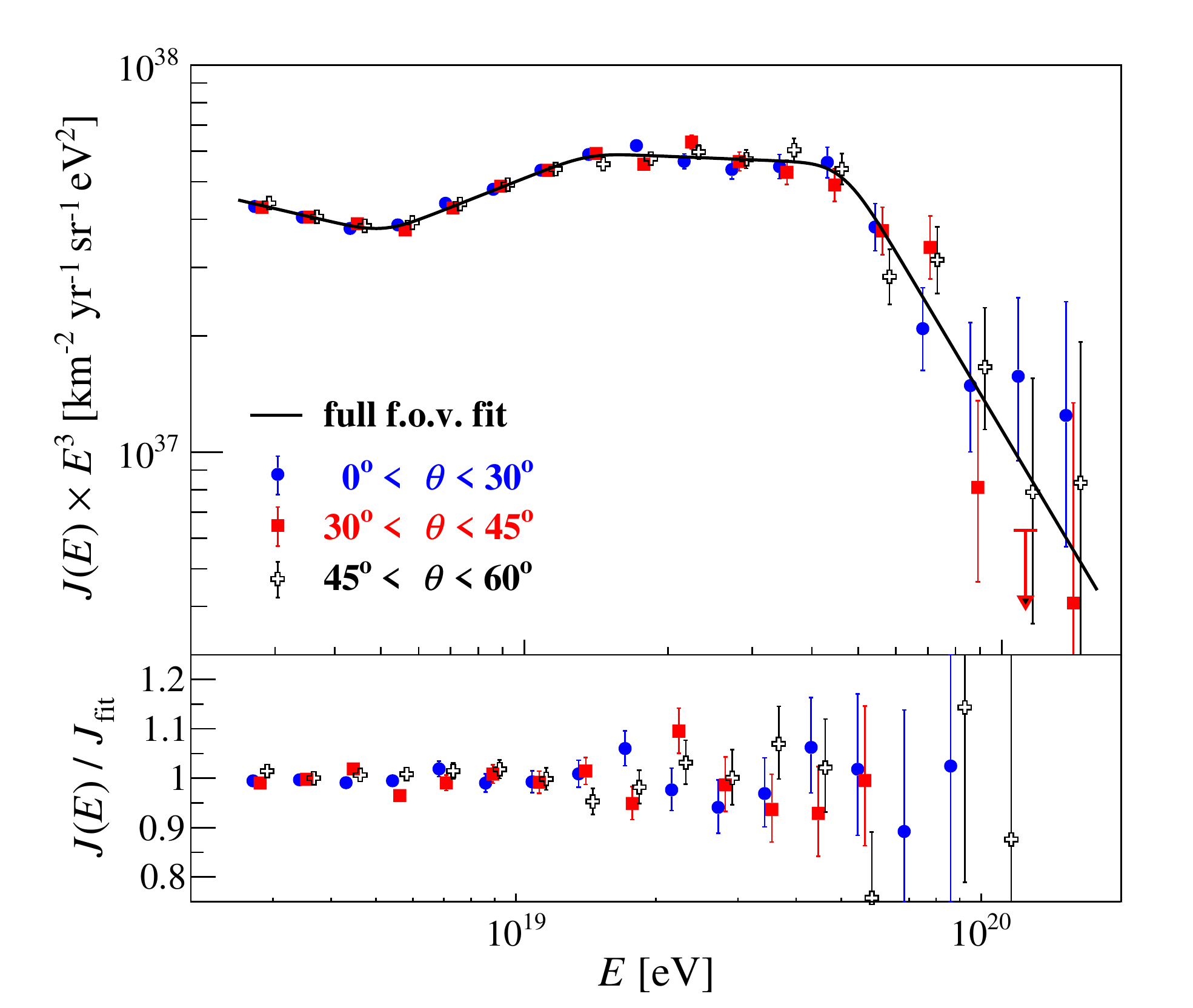}
	\caption{\small{Top panel: energy spectrum scaled by $E^3$ in three zenithal ranges of equal exposure. The solid line shows the results of the fit in the full f.o.v. presented in Sec.~\ref{sec:unfolded}.
	Bottom panel: relative difference between the spectra in the three zenithal ranges and the fitted spectrum in the full f.o.v..
	An artificial shift of $\pm 3.5\%$ is applied to the energies in the $x-$axis for the spectra obtained with the most and less inclined 
	showers to make easier to identify the different data points.
	}} 
\label{fig:J_vs_theta}
\end{figure}

We have performed several tests to assess the 
robustness  of  the  measurement. The spectrum, scaled by $E^3$, is shown in top panel of Fig.~\ref{fig:J_vs_theta} for three zenith angle intervals. Each interval is of equal size in $\sin^2{\theta}$ such that the exposure is the same, one third of the total one. The ratio of the three spectra to the results of the fit performed in the full field of view presented in Sec.~\ref{sec:unfolded} is shown in the bottom panel of the same figure.
The three estimates of the spectrum are in statistical agreement. In the region below 
$2 {\times} 10^{19}$~eV, where there are large numbers of events, the dependence on zenith angle is below 5\%. This is a robust demonstration of the efficacy of our methods.

We have also searched for systematic effects that might be seasonal to test the effectiveness of the corrections applied to $S(1000)$ to account for the influence 
of the changes in atmospheric temperature and pressure on the shower structure~\cite{AugerJINST2017}, and also searched for temporal effects as the data have been collected over a period of 14 years.  
Such tests have been performed by keeping the energy calibration curve determined in the full data taking period, as the systematic uncertainty associated with a non-perfect monitoring in time 
of the calibration of the FD telescopes is included in the overall $\pm 14\%$ uncertainty in the energy scale. 
The integral intensities above 10$^{19}$ eV for the four seasons are (0.271, 0.279, 0.269, 0.272)$\pm$0.004  km$^{-2}$ sr$^{-1}$ yr$^{-1}$  for winter, spring, summer and autumn respectively. The largest deviation with respect to the average of 0.273 $\pm$ 0.002 km$^{-2}$ sr$^{-1}$ yr$^{-1}$ is around 2\% for spring.  To look for long term effects we have divided the data into 5 sub--samples of equal number of events ordered in time. The integrated intensities above $10^{19}$~eV (corresponding to 16737 raw events) are (0.258, 0.272, 0.280, 0.280, 0.275)$\pm$0.005 km$^{-2}$ sr$^{-1}$ yr$^{-1}$, with a maximum deviation of 5\% with respect to the average value ($=$ 0.273 $\pm$ 0.002 km$^{-2}$ sr$^{-1}$ yr$^{-1}$). The largest deviation is in the first period (Jan 2004 -- Nov 2008) when the array was still under construction. 

The total systematic uncertainty, which is dominated by the uncertainty on the energy scale, is obtained by the quadratic sum of the described contributions and is depicted as a dashed red line in Fig.~\ref{fig:Syst}. 

The systematic uncertainties on the spectral  parameters  are also obtained adding in quadrature all the contributions  above described, and are shown in Table~\ref{tab:pars}. The uncertainties in the energy of the features ($E_{ij}$) and in the normalization parameter ($J_{0}$) are dominated by the uncertainty in the energy scale. On the other hand, those on the spectral indexes are also impacted by the other sources of systematic uncertainties.

\subsection{\label{sec:features} Discussion of the spectral features}
The unfolded spectrum shown in Fig.~\ref{fig:Spectrum} can be described using four power laws as detailed in Table~\ref{tab:pars} and equation~\eqref{eqn:J2}.  The well-known features of the ankle and the steepening are very clearly evident. The spectral index, $\gamma_3$, used to describe the new feature identified above $1.3 {\times} 10^{19}$~eV, differs from the index at lower energies, $\gamma_2$, by $\approx 4\sigma$ and from that in the highest energy region, $\gamma_4$, by $\approx 5\sigma$.

The representation of our data, and similar sets of spectral data, using spectral indices is long-established although, of course, it is unlikely that Nature generates exact power laws.  Furthermore these quantities are not usually derived from phenomenologically-based predictions. Rather it is customary to compare measurements to such outputs on a point-by-point basis (e.g.\ \cite{PhysRevD.74.043005,Aab_2017}). Accordingly, the data of Fig.~\ref{fig:Spectrum} are listed in Table~\ref{tab:Jdata}.  

\begin{figure}[t]
	\centering
	\includegraphics[width=0.5\textwidth]{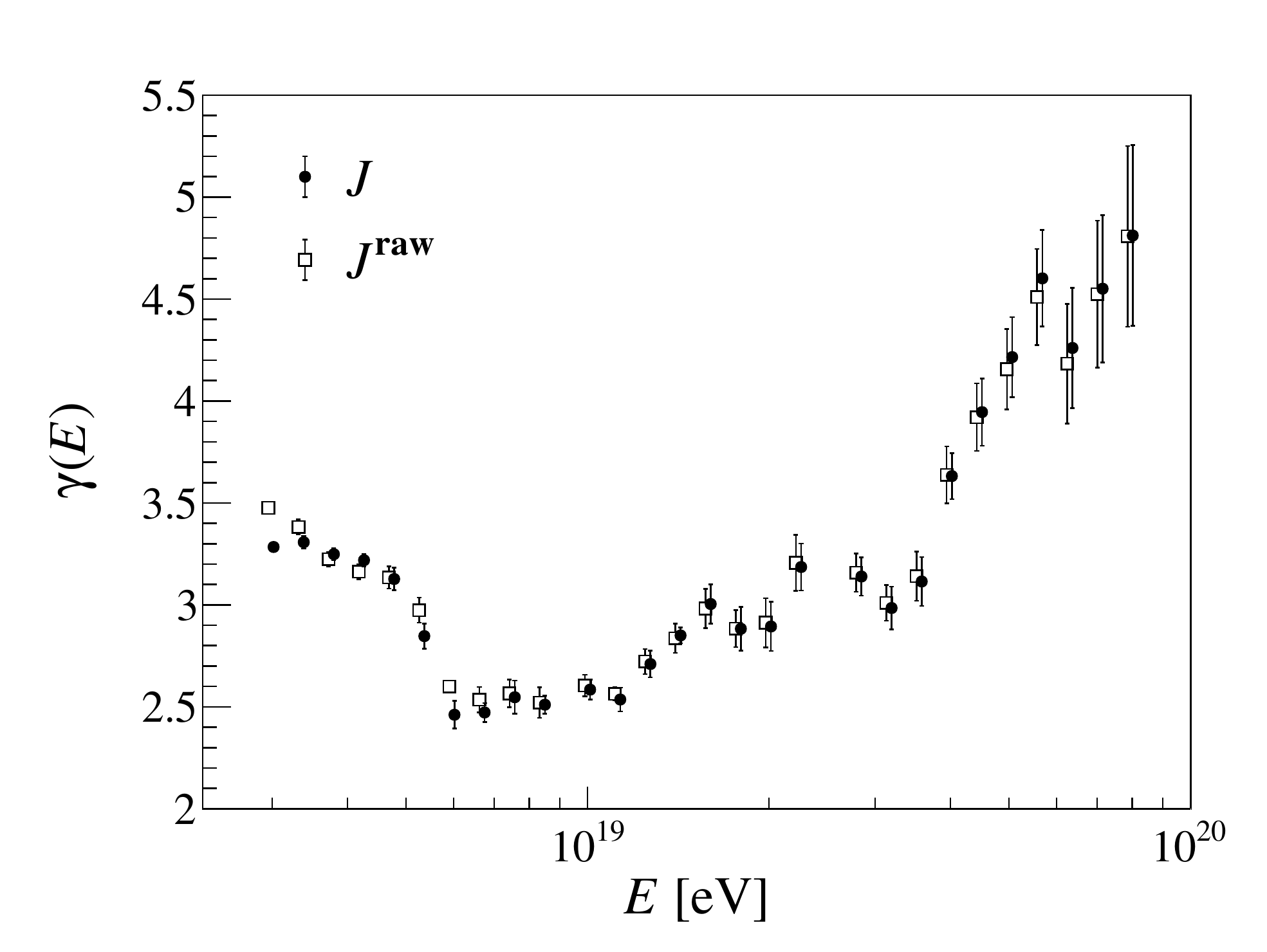}
	\caption{\small{Evolution of the spectral index as a function of energy. The spectral indices are derived from power-law fits to the raw and unfolded spectra performed over sliding windows in energy. Each slope is calculated using bins 
	$\Delta\log_{10}E=0.05$.  The width of the sliding windows are 3 bins at the lower energies and, to reduce the statistical fluctuations, are increased to 6 bins at the highest energies.
	}
	} 
	\label{fig:GammasVsEnergy}
\end{figure}

An alternative manner of presentation of the data is shown in Fig.~\ref{fig:GammasVsEnergy} where spectral indices have been computed over small ranges of energy (each point is computed for 3 bins at low energies growing to 6 at the highest energies).  The impact of the unfolding procedure is most clearly seen at the lowest energies (where the energy resolution is less good): the effect of the unfolding procedure is to sharpen the ankle feature.  It is also clear from Fig.~\ref{fig:GammasVsEnergy} that slopes are constant only over narrow ranges of energy, one of which embraces the new feature starting just beyond $10^{19}$~eV.  Above $\approx 5 {\times} 10^{19}$~eV, where the spectrum begins to soften sharply, it appears that $\gamma$ rises steadily up to the highest energies observed.  However, as beyond this energy there are only 278 events, an understanding of the detailed behaviour of the slope with energy must await further exposure.

\section{\label{sec:declination} The declination dependence of the energy spectrum}

\begin{figure*}[t]
        \centering
        \includegraphics[width=0.49\textwidth]{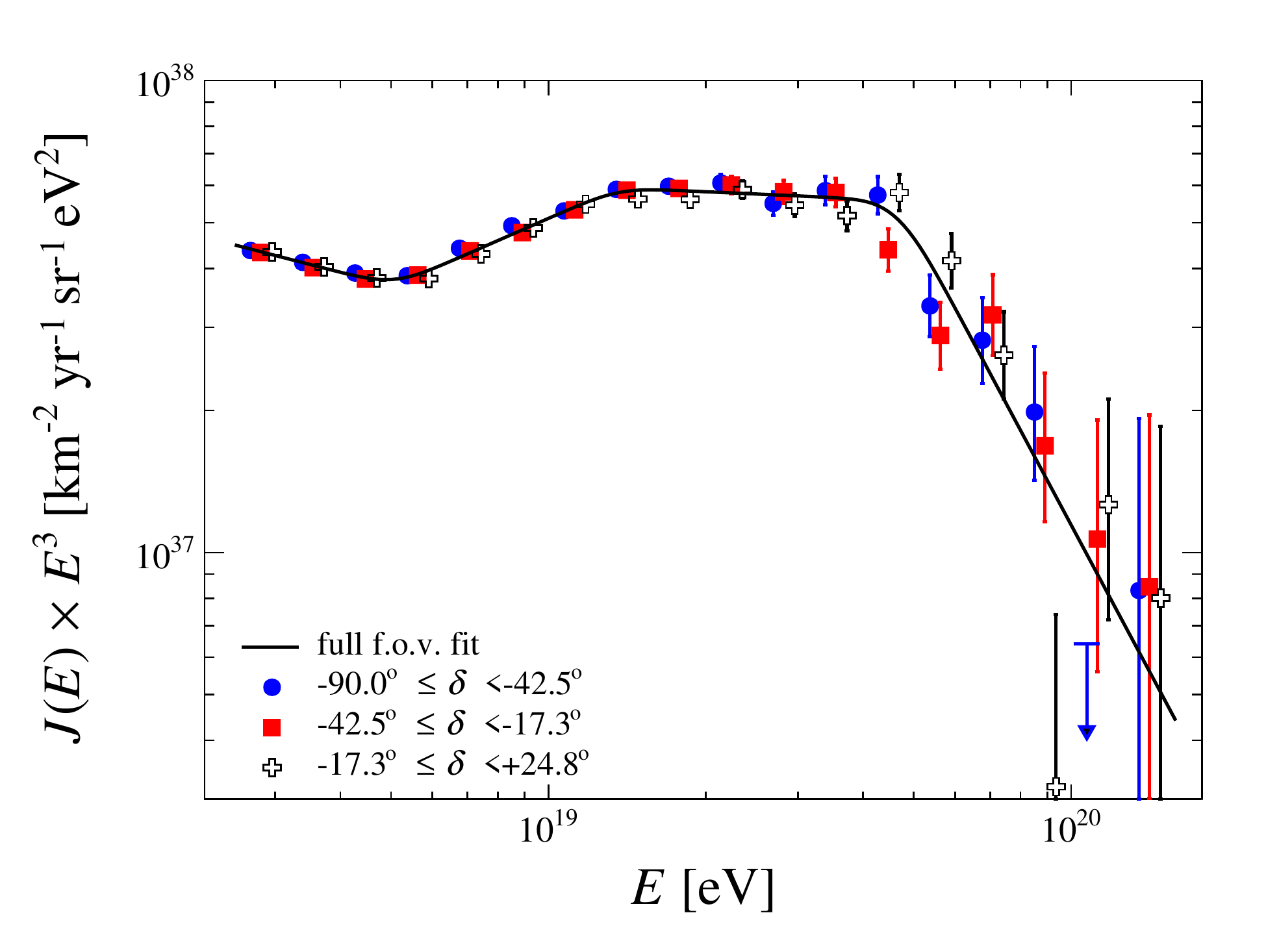}
        \includegraphics[width=0.49\textwidth]{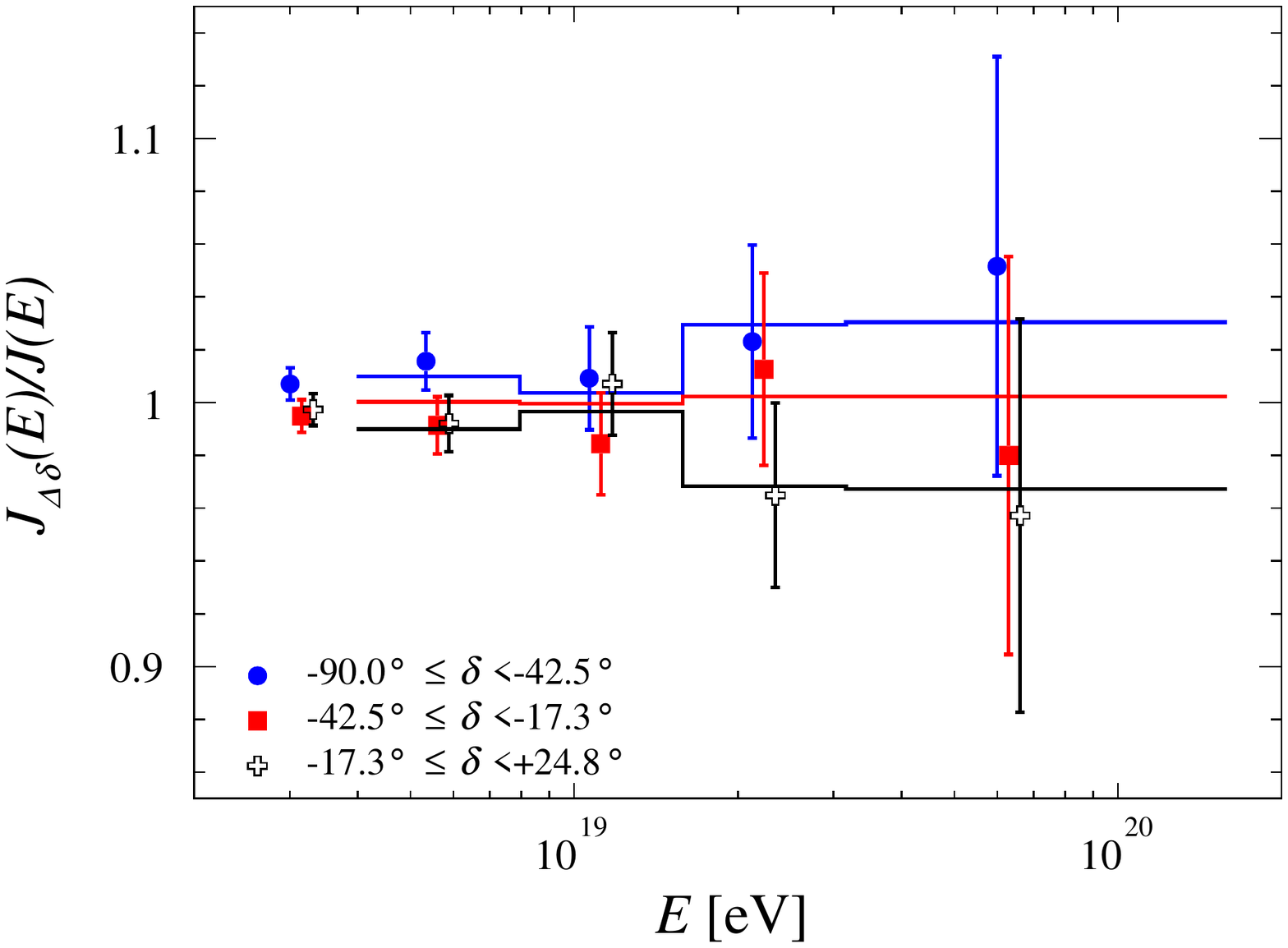}
        \caption{\small{Left: Energy spectra in three declination bands of equal exposure. Right: Ratio of the declination-band spectra to that of the full field-of-view. The horizontal lines show the expectation from the observed dipole~\cite{AugerAnis2018}. An artificial shift of $\pm 5\%$ is applied to the energies in the $x-$axis of the northernmost/southernmost declination spectra to make it easier to identify the different data points.}}
        \label{fig:DeclinationSpectrum}
\end{figure*}

In the previous section, the energy spectrum was estimated over the entire field of view, using the local horizon and zenith at the Observatory site to define the local zenithal and azimuth angles $(\theta,\varphi)$. Alternatively, we can make use of the fixed equatorial coordinates, right ascension and declination $(\alpha,\delta)$, aligned with the equator and poles of the Earth, for the same purpose. The wide range of declinations covered by using events with zenith angles up to $60^\circ$, from $\delta=-90^\circ$ to $\delta\simeq +24.8^\circ$ (covering 71\% of the sky), allows a search for dependencies of the energy spectrum on declination. We present below the determination of the energy spectrum in three declination bands and discuss the results. 

For each declination band under consideration, labelled as $k$, the energy spectrum is estimated as
\begin{equation}
J_{ik}=\frac{N_{ik}c_{ik}}{\mathcal{E}_k~\Delta E_i},
\label{eqn:J}
\end{equation}
where $N_{ik}$ and $c_{ik}$ stand for the number of events and the correction factors in the energy bin $\Delta E_i$ and in the declination band considered $k$, and  $\mathcal{E}_k$ is the exposure restricted to the declination band $k$. For this study, the observed part of the sky is divided into declination bands with equal exposure, $\mathcal{E}_k=\mathcal{E}/3$. The correction factors are inferred from a forward-folding procedure  identical to that described in section~\ref{sec:EnSp}, except that the response matrix is adapted to each declination band (for details see Appendix~\ref{app:unfold}).

The intervals in declination that guarantee that the exposure of the bands are each  $\mathcal{E}/3$ are determined by integrating the  directional exposure function, $\omega(\delta)$, derived in Appendix~\ref{app:direxp}, over the declination so as to satisfy
\begin{equation}
\label{eqn:exposure_domega}
\frac{\int_{\delta_{k-1}}^{\delta_k} \mathrm{d}\delta\cos{\delta}~\omega(\delta)}{\int_{\delta_0}^{\delta_3} \mathrm{d}\delta\cos{\delta}~\omega(\delta)}=\frac{1}{3},
\end{equation}
where $\delta_0=-\pi/2$ and $\delta_3=+24.8^\circ$. Numerically, it is found that $\delta_1=-42.5^\circ$ and $\delta_2=-17.3^\circ$.

The resulting spectra (scaled by $E^3$) are shown in the left panel of Fig.~\ref{fig:DeclinationSpectrum}. For reference, the best fit of the spectrum obtained in section~\ref{sec:unfolded} is shown as the black line. No strong dependence of the fluxes on declination is observed. 

\begin{table}[h]
\caption{Integral intensity above $8{\times} 10^{18}~$eV in the three declination bands considered.}
\label{tab:decintensity}
\begin{ruledtabular}
\begin{tabular}{l c}
 declination band & integral intensity [km$^{-2}$~yr$^{-1}$~sr$^{-1}$] \\
\colrule
 $-90.0^\circ\leq\delta < -42.5^\circ$ & $(4.17\pm0.04){\times}10^{-1}$ \\
 $-42.5^\circ\leq\delta < -17.3^\circ$ & $(4.11\pm0.04){\times}10^{-1}$ \\
 $-17.3^\circ\leq\delta < +24.8^\circ$ & $(4.11\pm0.04){\times}10^{-1}$ 
\end{tabular}
\end{ruledtabular}
\end{table}

To examine small differences, a ratio plot is shown in the right panel by taking the energy spectrum observed in the whole field of view as the reference. 
A weighted-average over wider energy bins is performed to avoid large statistical fluctuations preventing an accurate visual appreciation. For each energy, the data points are observed to be in statistical agreement with each other. Note that the same conclusions hold when analyzing data in terms of integral intensities, as evidenced for instance in table~\ref{tab:decintensity} above $8{\times}10^{18}~$eV. Similar statistical agreements are found above other energy thresholds. Hence this analysis provides no evidence for a strong declination dependence of the energy spectrum. 

A 4.6\% first-harmonic variation in the flux in right ascension has been observed in the energy bins above $8{\times} 10^{18}~$eV shown in the right panel of Fig.~\ref{fig:DeclinationSpectrum}~\cite{AugerAnis2018}. It is thus worth relating the data points reported here to these measurements that are interpreted as dipole anisotropies. The technical details to establish these relationships are given in Appendix~\ref{app:direxp}. 

The energy-dependent lines drawn in Fig.~\ref{fig:DeclinationSpectrum}-right show the different ratios of intensity expected from the dipolar patterns in each declination band relative to that across the whole field of view. The corresponding data points are observed, within uncertainties, to be in fair agreement with these expectations. 

Overall, there is thus no significant variation of the spectrum as a function of the declination in the field of view scrutinized here. A trend for a small declination dependence, with the flux being higher in the Southern hemisphere, is observed consistent with the dipolar patterns reported in~\cite{AugerAnis2018}. At the highest energies, the event numbers are still too small to identify any increase or decrease of the flux with the declinations in our field of view.

\section{\label{sec:ta} Comparison with other measurements}

\begin{figure}[h]
        \centering
        \includegraphics[width=0.5\textwidth]{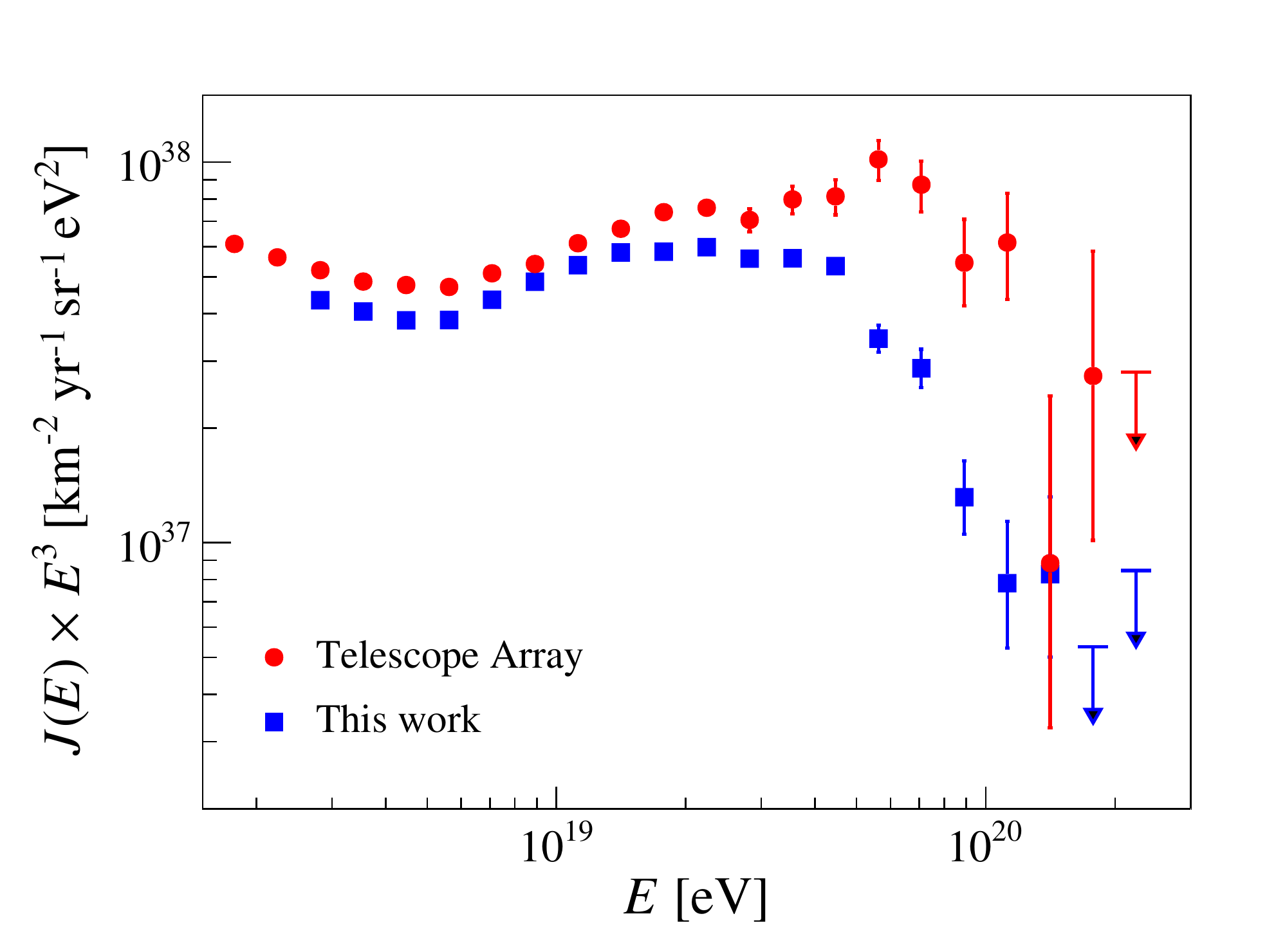}
        \caption{\small{Comparison between the $E^3$-scaled spectrum derived in this work and the one derived at the Telescope Array.}}
        \label{fig:AugerTA}
\end{figure}

Currently, the Telescope Array (TA) is the leading experiment dedicated to observing UHECRs in the northern hemisphere. As already pointed out, TA is also a hybrid detector making use of a 700~km$^2$ array of SD scintillators overlooked by fluorescence telescopes located at three sites. Although the techniques for assigning energies to events are similar, there are differences as to how the primary energies are derived, which result in differences in the spectral estimates, as can be appreciated in Fig.~\ref{fig:AugerTA} where the $E^3$-scaled spectrum derived in this work and the one derived by the TA  Collaboration~\cite{IvanovICRC19} are shown.

A useful way to appraise such differences is to make a comparison of the observations at the position of the ankle. Given the lack of anisotropy in this energy range, this spectral feature must be quasi-invariant with respect to direction on the sky. The energy at the ankle measured using the TA data is found to be $(4.9\pm0.1~(\mathrm{stat.})){\times} 10^{18}~$eV, with an uncertainty of 21\% in the energy scale~\cite{TA-escale}
in good agreement with the one reported here ($(5.0\pm0.1~(\mathrm{stat.})\pm0.8~(\mathrm{sys.})){\times} 10^{18}~$eV). 
Consistency between the two spectra can be obtained in the ankle-energy region up to $\simeq 10^{19}~$eV by rescaling the energies by $+5.2\%$ for Auger and $-5.2\%$ for TA. The factors are smaller than the current systematic uncertainties in the energy scale of both experiments. These values encompass the different fluorescence yields adopted by the two Collaborations, the uncertainties in the absolute calibration of the fluorescence telescopes, the influence of the atmospheric transmission used in the reconstruction, the uncertainties in the shower reconstruction, and the uncertainties in the correction factor for the invisible energy. It is worth noting that better agreement can be obtained if the same models are adopted for the fluorescence yield and for the invisible energy correction. Detailed discussions on these matters can be found in~\cite{VIT2018}. 

However, even after the rescaling, differences persist above $\simeq 10^{19}~$eV. At such high energies, anisotropies might increase in size and induce differences in the energy spectra detected in the northern and southern hemispheres. To disentangle possible anisotropy issues from systematic effects, a detailed scrutiny of the spectra in the declination range accessible to both observatories has been carried out~\cite{VerziUHECR16}. A further empirical, energy-dependent, systematic shift of $+10\%$ ($-10\%$) per decade for Auger (TA) is required to bring the spectra into agreement. A comprehensive search for energy-dependent systematic uncertainties in the energies has resulted in possible non-linearities in this decade amounting to $\pm 3\%$ for Auger and $(-0.3\pm9)\%$ for TA, which are insufficient to explain the observed effect~\cite{IvanovUHECR18}. A joint effort is underway to understand further the sources of the observed differences and to study their impact on the spectral features~\cite{DelignyICRC19}.

\section{Summary}
\label{sec:discussion}

We have presented a measurement of the energy spectrum of cosmic rays for energies above $2.5{\times}10^{18}~$eV based on 215,030 events recorded with zenith angles below $60^\circ$. The corresponding exposure of 60,400 km$^2$~yr~sr, calculated in a purely geometrical manner, is independent of any assumption on unknown hadronic physics or primary mass composition. This measurement relies on estimates of the energies that are similarly independent of such assumptions. This includes the analysis that minimizes the model/mass dependence of the invisible energy estimation as presented in~\cite{InvisibleEnergy2019}.
In the same manner, the flux correction for detector effects is evaluated using a data-driven analysis.
Thus the approach adopted differs from that of \textit{all} other spectrum determinations above $\simeq 5{\times}10^{14}~$eV where the air-shower phenomenon is used to obtain information. 

The measurement reported above is the most precise made hitherto and is dominated by systematic uncertainties except at energies above $\simeq 5{\times}10^{19}~$eV. The systematic uncertainties have been discussed in detail and it is shown that the dominant one ($\simeq 14$\%) comes from the energy scale assigned using measurements of the energy loss by ionisation in the atmosphere inferred using the fluorescence technique. 

In summary, the principal conclusions that can be drawn from the measurement are:
\begin{enumerate}
\item The flattening of the spectrum near $5{\times}10^{18}~$eV, the so-called ``ankle'', is confirmed.
\item The steepening of the spectrum at around $\simeq 5{\times}10^{19}~$eV is substantiated.  
\item A new feature has been identified in the spectrum: in the region above the ankle the spectral index changes from $2.51 \pm 0.03~{\rm (stat.)} \pm 0.05~{\rm (sys.)}$ to $3.05 \pm 0.05~{\rm (stat.)} \pm 0.10~{\rm (sys.)}$ before increasing sharply to $5.1 \pm 0.3~{\rm (stat.)} \pm 0.1~{\rm (sys.)}$ above $5 {\times} 10^{19}~$eV.
\item No evidence for any dependence of the energy spectrum on declination has been found other than a mild excess from the Southern Hemisphere that is consistent with the anisotropy observed above $8{\times}10^{18}~$eV.
\end{enumerate}
A discussion of the significance of these measurements from astrophysical perspectives can be found in~\cite{AugerPRL2019}.


\appendix 

\section{Expected distribution of cosmic rays in local zenith angle}
\label{app:sin2} 

To derive the attenuation curves in a purely data-driven way, we have required the same intensity in equal intervals of $\sin^2{\theta}$. This requirement relies on the high level of isotropy of the cosmic-ray intensity. In this appendix it is shown that neither the small anisotropies at the energy thresholds of interest, nor the slight zenithal dependencies of the response function of the SD array, alter the constancy of the intensity in terms of $\sin^2{\theta}$, by more than $\simeq 0.5$\% at $3{\times}10^{18}$~eV.  This is less than the magnitude of the statistical fluctuations at this energy.

\begin{figure*}[!t]
        \centering
        \includegraphics[width=0.32\textwidth]{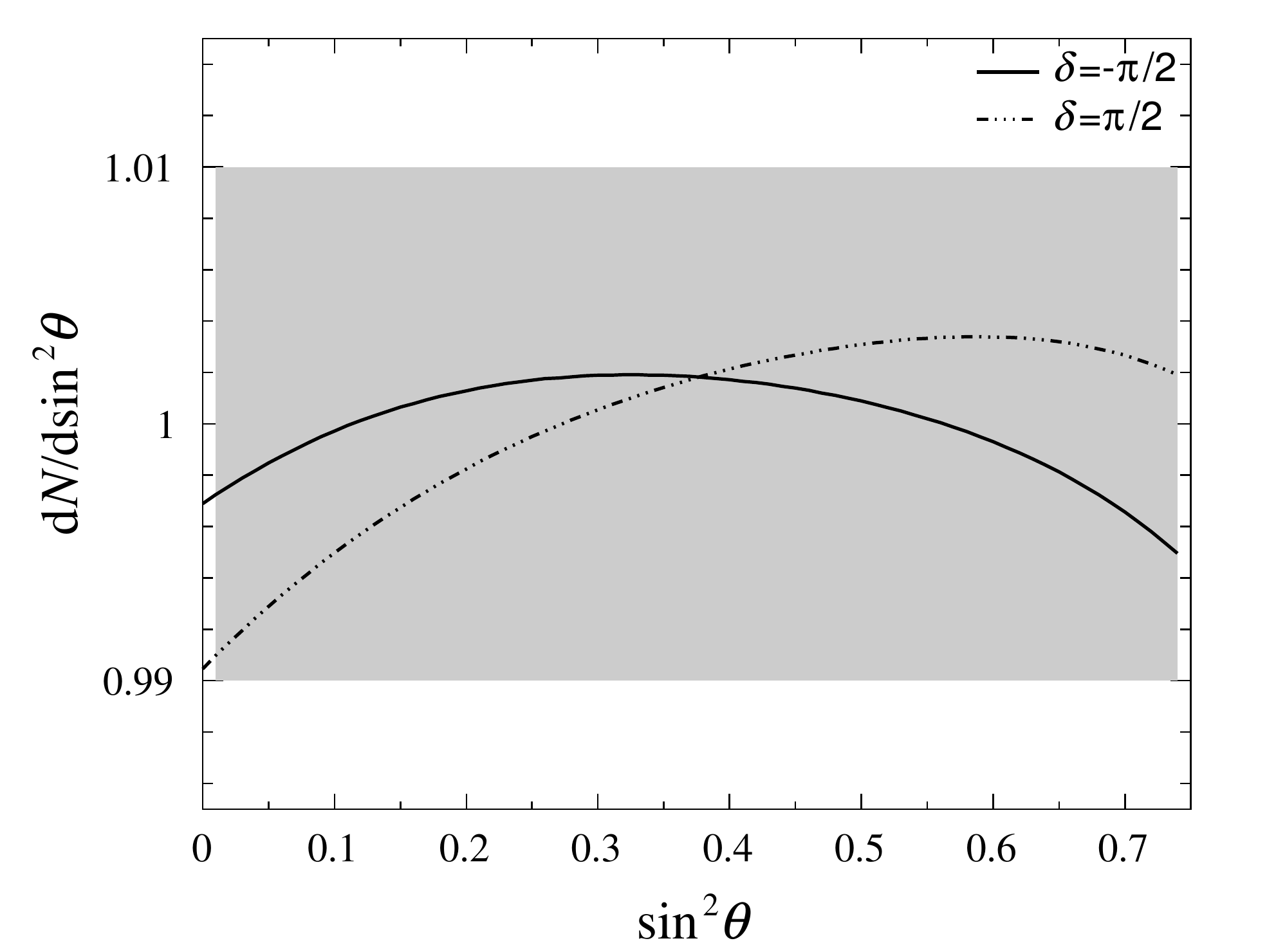}
        \includegraphics[width=0.32\textwidth]{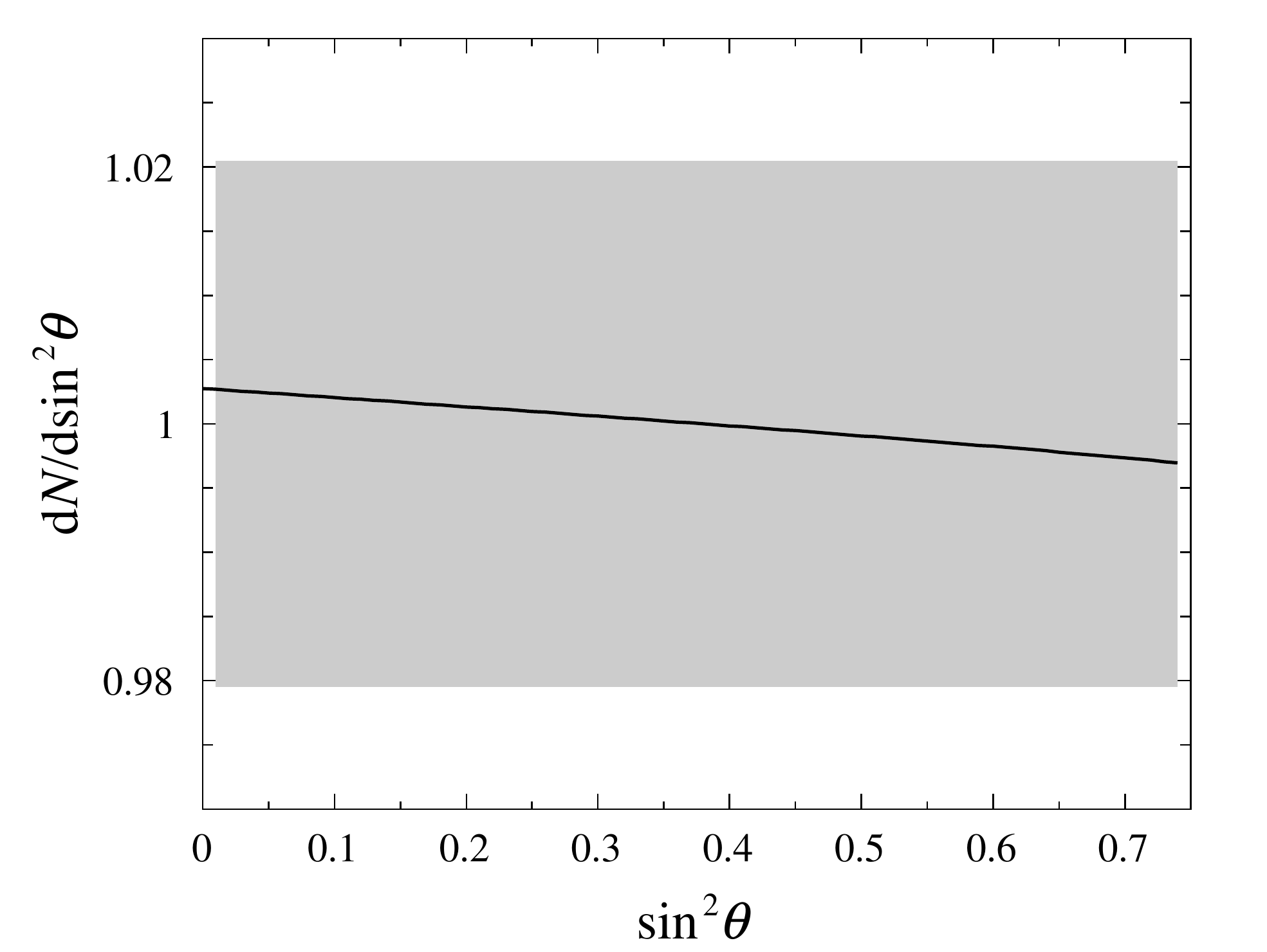}
        \includegraphics[width=0.32\textwidth]{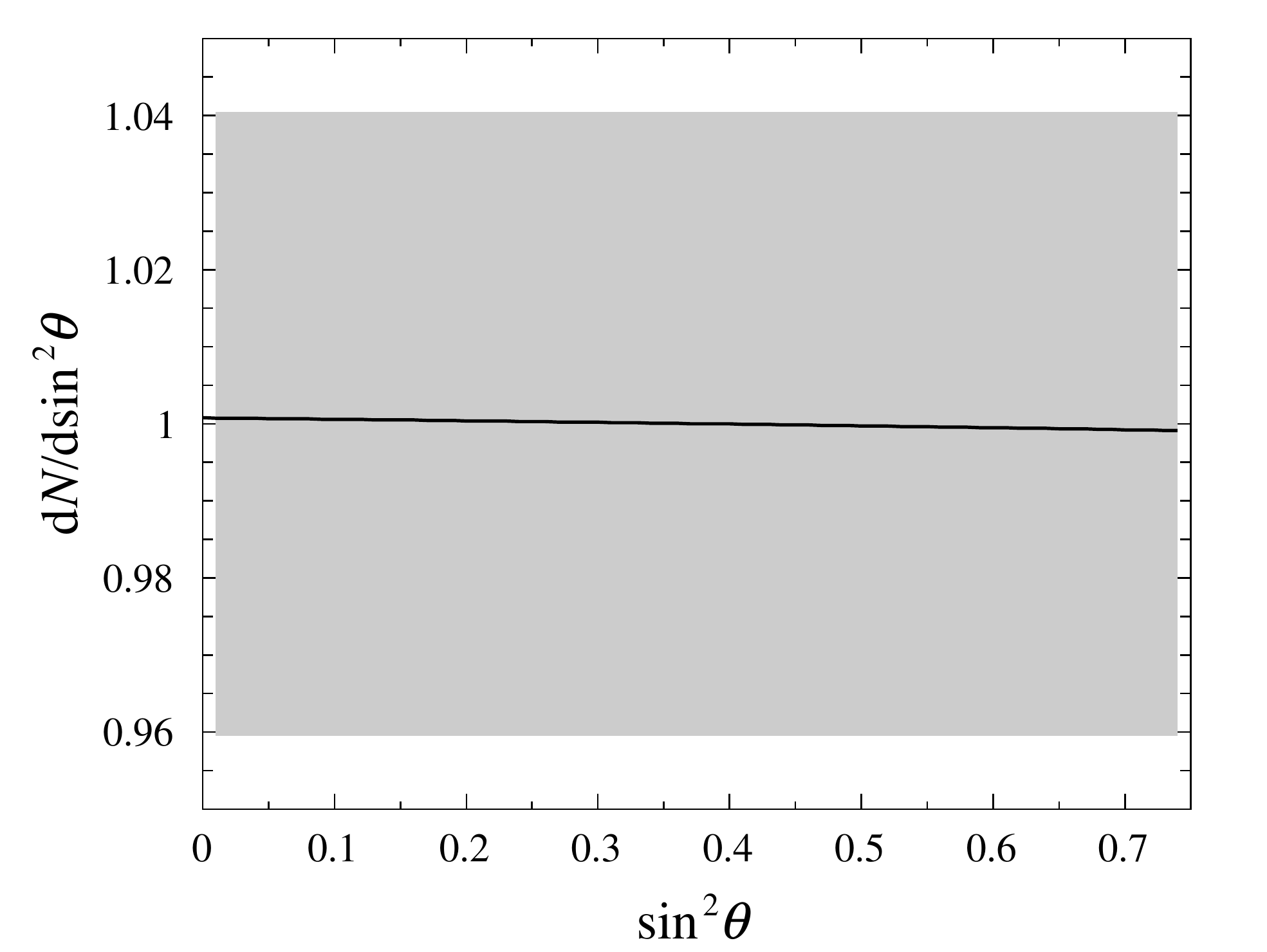}
        \caption{\small{Expected distribution in $\dif N(>E_0)/\dif\sin^2{\theta}$, normalized to its average value, taking into account the effects of the response function of the SD array and of anisotropies. The shaded band stands for the statistical fluctuations in each bin of $\sin^2{\theta}$ above the energy threshold used, given the number of events at our disposal. Left: $E_0=3{\times}10^{18}~$eV, using a conservative estimate of a dipolar anisotropy in directions maximizing the non-uniformity in $\dif N/\dif\sin^2{\theta}$ (see text). Middle: $E_0=8{\times}10^{18}$~eV, using the measured dipolar anisotropy. Right: $E_0=1.6{\times}10^{19}$~eV, using the measured dipolar anisotropy.}}
        \label{fig:dNdsin2}
\end{figure*}

Consider a possible directional dependence for the energy spectrum, $J(E,\theta(\alpha,\delta,t),\varphi(\alpha,\delta,t))$. Accounting for the energy resolution, the expected event number per steradian and per unit time above any energy threshold $E_0$ is given by\footnote{Note that we neglect here the time dependence of the detection efficiency in the regime $\epsilon<1$, induced by the modulations of the energies through the changes of atmospheric conditions at the time of detection of the events. This is reasonable since, for the energy thresholds $E_0$ explored here, the values of the detection efficiencies occurring throughout the integration in Eq.~\eqref{eqn:d2N_dn_dt} are always larger than $\simeq 0.9$. The time-dependent relative changes are thus negligible.}

\begin{widetext}
\begin{equation}
\label{eqn:d2N_dn_dt}
\frac{\dif^2N(>E_0)}{\dif\Omega~\dif t}=n_\mathrm{cell}(t)A_\mathrm{cell}\cos{\theta}\int_{>E_0}\dif E_\textrm{SD}\int\dif E~\epsilon(E,\theta)~\kappa(E_\textrm{SD}|E;\theta)~J(E,\theta(\alpha,\delta,t),\varphi(\alpha,\delta,t)).
\end{equation}
To get the expected number of events in each of the $\sin^2{\theta}$ intervals, it is convenient to consider the left hand side of Eq.~\eqref{eqn:d2N_dn_dt} expressed in terms of local sidereal time through the transformation
\begin{equation}
\label{eqn:d2N_dn_da0}
\frac{\dif^2N(>E_0)}{\dif\Omega~\dif\alpha_0}=\int\dif t~\frac{\dif^2N(>E_0)}{\dif\Omega~\dif t}~\delta(\alpha_0-\alpha_0(t)).
\end{equation}
Inclusion of the Dirac function guarantees that the direction $\alpha_0(t)$ considered throughout the time integration corresponds to the local sidereal time $\alpha_0$ seen at time $t$. On inserting Eq.~\eqref{eqn:d2N_dn_dt} into Eq.~\eqref{eqn:d2N_dn_da0} and carrying out the integration over time, $\dif^2N/\dif\Omega\dif\alpha_0$ becomes
\begin{equation}
\label{eqn:d2N_dn_da0_bis}
\frac{\dif^2N(>E_0)}{\dif\Omega~\dif\alpha_0}\simeq\frac{N_\mathrm{cell}(\alpha_0)\Delta t}{2\pi}\int\dif E_\textrm{SD}\int\dif E~\epsilon(E,\theta)~A_\mathrm{cell}\cos{\theta}~\kappa(E_\textrm{SD}|E;\theta)~J(E,\theta(\alpha,\delta,\alpha_0),\varphi(\alpha,\delta,\alpha_0)),
\end{equation}

where the notation $N_\mathrm{cell}(\alpha_0)$ stands for the total number of active hexagonal cells during the integrated observation time for a flux of cosmic rays from each direction $\alpha_0$, $N_\mathrm{cell}(\alpha_0)\equiv\int\dif t~n_\mathrm{cell}(t)\delta(\alpha_0-\alpha_0(t))$. Due to the long period considered here ($\simeq 4{,}880$ sidereal days), an averaging takes place and this function is nearly uniform, $N_\mathrm{cell}(\alpha_0)=N_\mathrm{cell}^0+\delta N_\mathrm{cell}(\alpha_0)$, with $\delta N_\mathrm{cell}(\alpha_0)/N_\mathrm{cell}^0$ of the order of a few $10^{-3}$. The expected $\dif^2N/\dif\sin^2{\theta}$ distribution is then obtained by integrating Eq.~\eqref{eqn:d2N_dn_da0_bis} over azimuth $\varphi$ and local sidereal time $\alpha_0$. 

Characterizing anisotropies, to first order, by a dipole vector with amplitude $d$ and equatorial directions $(\alpha_\mathrm{d},\delta_\mathrm{d})$, an effective ansatz for the spectrum is then $J(E,\theta,\varphi,\alpha_0)=J_0(E)(1+\mathbf{d}(\alpha_\mathrm{d},\delta_\mathrm{d})\cdot\nn(\theta,\varphi))$, with $\nn(\theta,\varphi)$ the unit vector on the sphere. The integration over the azimuthal angle  in Eq.~\eqref{eqn:dN_dsin2} selects the coordinate of the dipole vector along the local $z-$axis defining the zenith angle. The remaining integration over the local sidereal time $\alpha_0$ cancels the harmonic contribution $\delta N_\mathrm{cell}(\alpha_0)$ coupled to the isotropic part of the spectrum, and for small anisotropies, the leading order of $\dif N/\dif\sin^2{\theta}$ reads as
\begin{equation}
\label{eqn:dN_dsin2}
\frac{\dif N(>E_0)}{\dif\sin^2{\theta}}\simeq2\pi N_\mathrm{cell}^0A_\mathrm{cell}\Delta t\int\dif E_\textrm{SD}\int\dif E~\epsilon(E,\theta)~J_0(E)~\kappa(E_\textrm{SD}|E;\theta)\left(1+d\sin\ell\sin\delta_\mathrm{d}\cos\theta\right),
\end{equation}
which is the desired expression.
\end{widetext}

Searches for anisotropies throughout the EeV energy range have been reported previously~\cite{AugerAnis2018}. We show in Fig.~\ref{fig:dNdsin2} the expected departures from a uniform behaviour for the $\dif N(>E_0)/\dif\sin^2{\theta}$ distribution normalized to its average value, above three different thresholds. The shaded band stands for the statistical fluctuations in each bin of $\sin^2{\theta}$ above the energy threshold under consideration, given the number of events available. Above $3{\times}10^{18}~$eV (left panel), the upper limit previously obtained with a smaller data set between $2{\times}10^{18}~$eV and $4{\times}10^{18}~$eV is used, and the two directions maximizing the departure from a uniform behavior are selected ($\delta_{\mathrm{d}}=\pm \pi/2$). Above $8{\times}10^{18}~$eV and $1.6{\times}10^{19}~$eV, dipole parameters reported in~\cite{AugerAnis2018} are adopted. One can see that the small departures from uniformity all lie within the limits set by the statistical uncertainties, validating the use of the CIC method to derive the attenuation curves at the different energy thresholds.


\section{Energy calibration of $S_{38}$}
\label{app:ES38fit} 

To ease the notations in this appendix, we denote here as $S$ the underlying shower-size parameter $S_{38}$, and as $S_\SD$ its estimator. The probability density function, $f_0(E,S)$, to detect an hybrid event of underlying energy $E$ and shower size $S$ at ground that would be expected without shower-to-shower fluctuations and with an infinite energy resolution is proportional to
\begin{equation}
f_0(E,S) \propto  P_{\SD}(S)~h(E)~\delta( S - S(E) ),
\end{equation}
where $P_{\SD}(S)$ is the SD detection efficiency expressed in terms of $S$ and $h(E)$ is the energy distribution of the events, that is, the underlying energy spectrum multiplied by the effective exposure for the hybrid events. The Dirac function guarantees that the shower size can be modelled with a function $S(E)$ of the true energy. The probability density function for the detection of an hybrid shower that includes the finite energy resolution can be derived by folding $f_0(E,S)$ with both FD and SD resolution functions, taken as Gaussian distributions with standard deviations $\sigma_{\FD}$ and $\sigma_{\SD}$, respectively: 
\begin{widetext}
\begin{equation}
\label{SDCalib_pdf_1}
f(E_{\FD},S_{\SD}) \propto \int_{>0} \hspace{-0.2cm}\dif E  ~G(E_{\FD}|E,\sigma_{\FD}) ~G(S_{\SD}|S(E),\sigma_{\SD})~P_{\SD}(S(E))~h(E),
\end{equation}
where $E_{\FD}$ is the energy measured by the FD and $S_{\SD}$ is the shower size at ground measured by the SD. Here, $\sigma_{\SD}$ accounts for both the shower-to-shower fluctuations of the shower size, $\sigma_{\sh}$, and for the detector resolution, $\sigma_{\det}$:
\begin{equation}
\sigma_{\SD} = \sqrt{\sigma_{\det}^2 + \sigma_{\sh}^2 }.
\end{equation}

The function $h(E)$ is significantly less steep than the energy spectrum because the criteria to select high-quality FD events and to guarantee an unbiased $X_{\mathrm{max}}$ distribution are more effective at lower energies. Its estimation through Monte-Carlo simulations to the required accuracy is a difficult task. It is preferable to follow the strategy put forward in~\cite{Dembinski:2015wqa}, that consists of deriving $h(E)$ from the data distribution using the boostrap approximation 
\begin{equation}
\label{SDCalib_bootstrap}
 h(E) \approx \frac{1}{N} ~\sum_i \delta(E-E_{\FD i}),
\end{equation}
where the sum runs over the hybrid events. The hybrid events are selected according to the criteria reported in section~\ref{sec:EnScale}, at energies 
where the SD is almost fully efficient. This allows us to neglect the effect of $P_{\SD}(S(E))$. Inserting Eq.~\eqref{SDCalib_bootstrap} into~\eqref{SDCalib_pdf_1} leads to
\begin{equation}
\label{SDCalib_pdf_3}
 f(E_{\FD},S_{\SD}) \approx \frac{1}{N} ~\sum_i G(E_{\FD}|E_{\FD i},\sigma_{\FD i}) ~ G(S_{\SD} |  S(E_{\FD i}), \sigma_{\SD i} )   
\end{equation}
where $\sigma_{\FD i}$ and $\sigma_{\SD i}$ are the uncertainties on the measurements evaluated on an event-by-event basis. 

Finally the calibration parameters $A$ and $B$ that enter in  $f(E_{\FD},S_{\SD})$ through the relationship $S(E) = (E/A)^{1/B}$ are determined maximizing the following likelihood: 
\begin{equation}
\ln \mathcal{L}  =  \sum_k \ln f(E_{\FD k},S_{\SD k})  =  \sum_k \ln \left( \frac{1}{N} \sum_i G(E_{\FD k}|E_{\FD i},\sigma_{\FD i}) ~ G(S_{\SD k} |  S(E_{\FD i}), \sigma_{\SD i} ) \right) 
\end{equation}
where the sum with index $k$ runs over the hybrid events selected for the energy calibration. These events have $E_{\FD k} > 3 {\times} 10^{18}$~eV and the SD is fully efficient. The sum over $i$, that defines the probability density function, extends to lower energies to capture the fluctuations of the energy estimators. It is sufficient to select events with energies $E_{\FD i} > 2.5 {\times} 10^{18}$ eV, for which the detection efficiency is still very close to 100\% (see section~\ref{sec:raw}) and the power law $E = A S^B$ is still valid (see section~\ref{sec:EnCalib}). Given the good FD energy resolution ($\approx 7.4\%$), events below this energy would give a negligible contribution to the likelihood. 

Once the parameters that best describe the data are determined, the goodness of the fit is estimated through the following deviance definition,
\begin{equation}
D = 2~\left( - \sum_k \ln f(E_{\FD k},S_{\SD k}) + \sum_k \ln f(E_{\FD k},S(E_{\FD k})) \right) \ ,
\end{equation}
where the second term represents an ideal model in which the shower size distribution   is perfectly described by the fitted power law.

\end{widetext}
\section{Details of the forward-folding procedure}
\label{app:unfold} 
\begin{table*}[t]
\tiny
\begin{center}
 \resizebox{\textwidth}{!}{%
\begin{tabular}{c|c|c|c|c|c|c|c|c|c|c|c|c|c|c|c|c|c|c|c|c|c|c|c|c|c}
 \diagbox{$i$}{$j$} 
 & $18.05$ & $18.15$ & $18.25$ & $18.35$& $18.45$ & $18.55$ & $18.65$ & $18.75$ & $18.85$ & $18.95$ & $19.05$ & $19.15$ & $19.25$ & $19.35$ & $19.45$ & $19.55$ & $19.65$ & $19.75$ & $19.85$ & $19.95$ & $20.05$ & $20.15$ & $20.25$ & $20.35$ & $20.45$    \\
\hline
18.05  & 0.244  & 0.127  & 0.038  & 0.008  & 0.001  & 0.  & 0.  & 0.  & 0.  & 0.  & 0.  & 0.  & 0.  & 0.  & 0.  & 0.  & 0.  & 0.  & 0.  & 0.  & 0.  & 0.  & 0. & 0.  & 0. \\
18.15  & 0.237  & 0.310  & 0.165  & 0.048  & 0.008  & 0.  & 0.  & 0.  & 0.  & 0.  & 0.  & 0.  & 0.  & 0.  & 0.  & 0.  & 0.  & 0.  & 0.  & 0.  & 0.  & 0.  & 0.  & 0.  & 0.\\
18.25  & 0.051  & 0.260  & 0.368  & 0.204  & 0.053  & 0.006  & 0.  & 0.  & 0.  & 0.  & 0.  & 0.  & 0.  & 0.  & 0.  & 0.  & 0.  & 0.  & 0.  & 0.  & 0.  & 0.  & 0.  & 0.  & 0.\\
18.35  & 0.001  & 0.044  & 0.260  & 0.412  & 0.230  & 0.047  & 0.004  & 0.  & 0.  & 0.  & 0.  & 0.  & 0.  & 0.  & 0.  & 0.  & 0.  & 0.  & 0.  & 0.  & 0.  & 0.  & 0.  & 0.  & 0.\\
18.45  & 0.  & 0.  & 0.033  & 0.238  & 0.439  & 0.234  & 0.039  & 0.002  & 0.  & 0.  & 0.  & 0.  & 0.  & 0.  & 0.  & 0.  & 0.  & 0.  & 0.  & 0.  & 0.  & 0.  & 0.   & 0.  & 0.\\
18.55  & 0.  & 0.  & 0.  & 0.021  & 0.222  & 0.468  & 0.235  & 0.030  & 0.  & 0.  & 0.  & 0.  & 0.  & 0.  & 0.  & 0.  & 0.  & 0.  & 0.  & 0.  & 0.  & 0.  & 0.  & 0.  & 0.\\
18.65  & 0.  & 0.  & 0.  & 0.  & 0.016  & 0.219  & 0.497  & 0.228  & 0.021  & 0.  & 0.  & 0.  & 0.  & 0.  & 0.  & 0.  & 0.  & 0.  & 0.  & 0.  & 0.  & 0.  & 0.  & 0.  & 0.\\
18.75  & 0.  & 0.  & 0.  & 0.  & 0.  & 0.012  & 0.212  & 0.529  & 0.220  & 0.013  & 0.  & 0.  & 0.  & 0.  & 0.  & 0.  & 0.  & 0.  & 0.  & 0.  & 0.  & 0.  & 0.  & 0.  & 0.\\
18.85  & 0.  & 0.  & 0.  & 0.  & 0.  & 0.  & 0.008  & 0.205  & 0.564  & 0.210  & 0.008  & 0.  & 0.  & 0.  & 0.  & 0.  & 0.  & 0.  & 0.  & 0.  & 0.  & 0.  & 0.  & 0.  & 0.\\
18.95  & 0.  & 0.  & 0.  & 0.  & 0.  & 0.  & 0.  & 0.005  & 0.191  & 0.600  & 0.198  & 0.004  & 0.  & 0.  & 0.  & 0.  & 0.  & 0.  & 0.  & 0.  & 0.  & 0.  & 0.  & 0.  & 0.\\
19.05  & 0.  & 0.  & 0.  & 0.  & 0.  & 0.  & 0.  & 0.  & 0.003  & 0.174  & 0.637  & 0.187  & 0.002  & 0.  & 0.  & 0.  & 0.  & 0.  & 0.  & 0.  & 0.  & 0.  & 0.  & 0.  & 0.\\
19.15  & 0.  & 0.  & 0.  & 0.  & 0.  & 0.  & 0.  & 0.  & 0.  & 0.001  & 0.157  & 0.669  & 0.178  & 0.001  & 0.  & 0.  & 0.  & 0.  & 0.  & 0.  & 0.  & 0.  & 0.  & 0.  & 0.\\
19.25  & 0.  & 0.  & 0.  & 0.  & 0.  & 0.  & 0.  & 0.  & 0.  & 0.  & 0.  & 0.139  & 0.694  & 0.169  & 0.  & 0.  & 0.  & 0.  & 0.  & 0.  & 0.  & 0.  & 0.  & 0.  & 0.\\
19.35  & 0.  & 0.  & 0.  & 0.  & 0.  & 0.  & 0.  & 0.  & 0.  & 0.  & 0.  & 0.  & 0.126  & 0.712  & 0.163  & 0.  & 0.  & 0.  & 0.  & 0.  & 0.  & 0.  & 0.  & 0.  & 0.\\
19.45  & 0.  & 0.  & 0.  & 0.  & 0.  & 0.  & 0.  & 0.  & 0.  & 0.  & 0.  & 0.  & 0.  & 0.118  & 0.722  & 0.161  & 0.  & 0.  & 0.  & 0.  & 0.  & 0.  & 0. & 0.  & 0. \\
19.55  & 0.  & 0.  & 0.  & 0.  & 0.  & 0.  & 0.  & 0.  & 0.  & 0.  & 0.  & 0.  & 0.  & 0.  & 0.114  & 0.727  & 0.165  & 0.  & 0.  & 0.  & 0.  & 0.  & 0.  & 0.  & 0.\\
19.65  & 0.  & 0.  & 0.  & 0.  & 0.  & 0.  & 0.  & 0.  & 0.  & 0.  & 0.  & 0.  & 0.  & 0.  & 0.  & 0.111  & 0.730  & 0.177  & 0.  & 0.  & 0.  & 0.  & 0.  & 0.  & 0.\\
19.75  & 0.  & 0.  & 0.  & 0.  & 0.  & 0.  & 0.  & 0.  & 0.  & 0.  & 0.  & 0.  & 0.  & 0.  & 0.  & 0.  & 0.104  & 0.727  & 0.178  & 0.  & 0.  & 0.  & 0.  & 0.  & 0.\\
19.85  & 0.  & 0.  & 0.  & 0.  & 0.  & 0.  & 0.  & 0.  & 0.  & 0.  & 0.  & 0.  & 0.  & 0.  & 0.  & 0.  & 0.  & 0.095  & 0.727  & 0.178  & 0.  & 0.  & 0.  & 0.  & 0.\\
19.95  & 0.  & 0.  & 0.  & 0.  & 0.  & 0.  & 0.  & 0.  & 0.  & 0.  & 0.  & 0.  & 0.  & 0.  & 0.  & 0.  & 0.  & 0.  & 0.094  & 0.727  & 0.178  & 0.  & 0.  & 0.  & 0.\\
20.05  & 0.  & 0.  & 0.  & 0.  & 0.  & 0.  & 0.  & 0.  & 0.  & 0.  & 0.  & 0.  & 0.  & 0.  & 0.  & 0.  & 0.  & 0.  & 0.  & 0.094  & 0.727  & 0.178  & 0.  & 0.  & 0.\\
20.15  & 0.  & 0.  & 0.  & 0.  & 0.  & 0.  & 0.  & 0.  & 0.  & 0.  & 0.  & 0.  & 0.  & 0.  & 0.  & 0.  & 0.  & 0.  & 0.  & 0.  & 0.094  & 0.727  & 0.178  & 0.  & 0.\\
20.25  & 0.  & 0.  & 0.  & 0.  & 0.  & 0.  & 0.  & 0.  & 0.  & 0.  & 0.  & 0.  & 0.  & 0.  & 0.  & 0.  & 0.  & 0.  & 0.  & 0.  & 0.  & 0.094  & 0.727  & 0.178  & 0.\\
20.35  & 0.  & 0.  & 0.  & 0.  & 0.  & 0.  & 0.  & 0.  & 0.  & 0.  & 0.  & 0.  & 0.  & 0.  & 0.  & 0.  & 0.  & 0.  & 0.  & 0.  & 0.  & 0.  & 0.094  & 0.727  & 0.178\\
20.45  & 0.  & 0.  & 0.  & 0.  & 0.  & 0.  & 0.  & 0.  & 0.  & 0.  & 0.  & 0.  & 0.  & 0.  & 0.  & 0.  & 0.  & 0.  & 0.  & 0.  & 0.  & 0.  & 0.  & 0.094  & 0.727\\
20.55  & 0.  & 0.  & 0.  & 0.  & 0.  & 0.  & 0.  & 0.  & 0.  & 0.  & 0.  & 0.  & 0.  & 0.  & 0.  & 0.  & 0.  & 0.  & 0.  & 0.  & 0.  & 0.  & 0.  & 0.  & 0.094\\
\end{tabular}
}
\caption{\small{Elements of the response matrix $R_{ij}$.}}
\label{tab1:Rij}
\end{center}
\end{table*}%

The response matrix elements, $R_{ij}$, are the conditional probabilities that the reconstructed energy $E_\textrm{SD}$ of an event falls into the bin $i$ given that the true energy $E$ should have been in the bin $j$:
\begin{equation}
R_{ij}= \hspace{6.5cm}
\label{eqn:Rij}
\end{equation}
\begin{equation*}
\frac{\int_{\Delta E_i}\dif E_\textrm{SD}\int_{\Delta E_j}\dif E\int\dif\Omega\cos\theta\kappa(E_\textrm{SD}|E,\theta)\epsilon(E,\theta)J(E;\mathbf{s})}{\int_{\Delta E_j}\dif E\int\dif\Omega\cos\theta~J(E;\mathbf{s})}
\end{equation*}
where the zenithal part of the angular integration is performed up to $\thetamax=60^\circ$. 
It is used to calculate  
the number of events which is expected between $E_i$ and $E_i+\Delta E_i$, $\nu_i=\sum_j R_{ij}\mu_j$, where $\mu_j$
are the ones expected without detector effects between $E_j$ and $E_j+\Delta E_j$.

For sufficiently small bin sizes so that the values of $\kappa$ and, below full efficiency, $\epsilon$ are approximately constant, the dependence on $J$ cancels out. 
The forward-folding fit is thus performed under this approximation (we use $\mathrm \Delta \log_{10} E = 0.01$), allowing the re-calculation of the matrix elements to
 be avoided at each step of the fit (leading to the definition of a matrix $R'_{kl}$). We then integrate the expected number of events $\mu_l$ and $\nu_k$ ($\nu_k= \sum_l R'_{kl}\mu_l$) obtained with the best fit parameters to calculate the number of events in the $\mathrm \Delta \log_{10} E = 0.1$ bins and, from them, the correction coefficients.
The $R_{ij}$  matrix calculated in the $\mathrm \Delta \log_{10} E = 0.1$ bins according to Eq.~\eqref{eqn:Rij} is reported in Table~\ref{tab1:Rij}. It can be used for testing any model $J(E;\mathbf{s})$ that fits reasonably well the data.

The observed number of events as a function of energy is a single measurement of a random process for which the p.d.f. for observing the set of values $N_i$ given a set of expectations $\nu_i$ follows a multinomial distribution. The total number of events $N=\sum_iN_i$ being itself a random variable from a Poisson process, the resulting joint p.d.f. for the histogram is the product, over the energy bins considered, of the Poisson probabilities to observe $N_i$ events in each bin given an expectation $\nu_i(\mathbf{s})$. Dropping the constant terms with respect to $\mathbf{s}$, the likelihood function to be minimized then reads:
\begin{equation}
-\ln{\mathcal{L}(\mathbf{s})}=\sum_i\left(\nu_i(\mathbf{s})-N_i\ln{\nu_i(\mathbf{s})}\right).
\label{eqn:logL}
\end{equation}

Once the best-fit parameters $\mathbf{s}$ are derived, the correction factors $c_i$ are then obtained from the estimates of $\mu_i$ and $\nu_i$ as $c_i=\mu_i/\nu_i$. The goodness-of-fit statistic is provided by its deviance, 
\begin{eqnarray}
D&=&- 2 \ln{\frac{\mathcal{L}(\mathbf{N}|\boldsymbol{\nu}(\mathbf{s}))}{\mathcal{L}(\mathbf{N}|\mathbf{N})}} \\ \nonumber 
&=& 2 \sum_i \left(\nu_i(\mathbf{s})-N_i+N_i\ln{\left(\frac{N_i}{\nu_i(\mathbf{s})}\right)}\right),
\label{eqn:logLratio}
\end{eqnarray}
which asymptotically behaves as a $\chi^2$ variable with $k-s$ degrees of freedom, where $k$ is the number of measurements (the number of energy bins) and $s$ the number of model parameters~\cite{BC1984}. 

The uncertainties in the energy spectrum that is recovered follow from the covariance matrix of the $\mu_i$ estimators, which for a Poisson process is given by
\begin{equation}
\mathrm{cov}[\mu_i,\mu_j]=\mathrm{cov}[c_i\nu_i,c_j\nu_j]=c_i\nu_i\delta_{ij},
\label{eqn:cov_mu}
\end{equation}
so that the uncertainties $\sigma_{J_i}$ scale as $\sqrt{c_iN_i}/(\mathcal{E}~\Delta E_i)$. The confidence intervals reported in this paper are then estimated by calculating the 2-sided 16\% quantiles of the underlying p.d.f while the $90\%$ confidence-level limits are calculated according to~\cite{FC1988}. 

In section~\ref{sec:declination}, the energy spectrum is reported for specific ranges of declinations. The forward-folding procedure used to infer the different spectra is then identical, adapting the response matrix to the declination range under consideration in the following way. The directional raw energy spectrum, $J^{\mathrm{raw}}(E_\textrm{SD},\alpha,\delta;\mathbf{s})$, is related to the underlying energy spectrum through
\begin{widetext} 
\begin{equation}
J^{\mathrm{raw}}(E_\textrm{SD},\alpha,\delta;\mathbf{s})=\frac{1}{\Delta t}\int\dif t\dif E\dif\Omega\cos\theta~\epsilon(E,\theta)\kappa(E_\textrm{SD}|E,\theta)J(E,\alpha,\delta;\mathbf{s})\delta(\Omega-\Omega(\alpha,\delta,t)),
\label{eqn:Jfolded_neq}
\end{equation}
where the Dirac function guarantees that only the local angles $(\theta,\varphi)$ pointing to the $(\alpha,\delta)$ considered contribute to the integration at time $t$. By applying the Dirac function, and by using Eq.~\eqref{eqn:local-eq-trans3}, the response matrix elements can be shown to be
\begin{equation}
R'_{ij}=\frac{\int_{\Delta E_i}\dif E_\textrm{SD}\int_{\Delta E_j}\dif E\int\dif h\int\dif\delta\cos{\delta}\cos{(\theta(\delta,h))}\epsilon(E,\theta(\delta,h))\kappa(E_\textrm{SD}|E,\theta(\delta,h))}{\int\dif h\int\dif\delta\cos{\delta}\cos{(\theta(\delta,h))}H(\cos{(\theta(\delta,h))}-\cos\thetamax)\Delta E_j},
\label{eqn:Rijprime}
\end{equation}
\end{widetext}
where the time integration has been substituted for an integration over $h=\alpha_0(t)-\alpha$ between $-\pi$ and $\pi$, and the Heaviside function, $H$, guarantees that only the effective zenithal range $[0,\thetamax]$ contributes to the integrations. Note that the integration over declination covers only the range under consideration in the numerator, while it covers the whole field of view in the denominator. 

The functional shape $J(E;\mathbf{s})$ that best describes the energy spectrum is selected by adopting for the test statistics (TS) 
the likelihood ratio between an alternative model and a reference one. For each model, the forward-folding fit is carried out and the corresponding likelihood value is recorded. 
The TS values are first converted into $p$-values by integrating the distributions of the TS for the reference 
model above the value obtained in data. The $p$-values are then converted into significances assuming 1-sided Gaussian distributions. 

The reference model, Eq.~\eqref{eqn:J1}, is the one that we have used for over a decade. With the new model, Eq.~\eqref{eqn:J2}, which has two additional parameters to define the feature at 
$1.3 \times 10^{19}$ eV, TS $\simeq 20$.
As the two hypotheses are not nested, the likelihood ratio distribution is built by Monte-Carlo to calculate the corresponding $p$-value. 
Mock samples of reconstructed energies were simulated 
\begin{figure}[t]
        \centering
        \hspace{-0.7cm} \includegraphics[width=0.5\textwidth]{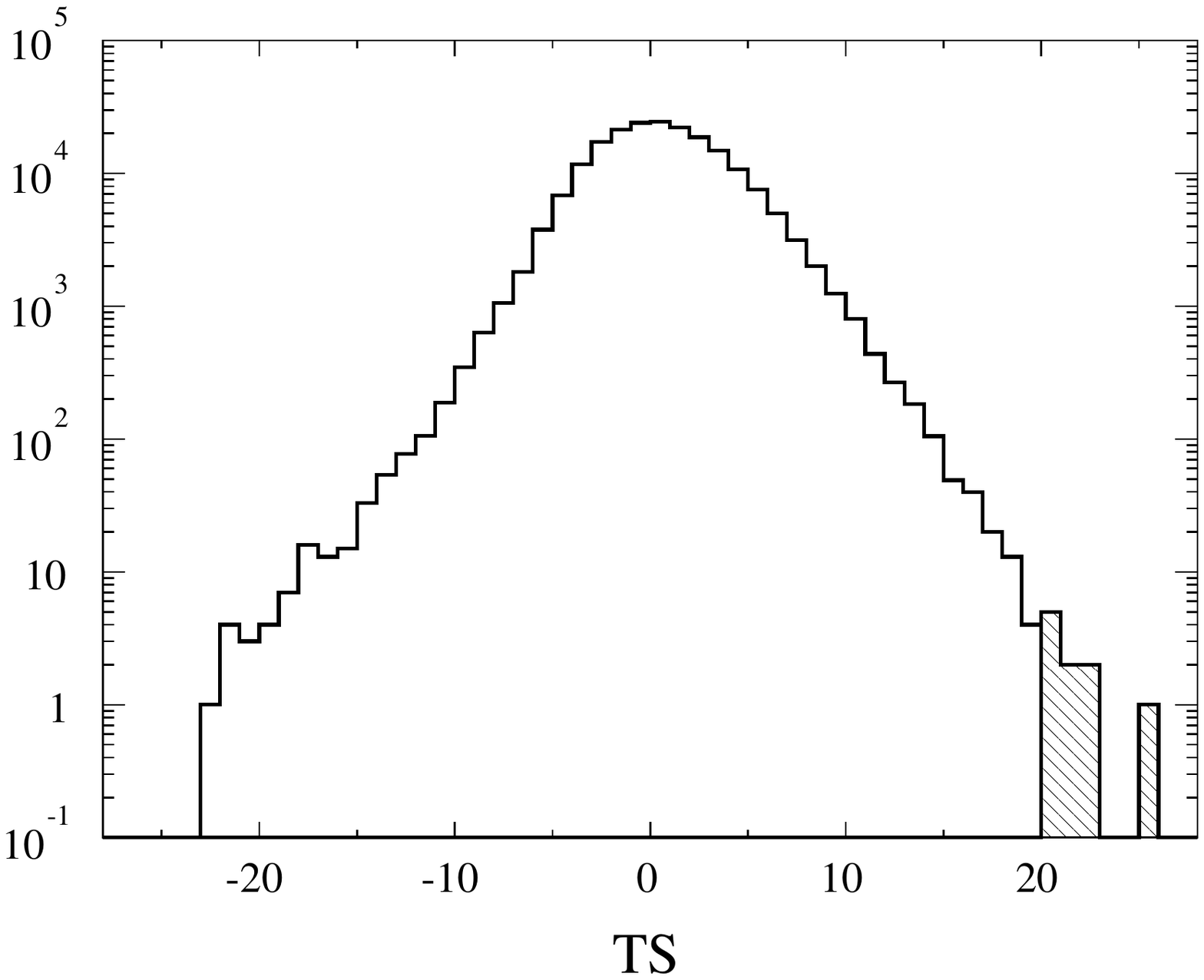}
        \vspace{-0.3cm}
	\caption{\small{TS distribution.
	}} 
\label{fig:TS}
\end{figure}
(with the reference model for $J(E;\mathbf{s}))$ by drawing at random a total number of events similar to that of the data. 
The spectra of these samples were then reconstructed according to the method presented in Sec.~IV.B with both the reference model and the alternative one. For each sample, 
the TS has been recorded. The distribution obtained is shown in Fig.~\ref{fig:TS}.  There are only 10 counts out of 210,000 above TS $\simeq 20$, thus enabling us to reject the reference model at the $3.9 \sigma$ confidence level. 

The test statistic has also been adopted to test our sensitivity to the speed of the transitions. In this case the reference model is Eq.~\eqref{eqn:J2}, 
with all paramaters $w_{ij}$ fixed to 0.05, and the alternative one obtained leaving them free in the fit.  Since the hypotheses are in this case nested, Wilks theorem applies. For each of the models tested,  the increase in TS is less than $2 \sigma$.


\section{Spectrum data list}
\label{app:spectrumdata}

The spectrum data points with their statistical uncertainties are collected in Table~\ref{tab:Jdata} together with the number of events ($N$) and the corrected number of events ($N_{\rm corr}$). 
Upper limits are at the $90\%$ confidence level. 

\begin{widetext}

\begin{table}[h]
\caption{Spectrum data. The corrected number of events are rounded to the closest integers.}
\label{tab:Jdata}
\begin{ruledtabular}
\begin{tabular}{l c c c}
$\mathrm{log}_{10}\left(E/{\rm eV}\right)$ & $N$ & $N_{\rm corr}$ & $J \pm \sigma_{\rm stat}$  \\
   &  & &  [${\rm km}^{-2}~{\rm yr}^{-1}~{\rm sr}^{-1}~{\rm eV}^{-1}$]  \\
\colrule
&&&\\[-0.8em]
18.45 & 83143 &  76176 & $\left( 1.9383 ~^{+0.0067}_{-0.0067} \right) {\times} 10^{-18}$ \\
&&&\\[-0.8em]
18.55 & 47500  &  44904 & $\left( 9.076 ~^{+0.042}_{-0.041} \right) {\times} 10^{-19}$ \\
&&&\\[-0.8em]
18.65 & 28657 &  26843 & $\left( 4.310 ~^{+0.025}_{-0.025} \right) {\times} 10^{-19}$ \\
&&&\\[-0.8em]
18.75 & 17843 &  16970 & $\left( 2.164 ~^{+0.016}_{-0.016} \right) {\times} 10^{-19}$ \\
&&&\\[-0.8em]
18.85 & 12435 &  12109 & $\left( 1.227 ~^{+0.011}_{-0.011} \right) {\times} 10^{-19}$ \\
&&&\\[-0.8em]
18.95 & 8715  &  8515 & $\left( 6.852 ~^{+0.074}_{-0.073} \right) {\times} 10^{-20}$ \\
&&&\\[-0.8em]
19.05 & 6050  &  5939 & $\left( 3.796 ~^{+0.049}_{-0.049} \right) {\times} 10^{-20}$ \\
&&&\\[-0.8em]
19.15 & 4111  &  4048 & $\left( 2.055 ~^{+0.032}_{-0.032} \right) {\times} 10^{-20}$  \\
&&&\\[-0.8em]
19.25 & 2620 &  2567  & $\left( 1.035 ~^{+0.021}_{-0.020} \right) {\times} 10^{-20}$ \\
&&&\\[-0.8em]
19.35 & 1691 &  1664 & $\left( 0.533 ~^{+0.013}_{-0.013} \right) {\times} 10^{-20}$ \\
&&&\\[-0.8em]
19.45 & 991 &  979   & $\left( 2.492 ~^{+0.081}_{-0.079} \right) {\times} 10^{-21}$ \\
&&&\\[-0.8em]
19.55 & 624 &  619   & $\left( 1.252 ~^{+0.052}_{-0.050} \right) {\times} 10^{-21}$ \\
&&&\\[-0.8em]
19.65 & 372 &  373   & $\left( 5.98 ~^{+0.32}_{-0.31} \right) {\times} 10^{-22}$  \\
&&&\\[-0.8em]
19.75 & 156 &  152  & $\left( 1.93 ~^{+0.17}_{-0.15} \right) {\times} 10^{-22}$ \\
&&&\\[-0.8em]
19.85 & 83 &  80   & $\left( 8.10 ~^{+0.99}_{-0.88} \right) {\times} 10^{-23}$\\
&&&\\[-0.8em]
19.95 & 24 &  23  & $\left( 1.86 ~^{+0.46}_{-0.38} \right) {\times} 10^{-23}$\\
&&&\\[-0.8em]
20.05 & 9 &  9   & $\left( 5.5 ~^{+2.5}_{-1.8} \right) {\times} 10^{-24}$\\
&&&\\[-0.8em]
20.15 & 6 &  6   & $\left( 2.9 ~^{+1.7}_{-1.2}\right) {\times} 10^{-24}$\\ 
&&&\\[-0.8em]
20.25 & 0 &  0   & $< 9.5 {\times} 10^{-25}$\\ 
&&&\\[-0.8em]
20.35 & 0 &  0   & $< 7.5 {\times} 10^{-25}$ 
\end{tabular}
\end{ruledtabular}
\end{table}

\end{widetext}


\section{Directional exposure and expectations from anisotropies for the declination dependence of the spectrum}
\label{app:direxp}

We give in this appendix the technical details used in section~\ref{sec:declination} to derive the directional exposure in equatorial coordinates and to infer the expected spectra from the anisotropy measurements. 

The directional exposure results from the time-integration of the directional aperture of an active region of the array, which is considered here as running constantly. To first order, it is well approximated by $A_{\mathrm{cell}}\cos\theta$, but it is actually slightly larger for showers arriving from the downhill direction due to the small tilt of the Observatory towards the south-east. Although small, it is important to account here for this effect because the directional exposure estimated by neglecting it would distort the cosmic-ray flux in an overall dipolar-shaped way with an amplitude of $\simeq 0.5\%$ along the declination coordinate. That would consequently bias the energy spectrum estimates in different declination bands. For an angle of incidence $(\theta,\varphi)$, the directional aperture per cell is thus given, on average, by 
\begin{equation}
\label{eqn:cell_aperture}
A_\mathrm{cell}(\theta,\varphi)\simeq 1.95~\mathrm{km}^2\left(1+\theta_{\mathrm{tilt}}\tan{\theta}\cos{(\varphi-\varphi_{\mathrm{tilt}})}\right)\cos{\theta},
\end{equation}
with $\theta_{\mathrm{tilt}}=0.2^\circ$ the average inclination to the vertical and $\varphi_{\mathrm{tilt}}=-30^\circ$ the direction in azimuth counter-clockwise from east. From this expression, the directional exposure is finally estimated by making use of the time-dependent transformation rules relating the equatorial coordinates $(\alpha,\delta)$ to the corresponding local ones  $(\theta(\alpha,\delta,t),\varphi(\alpha,\delta,t))$ at time $t$,
\begin{eqnarray}
\label{eqn:local-eq-trans1}
\hspace{-0.5cm} \sin{\theta}\cos{\varphi}&=&-\cos{\delta}\sin{(\alpha_0(t)-\alpha)},  \\
\label{eqn:local-eq-trans2}
\hspace{-0.5cm} \sin{\theta}\sin{\varphi}&=&\cos{\ell}\sin{\delta}-\sin{\ell}\cos{\delta}\cos{(\alpha_0(t)-\alpha)}, \\
\label{eqn:local-eq-trans3}
\hspace{-0.5cm} \cos{\theta}&=& \cos{\ell}\cos{\delta}\cos{(\alpha_0(t)-\alpha)}+\sin{\ell}\sin{\delta},
\end{eqnarray}
where $\alpha_0(t)$ is the local sidereal time and $\ell$ the latitude of the Observatory. Since the time integration of $A_\mathrm{cell}(\theta(\alpha,\delta,t),\varphi(\alpha,\delta,t))$ depends here only on the difference $\alpha_0(t)-\alpha$, it can be substituted for an integration over the hour angle $h=\alpha_0(t)-\alpha$. Following~\cite{Sommers:2000us}, the constraining $\theta_{\text{max}}$ ($60^\circ$ here) translates to an integration over $h$ ranging from $-h_{\text{m}}$ to $h_{\text{m}}$ with $h_{\text{m}}=\arccos{[(\cos{\theta_{\text{max}}-\sin{\ell}\sin{\delta}})/(\cos{\ell}\cos{\delta})]}$, with the additional constraining that $h_{\text{m}}=0$ for declinations giving rise to an argument greater than $1$ in the arccos function, and $h_{\text{m}}=\pi$ for declinations giving rise to an argument smaller than $-1$. This leads to the expression
\begin{widetext}
\begin{eqnarray}
\label{eqn:omega}
\omega(\delta)\propto\cos{\ell}\cos{\delta}\sin{h_{\text{m}}(\delta)} + h_{\text{m}}(\delta)\sin{\ell}\sin{\delta} 
+\theta_{\mathrm{tilt}}\sin{\varphi_{\mathrm{tilt}}} (h_{\text{m}}(\delta)\cos{\ell}\sin{\delta}-\sin{\ell}\cos{\delta}\sin{h_{\text{m}}(\delta)}),
\end{eqnarray}
which is then used to determine numerically the $\delta_k$ values in Eq.~\eqref{eqn:exposure_domega}. The normalization of $\omega$ is such that $\mathcal{E}=2\pi A_0\Delta t\int_{\delta_0}^{\delta_3}\dif\delta\cos{\delta}~\omega(\delta)$.

To compare the unfolded spectra obtained for each declination band with those expected from the anisotropy measurements characterized by a vector, $\mathbf{d}$, with amplitude $d$ and equatorial directions $(\alpha_{\mathrm{d}},\delta_{\mathrm{d}})$, we make use of the ansatz
\begin{eqnarray}
\label{eqn:dipansatz}
\frac{\dif N_{i}}{\dif\Omega'}=A_0\Delta t~ \omega(\delta)\left(1+\mathbf{d}(\alpha_{\mathrm{d}},\delta_{\mathrm{d}})\cdot\mathbf{n}(\alpha,\delta)\right)\int_{\Delta E_i}\dif E~J_0(E),
\end{eqnarray}
considering $d$ constant within $\Delta E$, and with $\dif\Omega'=\cos\delta\dif\delta\dif\alpha$. 
The choice of the energy bins, $\Delta E_i$, follows from that performed in~\cite{AugerAnis2018} in which the values of $\mathbf{d}$ are reported. The integration over right ascension selects the isotropic component and the dipole component along the $z-$axis defining the declination:
\begin{eqnarray}
\label{eqn:dipansatz2}
\frac{\dif N_{i}}{\dif\sin\delta}=2\pi A_0\Delta t~ \omega(\delta)\left(1+d\sin\delta_{\mathrm{d}}\sin\delta\right)\int_{\Delta E_i}\dif E~J_0(E).
\end{eqnarray}
Finally, the expected number of events in a given declination band is obtained through a final integration over declination:
\begin{eqnarray}
\label{eqn:dipansatz3}
N_{ik}=\left(\frac{\mathcal{E}}{3}+2\pi A_0\Delta t~d\sin\delta_{\mathrm{d}}\int_{\delta_{k-1}}^{\delta_k}\dif\delta ~\cos\delta\sin\delta\omega(\delta)\right)\int_{\Delta E_i}\dif E~J_0(E),
\end{eqnarray}
where the division in three declination bands is explicitly used. The expected intensity is then $J_{ik}=N_{ik}/(\mathcal{E}/3)/\Delta E_i$ in each energy bin. The ratios of intensities, $r_k$, depicted by the lines in Fig.~\ref{fig:DeclinationSpectrum}-right are thus drawn from
\begin{equation}
\label{eqn:j_dec}
r=3~\frac{\int_{\delta_{k-1}}^{\delta_k}\dif\delta\cos{\delta}~\omega(\delta)+d\sin{\delta_{\mathrm{d}}} \int_{\delta_{k-1}}^{\delta_k}\dif\delta\cos{\delta}\sin{\delta}~\omega(\delta)}{\int_{\delta_0}^{\delta_3}\dif\delta\cos{\delta}~\omega(\delta)+d\sin{\delta_{\mathrm{d}}} \int_{\delta_0}^{\delta_3}\dif\delta\cos{\delta}\sin{\delta}~\omega(\delta)}.
\end{equation}
\end{widetext}

Note that in these calculations, the dead times of the SD array have not been considered. As discussed in Appendix~\ref{app:sin2}, they lead to variations of the event rate of order of a few $10^{-3}$, imprinting small harmonic dependencies in right ascension to the directional exposure, $\omega'(\alpha,\delta)=\omega(\delta)(1+\delta\omega(\alpha,\delta))$. Throughout the integrations over $2\pi$ in right ascension, these harmonic dependencies cancel exactly when coupled to the isotropic component of the intensity, and give rise to second-order corrections when coupled to the dipolar fraction of amplitude $d$. The term $\delta\omega(\alpha,\delta)$ can thus be safely neglected here. 

\input{acknowledgments.tex}


\end{document}

%% file: revtex_authorlist.tex

\author{A.~Aab}
\affiliation{IMAPP, Radboud University Nijmegen, Nijmegen, The Netherlands}

\author{P.~Abreu}
\affiliation{Laborat\'orio de Instrumenta\c{c}\~ao e F\'\i{}sica Experimental de Part\'\i{}culas -- LIP and Instituto Superior T\'ecnico -- IST, Universidade de Lisboa -- UL, Lisboa, Portugal}

\author{M.~Aglietta}
\affiliation{Osservatorio Astrofisico di Torino (INAF), Torino, Italy}
\affiliation{INFN, Sezione di Torino, Torino, Italy}

\author{J.M.~Albury}
\affiliation{University of Adelaide, Adelaide, S.A., Australia}

\author{I.~Allekotte}
\affiliation{Centro At\'omico Bariloche and Instituto Balseiro (CNEA-UNCuyo-CONICET), San Carlos de Bariloche, Argentina}

\author{A.~Almela}
\affiliation{Instituto de Tecnolog\'\i{}as en Detecci\'on y Astropart\'\i{}culas (CNEA, CONICET, UNSAM), Buenos Aires, Argentina}
\affiliation{Universidad Tecnol\'ogica Nacional -- Facultad Regional Buenos Aires, Buenos Aires, Argentina}

\author{J.~Alvarez Castillo}
\affiliation{Universidad Nacional Aut\'onoma de M\'exico, M\'exico, D.F., M\'exico}

\author{J.~Alvarez-Mu\~niz}
\affiliation{Instituto Galego de F\'\i{}sica de Altas Enerx\'\i{}as (IGFAE), Universidade de Santiago de Compostela, Santiago de Compostela, Spain}

\author{R.~Alves Batista}
\affiliation{IMAPP, Radboud University Nijmegen, Nijmegen, The Netherlands}

\author{G.A.~Anastasi}
\affiliation{Universit\`a Torino, Dipartimento di Fisica, Torino, Italy}
\affiliation{INFN, Sezione di Torino, Torino, Italy}

% ...

\collaboration{The Pierre Auger Collaboration}
\email{auger_spokespersons@fnal.gov}
\homepage{http://www.auger.org}
\noaffiliation

%% file: acknowledgments.tex

\section*{Acknowledgments}

\begin{sloppypar}
The successful installation, commissioning, and operation of the Pierre
Auger Observatory would not have been possible without the strong
commitment and effort from the technical and administrative staff in
Malarg\"ue. We are very grateful to the following agencies and
organizations for financial support:
\end{sloppypar}

\begin{sloppypar}
Argentina -- Comisi\'on Nacional de Energ\'\i{}a At\'omica; Agencia Nacional de
Promoci\'on Cient\'\i{}fica y Tecnol\'ogica (ANPCyT); Consejo Nacional de
Investigaciones Cient\'\i{}ficas y T\'ecnicas (CONICET); Gobierno de la
Provincia de Mendoza; Municipalidad de Malarg\"ue; NDM Holdings and Valle
Las Le\~nas; in gratitude for their continuing cooperation over land
access; Australia -- the Australian Research Council; Brazil -- Conselho
Nacional de Desenvolvimento Cient\'\i{}fico e Tecnol\'ogico (CNPq);
Financiadora de Estudos e Projetos (FINEP); Funda\c{c}\~ao de Amparo \`a
Pesquisa do Estado de Rio de Janeiro (FAPERJ); S\~ao Paulo Research
Foundation (FAPESP) Grants No.~2019/10151-2, No.~2010/07359-6 and
No.~1999/05404-3; Minist\'erio da Ci\^encia, Tecnologia, Inova\c{c}\~oes e
Comunica\c{c}\~oes (MCTIC); Czech Republic -- Grant No.~MSMT CR LTT18004,
LM2015038, LM2018102, CZ.02.1.01/0.0/0.0/16{\textunderscore}013/0001402,
CZ.02.1.01/0.0/0.0/18{\textunderscore}046/0016010 and
CZ.02.1.01/0.0/0.0/17{\textunderscore}049/0008422; France -- Centre de Calcul
IN2P3/CNRS; Centre National de la Recherche Scientifique (CNRS); Conseil
R\'egional Ile-de-France; D\'epartement Physique Nucl\'eaire et Corpusculaire
(PNC-IN2P3/CNRS); D\'epartement Sciences de l'Univers (SDU-INSU/CNRS);
Institut Lagrange de Paris (ILP) Grant No.~LABEX ANR-10-LABX-63 within
the Investissements d'Avenir Programme Grant No.~ANR-11-IDEX-0004-02;
Germany -- Bundesministerium f\"ur Bildung und Forschung (BMBF); Deutsche
Forschungsgemeinschaft (DFG); Finanzministerium Baden-W\"urttemberg;
Helmholtz Alliance for Astroparticle Physics (HAP);
Helmholtz-Gemeinschaft Deutscher Forschungszentren (HGF); Ministerium
f\"ur Innovation, Wissenschaft und Forschung des Landes
Nordrhein-Westfalen; Ministerium f\"ur Wissenschaft, Forschung und Kunst
des Landes Baden-W\"urttemberg; Italy -- Istituto Nazionale di Fisica
Nucleare (INFN); Istituto Nazionale di Astrofisica (INAF); Ministero
dell'Istruzione, dell'Universit\'a e della Ricerca (MIUR); CETEMPS Center
of Excellence; Ministero degli Affari Esteri (MAE); M\'exico -- Consejo
Nacional de Ciencia y Tecnolog\'\i{}a (CONACYT) No.~167733; Universidad
Nacional Aut\'onoma de M\'exico (UNAM); PAPIIT DGAPA-UNAM; The Netherlands
-- Ministry of Education, Culture and Science; Netherlands Organisation
for Scientific Research (NWO); Dutch national e-infrastructure with the
support of SURF Cooperative; Poland -Ministry of Science and Higher
Education, grant No.~DIR/WK/2018/11; National Science Centre, Grants
No.~2013/08/M/ST9/00322, No.~2016/23/B/ST9/01635 and No.~HARMONIA
5--2013/10/M/ST9/00062, UMO-2016/22/M/ST9/00198; Portugal -- Portuguese
national funds and FEDER funds within Programa Operacional Factores de
Competitividade through Funda\c{c}\~ao para a Ci\^encia e a Tecnologia
(COMPETE); Romania -- Romanian Ministry of Education and Research, the
Program Nucleu within MCI (PN19150201/16N/2019 and PN19060102) and
project PN-III-P1-1.2-PCCDI-2017-0839/19PCCDI/2018 within PNCDI III;
Slovenia -- Slovenian Research Agency, grants P1-0031, P1-0385, I0-0033,
N1-0111; Spain -- Ministerio de Econom\'\i{}a, Industria y Competitividad
(FPA2017-85114-P and FPA2017-85197-P), Xunta de Galicia (ED431C
2017/07), Junta de Andaluc\'\i{}a (SOMM17/6104/UGR), Feder Funds, RENATA Red
Nacional Tem\'atica de Astropart\'\i{}culas (FPA2015-68783-REDT) and Mar\'\i{}a de
Maeztu Unit of Excellence (MDM-2016-0692); USA -- Department of Energy,
Contracts No.~DE-AC02-07CH11359, No.~DE-FR02-04ER41300,
No.~DE-FG02-99ER41107 and No.~DE-SC0011689; National Science Foundation,
Grant No.~0450696; The Grainger Foundation; Marie Curie-IRSES/EPLANET;
European Particle Physics Latin American Network; and UNESCO.
\end{sloppypar}